\pdfoutput=1
\documentclass{article}
\usepackage[T1]{fontenc}   %
\usepackage{arxiv}
\usepackage[numbers,sort&compress]{natbib}

\newcommand{\biboptions}[1]{}
\newcommand{\journal}[1]{}
\makeatletter\expandafter\gdef\csname ver@svg.sty\endcsname{2099/01/01}\makeatother %

\usepackage{amsmath}
\usepackage{amssymb}

\usepackage{graphicx}
\usepackage{svg}
\usepackage{booktabs}
\usepackage[table]{xcolor}
\usepackage{longtable}
\usepackage{multirow}
\usepackage{xspace}
\usepackage{multicol}
\usepackage{placeins}
\usepackage{float}
\usepackage{makecell}
\usepackage{pgfplots}
\usepgfplotslibrary{statistics}
\pgfplotsset{compat=1.18}
\usepackage{url}
\usepackage{hyperref}

\biboptions{sort&compress}

\newif\ifcombined
\combinedfalse
\newcommand{\ofsupplementary}{\ifcombined\else{} of the Supplementary Material\fi}

\def\locnext{LocNeXt}

\def\swinv2{\mbox{SwinUNETR-V2}} %
\newcommand{\nnresunet}{\mbox{nnResUNet}\xspace} %

\definecolor{citation}{rgb}{0.2, 0.2, 0.2}
\definecolor{gray}{rgb}{0.5, 0.5, 0.5}
\definecolor{lightgray}{rgb}{0.7, 0.7, 0.7}
\definecolor{bronze}{rgb}{0.8, 0.5, 0.2}
\definecolor{darkcerulean}{rgb}{0.03, 0.27, 0.49}
\definecolor{transparent}{rgb}{1.0, 1.0, 1.0}

\def\afterHLineSpacer{ & \vspace{-2.2mm} \\}
\def\vertspace{\vspace{1.2mm}}
\NewDocumentCommand{\subds}{m}{\textcolor{gray}{#1}}
\NewDocumentCommand{\bucketrow}{mmmm}{{\tiny(O:~#1~\textbar~G:~#2~\textbar~A:~#3~\textbar~F:~#4)}}

\journal{Knowledge-Based Systems}

\combinedtrue   %

\AtBeginDocument{\setkeys{Gin}{keepaspectratio, height=0.45\textheight}}

\hypersetup{colorlinks=true, linkcolor=blue!50!black, citecolor=blue!50!black, urlcolor=blue!50!black}

\makeatletter
\newcommand{\arxivignore}[2][]{}
\begingroup
  \let\author\arxivignore \let\affiliation\arxivignore \let\cortext\arxivignore \let\ead\@gobble
  
\gdef\papertitle{Beyond Segmentation: Enriching ImageCAS with Coronary Ostia Landmarks, Rooted Centerline Graphs, and Open-Weight Localization Models}

\author[polimi]{Matteo Leccardi\corref{cor1}}
\ead{matteo.leccardi@polimi.it}
\author[polimi]{Andrea Moglia}
\author[polimi]{Marco Morandini}
\author[polimi]{Michele Brembilla}
\author[polimi]{Alessandro Nicosia}
\author[polimi]{Andrea Penserini}
\author[polimi]{Christian Brandini}
\author[ethz]{Ender Konukoglu}
\author[polimi]{Marco Marcon}
\author[polimi]{Luca Mainardi}
\author[polimi,pavia]{Pietro Cerveri}

\affiliation[polimi]{
  organization={Department of Electronics, Information and Bioengineering,
                Politecnico di Milano},
  addressline={Via Ponzio 34/5},
  city={Milano},
  postcode={20133},
  country={Italy}}

\affiliation[ethz]{
  organization={Computer Vision Laboratory, Department of Information Technology and Electrical Engineering,
                ETH Zurich},
  addressline={Sternwartstrasse 7},
  city={Zurich},
  postcode={8092},
  country={Switzerland}}

\affiliation[pavia]{
  organization={Department of Industrial and Information Engineering,
                University of Pavia},
  addressline={Via A. Ferrata 5},
  city={Pavia},
  postcode={27100},
  country={Italy}}

\cortext[cor1]{Corresponding author.}

\endgroup
\makeatother

\title{\papertitle}
\renewcommand{\shorttitle}{Beyond Segmentation: Enriching ImageCAS with Coronary Ostia Landmarks}
\renewcommand{\headeright}{Preprint}
\renewcommand{\undertitle}{Preprint}
\date{\today}

\author{%
  \parbox{0.94\textwidth}{\centering\normalfont\normalsize
  Matteo Leccardi\textsuperscript{1,*}, Andrea Moglia\textsuperscript{1}, Marco Morandini\textsuperscript{1},
  Michele Brembilla\textsuperscript{1}, Alessandro Nicosia\textsuperscript{1}, Andrea Penserini\textsuperscript{1},
  Christian Brandini\textsuperscript{1}, Ender Konukoglu\textsuperscript{2}, Marco Marcon\textsuperscript{1},
  Luca Mainardi\textsuperscript{1}, Pietro Cerveri\textsuperscript{1,3}\\[0.9em]
  {\small
  \textsuperscript{1}Department of Electronics, Information and Bioengineering, Politecnico di Milano,
  Via Ponzio 34/5, 20133 Milano, Italy\\
  \textsuperscript{2}Computer Vision Laboratory, Department of Information Technology and Electrical Engineering,
  ETH Zurich, Sternwartstrasse 7, 8092 Zurich, Switzerland\\
  \textsuperscript{3}Department of Industrial and Information Engineering, University of Pavia,
  Via A. Ferrata 5, 27100 Pavia, Italy\\[0.4em]
  \textsuperscript{*}Corresponding author: matteo.leccardi@polimi.it}}%
}

\let\arxivappendix\appendix
\renewcommand{\appendix}{\arxivappendix
  \renewcommand{\thesection}{Appendix \Alph{section}}
  \renewcommand{\thetable}{\Alph{section}.\arabic{table}}
  \renewcommand{\thefigure}{\Alph{section}.\arabic{figure}}
  \renewcommand{\theequation}{\Alph{section}.\arabic{equation}}
  \counterwithin*{equation}{section}}

\begin{document}
\maketitle

\begin{abstract}
Coronary ostia and centerlines are key reference points for automated analysis of coronary computed tomography angiography (CCTA), from surgical planning to intraoperative vessel tracking. However, the only public ostium annotations cover 77 non-gated aortic scans, not CCTA, while CAT08 and ASOCA (32 and 40 CCTA scans) mark the ostia only implicitly, as centerline start points. This work enriches ImageCAS, the largest public dataset for coronary artery segmentation, with ostium annotations for its 1000 patients and centerline graphs rooted at these new ostia, making it, to the authors' knowledge, the largest public dataset for ostium localization and the only one with centerlines following the official ImageCAS masks and starting at annotated ostia. Three annotators marked each ostium in the first 500 volumes; in the next 500, the candidates of a model trained on the first batch were corrected (human-in-the-loop). Six 3D localization models, trained on ImageCAS, CAT08 and ASOCA, were benchmarked on 58 held-out volumes (50 ImageCAS, 4 CAT08, 4 ASOCA). \swinv2 and \nnresunet localized all ostia without failure (>5~mm error), with median errors of 0.70--0.87~mm.
Trained on CAT08 and ASOCA alone, \swinv2 fails on 34\% of right and 53\% of left ostia, whereas adding 250 ImageCAS volumes removes all failures.
The centerlines were extracted by a deterministic, quality-controlled minimal-path procedure; on ASOCA, they cover a median of 99.5\% of the reference length, with a median per-patient mean distance of 0.22~mm between matched reference points and centerlines.
Annotations, centerlines, code, and weights are available at \url{https://hal9000-lab.github.io/} .
\end{abstract}


\section{Introduction}

In cardiac imaging, the coronary ostia and the centerlines represent important landmarks from the clinical point of view. The former correspond to the points where the left and right coronary arteries originate from the aorta, while the latter are the topological skeleton of the coronary tree, rooted at the ostia, and representing the vessel path and branching.
In diagnosis, the position and proximal course of the ostia distinguish benign anatomical variants from anomalous coronary origins, a recognized cause of ischemia and sudden cardiac death~\citep{ImageALCAPA}. In surgical planning, the ostia must be located before coronary artery bypass grafting and aortic root surgery, their height relative to the valve annulus predicts the risk of coronary occlusion during transcatheter aortic valve implantation~\citep{aortic_valve_01,aortic_valve_02,aortic_valve_03}, and stenting of aorto-ostial lesions requires the ostium to be placed precisely~\citep{jokhi2009ostial}. In automated image analysis, the ostia are the natural seed points of vessel segmentation and centerline tracking algorithms~\citep{cat08,wolterink2019tracker,Leccardi2023}, and the resulting centerline graphs support stenosis grading~\citep{Cury2022,KIRISLI2013859}, computational hemodynamic modeling~\citep{TAYLOR20132233,doi:10.1148/radiol.14140992,torii2021ctbased}, and revascularization planning~\citep{doi:10.1161/CIR.0000000000001038}. Reliable, automatic localization of the ostia and extraction of the centerlines are therefore the building blocks of AI-based decision-support systems for the analysis of coronary computed tomography angiography (CCTA).

Manual identification of the coronary ostia is time-consuming and prone to intra- and inter-observer variability, since the exact point at which the aorta becomes a coronary artery is anatomically ambiguous; substantial anatomical variability across patients and differences in acquisition protocols add to the difficulty. Automated approaches evolved from probabilistic appearance and shape models~\citep{stat_land_detect_01,stat_land_detect_02,stat_land_detect_03} to deep networks that exploit the full three-dimensional context (Section~\ref{subsec:rw-landmarks}), but the few methods developed specifically for the coronary ostia were mostly trained and evaluated on private data (Section~\ref{subsec:rw-ostia}).

Centerline extraction faces an analogous pattern. Reference centerlines are produced through labor-intensive, multi-observer procedures~\citep{cat08,asoca1}, which have long kept the public datasets that provide them to a few tens of scans, and automated methods depend on the quality of a segmentation mask, on their starting points, or on reference centerlines to be trained (Section~\ref{subsec:rw-centerline-methods}). Extraction that is validated, applied at scale, and anchored at the coronary ostia therefore remains an open problem.

Despite the progress of the methods, public data for this task remain scarce. The only public annotations designed for coronary ostium localization cover 77 aortic CT angiography volumes of ImageTBAD~\citep{yao2021imagetbad}, released by \citet{gajowczyk2022ostia}. These scans were acquired without ECG gating to study aortic dissection, not the coronary arteries, and image the ostia at a lower quality than CCTA: the localizer developed on them reached median errors of 3.48~mm (left) and 2.97~mm (right) on these scans, against 1.14 and 0.98~mm on gated CCTA~\citep{gajowczyk2022ostia}. This work, which addresses ECG-gated CCTA, therefore does not use them (Section~\ref{sec:discussion-implications}). None of the most widely used public CCTA datasets (ASOCA~\citep{asoca1}, CAT08~\citep{cat08}, and ImageCAS~\citep{imagecas}) was curated for ostium localization: CAT08 and ASOCA mark the ostium only implicitly, as the start point of their reference centerlines, in 72 scans (32 and 40), and ImageCAS provides no ostium information. Moreover, CAT08 can no longer be obtained from an official source, which leaves the 40 annotated scans of ASOCA as the only openly accessible CCTA data with ostium information. Finally, none of the dedicated ostium localizers has been released with code or trained weights~\citep{wolterink2020landmarks,jang2021ostia,mostafa2021ostia,gajowczyk2022ostia}, and each was evaluated on its own data and with its own definition of the ostium; to the authors' knowledge, no public benchmark, that is, no shared CCTA test set with openly available models, therefore exists for this task. Likewise, no public coronary centerline graphs correspond to the official ImageCAS segmentation masks: the recent ImageCAS-X~\citep{imagecasx2026} re-annotates the masks of 800 of the ImageCAS scans and derives centerlines automatically from these new masks, starting them at skeleton endpoints near an automatically segmented aorta rather than at annotated ostia. These centerlines thus serve the users of the new masks, whereas a pipeline that starts from a localized ostium, or from the official masks, needs centerlines rooted at the ostium and derived from those masks: the start points of ImageCAS-X are heuristic, not the landmark that a localizer predicts, and the 200 scans it leaves out have no centerlines (Section~\ref{sec:discussion-centerlines}).

This work addresses these gaps. ImageCAS~\citep{imagecas}, to date the largest publicly available dataset for coronary artery segmentation, is extended with (i) left and right coronary ostium annotations for its complete 1000 CCTAs, obtained by multi-annotator consensus on the first 500 volumes and by a human-in-the-loop correction of model-generated candidates on the remaining 500, and (ii) coronary centerline graphs, for 999 of them, rooted at these ostia and extracted from its segmentation masks. On these annotations, together with the implicit ostia of CAT08 and ASOCA harmonized into the same landmark definition, ready-to-use, open-weight deep learning models for automatic coronary ostium localization are trained, benchmarked on held-out volumes of the three datasets, and released together with the annotations, ready to be used in downstream analysis pipelines. The annotations and centerlines are released as standalone files that refer to the original ImageCAS volumes, which users obtain from their original source; the images, masks and reference annotations of the three datasets are not redistributed (Section~\ref{sec:data-availability}). To the authors' knowledge, the resulting dataset is the largest publicly available for ostium localization, and its centerlines are the only public ones that follow the official ImageCAS masks and start at annotated ostia. %

The main contributions of this work are summarized as follows:
\begin{itemize}
    \item a dataset of left and right coronary ostium landmarks for all 1000 ImageCAS CCTAs: 500 curated by multi-annotator consensus and 500 through a human-in-the-loop protocol in which correcting the candidates of a model took about 5 person-hours, against about 144 for the first batch, in which every volume was annotated manually three times;
    \item a dataset of coronary centerline graphs for the same CCTAs (999 of the 1000 patients), together with the deterministic extraction procedure that produces them, rooted at the annotated ostia and validated against the reference centerlines of ASOCA;
    \item a benchmark of six ostium localization models (\swinv2, ResNet50, ResNet101, \nnresunet, MedNeXt-S and MedNeXt-B), with the weights of four of them (all but the MedNeXt models) available as ready-to-use localizers (Section~\ref{sec:data-availability});
    \item an evaluation framework that grades the localization errors against thresholds derived from the ostial anatomy and from the variability among annotators, proposed as a reporting standard, together with ablation studies showing that the new annotations are necessary for the task and giving practical guidance on the use of the models.
\end{itemize}

Figure~\ref{fig:overview-pipeline} gives an overview of the work, and the remainder of this paper is organized as follows. Section~\ref{sec:related-work} reviews the public datasets and the methods for coronary ostia localization and centerline extraction. Section~\ref{sec:methods} describes the ostium annotation protocol, the centerline extraction procedure, and the benchmark models. Section~\ref{sec:results} reports the annotation statistics, the benchmark results on ostia localization, and the validation of the centerlines, which are then discussed in Section~\ref{sec:discussion}. Section~\ref{sec:conclusions} concludes the manuscript. The availability of the data, the code, and the model weights is described in Section~\ref{sec:data-availability}. \ifcombined The appendices collect\else The Supplementary Material collects\fi{} further details on the data and the methods, the analysis of the ostium definition across datasets, the model of the second annotation stage, the ablation studies, the values published by previous works, and the comparison with the ImageCAS-X centerlines.

\section{Related Work}
\label{sec:related-work}

\subsection{Public datasets for coronary ostium localization}

Among the publicly available datasets, ASOCA \citep{asoca1}, CAT08 \citep{cat08}, and ImageCAS \citep{imagecas} are the most relevant CCTA data collections for coronary artery segmentation; their acquisition and image characteristics are detailed in~\ref{subsec:app-datasets}\ofsupplementary.
The ASOCA dataset by \citet{asoca1} is made up of 60 CCTA volumes, collected in Auckland (New Zealand) for the ASOCA Challenge \citep{asoca_challenge,asoca1}, of which 20 have no publicly available labels. The annotated dataset is split in half between diseased and healthy patients, and the coronary artery segmentation is the result of the majority vote between the labels of three independent annotators \citep{asoca2}.
The CAT08 dataset by \citet{cat08} is made up of 32 CCTA volumes acquired between 2005 and 2006 at the University Medical Center, Rotterdam (The Netherlands) for the MICCAI Coronary Artery Challenge \citep{cat08,cat08_2}, graded by an expert radiologist for image quality and for the presence of calcification. Although CAT08 was released as a public benchmark, its original distribution website is no longer accessible, and we could not locate any official source from which the data can currently be downloaded; the volumes used in this work were obtained while the dataset was still online.
The ImageCAS dataset by \citet{imagecas} is the largest publicly available one for coronary artery segmentation. It is composed of 1000 CCTA volumes, collected from 2012 to 2018 at the Guangdong Provincial People’s Hospital in Guangzhou (China). Left and right coronary arteries were labeled independently by two expert radiologists and then cross-validated; in case of disagreement, a third radiologist made the annotation, and the final agreement was reached by consensus \citep{imagecas}.
The only public annotations designed for coronary ostium localization were released by \citet{gajowczyk2022ostia} for ImageTBAD \citep{yao2021imagetbad}, a dataset of 100 aortic CT angiography volumes of patients with Type-B aortic dissection, acquired without ECG gating at the same hospital as ImageCAS. Both ostia were placed upon agreement of at least two radiologists in 77 volumes, after 23 had been discarded for insufficient quality or artefacts near the ostia; the annotations are distributed through the Synapse platform \citep{gajowczyk2022imagetbadostia}, although the publication gives no link to them.

\subsection{Anatomical landmark detection}
\label{subsec:rw-landmarks}

Initial deep learning approaches to anatomical landmark detection were largely 2D or 2.5D, analyzing orthogonal slices through a volume~\citep{landmark_femur} or using coarse-to-fine cascades of region-proposal and refinement networks~\citep{payer2019}, and could not fully exploit the three-dimensional nature of anatomical structures. Fully 3D approaches perform better, at the price of higher computational costs, memory requirements and training-data demands, which has motivated both patch-based sampling strategies~\citep{zheng2015,cranio2017zhang,landmarks_fetal} and deep reinforcement learning agents that navigate the volume toward the target landmark~\citep{landmark_rl_01,landmark_rl_02,landmark_rl_03}.
More recent works combine classification and regression within fully convolutional architectures: in cardiac imaging, multiple landmarks have been localized simultaneously via 3D displacement-vector regression with per-patch presence classification~\citep{wolterink2018landmarks,wolterink2020landmarks}, and heatmap regression has achieved detection rates comparable to inter-reader variability on other cardiac landmarks~\citep{landmark_cardiac_mri}; in other domains, two-stage detectors have combined region proposal with heatmap-based refinement~\citep{cranio2021chen}, and heatmap regression has been widely adopted, for instance for human pose estimation~\citep{pose_estim_oxford}. Single-stage, heatmap-based formulations, of the kind adopted in this work, avoid both the large tensors that vector-field regression processes across the entire volume and the architectural complexity and training overhead of multi-stage pipelines.

\subsection{Ostia localization}
\label{subsec:rw-ostia}

Multiple approaches to the localization of the coronary ostia have been proposed.
\citet{zheng2011ostia} localized the two native ostia jointly with Marginal Space Learning, by estimating a bounding box whose two opposite faces are centered on the ostia, and the bypass graft ostia on the surface of a segmented aorta.
\citet{liu2014ostia} registered a heart and coronary atlas onto the CCTA volume and refined the ostia with a learned detector based on intensity and vesselness features, using them as starting points of a minimal-path centerline extraction.
\citet{jang2021ostia} proposed a deep reinforcement learning agent that observes 2.5D historical patches, trained and evaluated on CAT08, with a coarse-to-fine search at inference.
\citet{mostafa2021ostia} appended a coordinate-regression head to a U-Net to localize the ostium as the starting point of coronary centerline extraction, on a private in-house dataset.
\citet{gajowczyk2022ostia} proposed a Residual U-Net with a non-trainable Differentiable Spatial to Numerical Transform (DSNT) layer, which extracts the coordinates from the predicted heatmap for a Euclidean loss, while a Jensen-Shannon (JS) divergence term regularizes the heatmap toward a Gaussian distribution; the training was performed on ImageTBAD \citep{yao2021imagetbad} and on a private dataset; the median errors were 1.14 and 0.98~mm (left and right ostia) on the private CCTA scans and 3.48 and 2.97~mm on ImageTBAD, a gap the authors attributed to the lower resolution and the artefacts of the non-gated scans.
More closely related to the present work, \citet{locnext_embc} proposed \locnext, a ConvNeXt-based architecture that reframes ostia localization as a per-channel 3D classification task over downsampled heatmaps, one channel per coronary ostium. \locnext\ was trained and evaluated on CAT08 and ASOCA, achieving a median localization error of 5.24~mm with a mean inference time of 24.29~s.
Coronary ostia have also been localized as part of multi-task models for transcatheter valve planning: \citet{tolle2025federated} trained 3D residual U-Nets in a federated manner across eight hospitals and distilled them into a single SwinUNETR \citep{swin-unetr}, in which the coronary ostia and the aortic valve hinge points were predicted by a shared head, and reported the localization errors of the two landmark types jointly (3.36\,$\pm$\,1.83\,mm on held-out centers).

\subsection{Public datasets for coronary centerlines}
\label{subsec:rw-centerline-datasets}

Reference coronary centerlines are publicly available for two of the datasets described above. In CAT08, the centerlines of four vessels per scan were annotated by three trained observers and averaged by a mean-shift algorithm for open curves, with the same approach applied to the radius \citep{cat08,cat08_2}; the complete reference centerlines were distributed for only 8 of the 32 scans, the other 24 being evaluated through the challenge website \citep{cat08}, which is no longer available. In ASOCA, the centerlines of the 40 annotated scans were computed with the Vascular Modeling Toolkit in 3D Slicer, between inlet and outlet points marked manually on the vessel surface \citep{asoca2}. ImageCAS provides segmentation masks but no centerlines \citep{imagecas}. Most recently, ImageCAS-X \citep{imagecasx2026} re-annotated the lumen of 800 of the 1000 ImageCAS scans, excluding the other 200 as non-diagnostic, and derived centerlines automatically from the new masks by thinning, with start points at the skeleton endpoints near an automatically segmented aorta; it provides no ostium landmarks.

\subsection{Methods for centerline extraction}
\label{subsec:rw-centerline-methods}

Methods for the extraction of vessel lumens and centerlines have been reviewed extensively \citep{lesage2009review}, and the CAT08 challenge established a common framework to evaluate coronary centerline extraction on CCTA \citep{cat08}. Three families of methods are most relevant to this work.
Skeletonization reduces a segmentation mask to its medial axis, typically by 3D thinning \citep{lee1994skeleton}. It needs no seed points, but it is sensitive to irregularities of the mask surface, and tends to produce spurious side branches and loops that must be repaired heuristically.
Minimal-path methods extract the path of minimal cumulative cost between two points \citep{cohen1997minimalpath}, with costs designed to keep the path centered in the lumen \citep{deschamps2001minimalpaths}; combined with a farthest-point strategy, they can grow a whole tree from a single root \citep{sato2000teasar}, and they have been started from automatically detected ostia to extract the coronary centerlines \citep{liu2014ostia}. The centerlines of the Vascular Modeling Toolkit follow the same principle, as minimal paths on the Voronoi diagram of the vessel surface \citep{antiga2003geometry}, but require the inlet and the outlets of the vessels to be specified.
Learning-based trackers, finally, follow the vessel from a seed point, predicting its local direction at each step with a convolutional network \citep{wolterink2019tracker,wolterink2019gcn,salahuddin2021multiresolution}, and deep networks have also been used to map the coronary centerlines directly from the image \citep{Leccardi2023}. Trackers can operate directly on the image, without a prior segmentation, but need reference centerlines to be trained and seed points to start from, which can be provided by ostium landmarks \citep{salahuddin2021multiresolution}.
Unlike trackers, skeletonization and the minimal-path methods applied to a segmentation mask inherit the defects of the mask, such as interrupted or missing vessel segments, which the extraction alone cannot recover.

\begin{figure*}[tb]
  \centering
  \includegraphics[width=\linewidth]{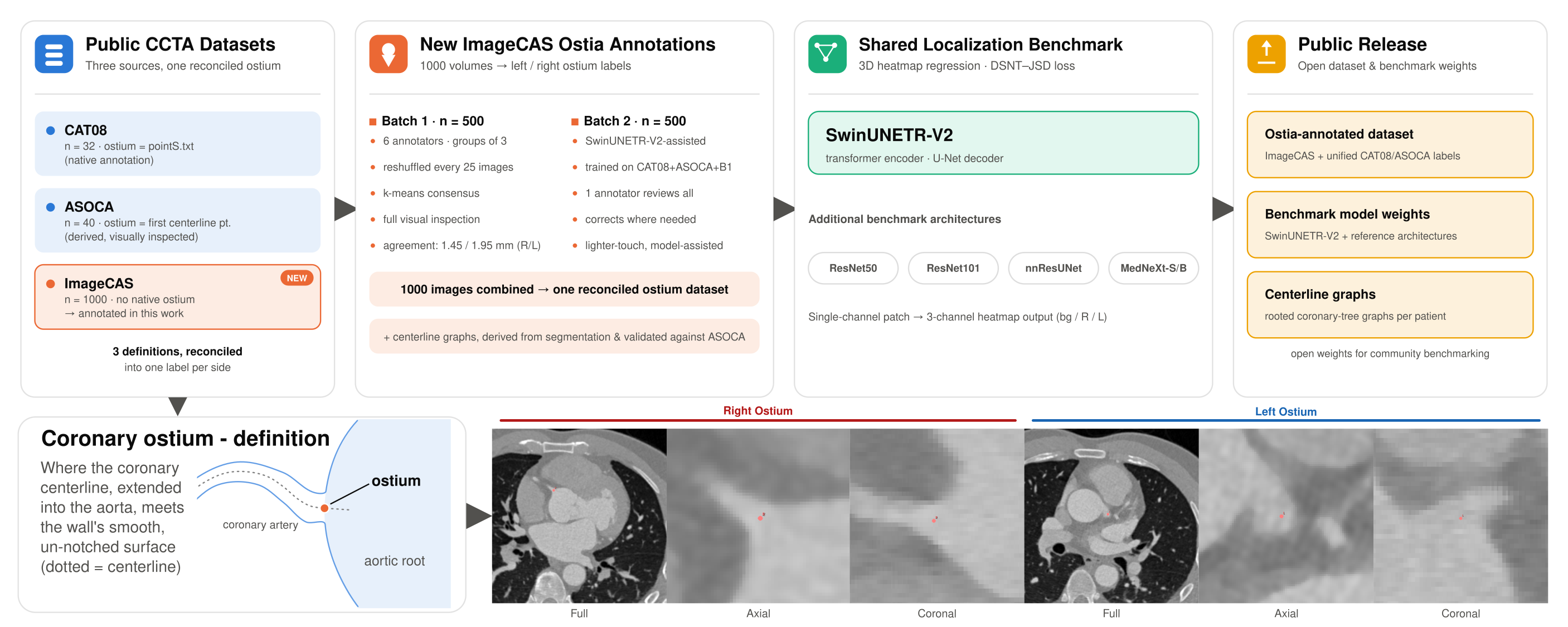}
  \caption{Overview of this work: the public CAT08, ASOCA, and ImageCAS datasets and their reconciled coronary ostium definitions; the two-stage annotation protocol used to extend ImageCAS with left/right ostium landmarks and centerline graphs; the shared benchmark of localization models trained on the combined dataset; and the resulting public release.}
  \label{fig:overview-pipeline}
\end{figure*}

\section{Methods}
\label{sec:methods}

\subsection{Extended ImageCAS dataset with ostia landmarks}
\label{subsec:ostia-dataset}

For the purposes of this work, the left and right coronary ostia are treated as two distinct, individually labeled landmarks per CCTA. Unlike CAT08 \citep{cat08}, which provides this information natively (via the centerline start point recorded in each case's \texttt{pointS.txt} file), and ASOCA \citep{asoca1}, from whose native centerlines it can be derived (via the first centerline point, after aggregation and visual inspection), ImageCAS provides no ostium annotations, and the only centerlines available for its scans, released with ImageCAS-X~\citep{imagecasx2026} for 800 of them, follow newly re-segmented masks rather than the original ones and start at automatically detected points near the aorta rather than at the ostia. 

In this paper, at the initial stage, the left and right coronary ostia were manually annotated in the first 500 ImageCAS images by six annotators, five biomedical engineering master's students, highly trained for this task, and the first author, each responsible for 250 images in total. The annotators followed a written definition of the ostium and learned from the ostium references of CAT08 and ASOCA as examples (\ref{subsec:app-annotation}\ofsupplementary). Rather than assigning a fixed subset of images to each annotator, annotators were organized into rotating groups of three, with group composition reshuffled every 25 images; each image was therefore assigned to three independent annotators, a group whose composition varied across the annotated cohort, reducing the risk that any single annotator's systematic bias or drift would propagate uniformly through the dataset. Annotations were placed directly on the CCTA volumes in 3D Slicer, using the axial, coronal and sagittal views together. For each image and each side, the consensus ostium location was computed as the centroid of the corresponding annotator markers, obtained via $k$-means clustering with $k=1$ (equivalent to their arithmetic mean in the marker coordinate space).
Following consensus computation, every image and its resulting marker were visually inspected by the annotator with the lowest mean distance from the other two annotators (Annotator 4 in Table~\ref{tab:annotator-stats}, Section~\ref{subsec:res-annotations}) to identify and correct gross annotation errors before the first batch was finalized.

In the second stage, the remaining 500 ImageCAS images were annotated semi-automatically. A separate \swinv2 model \citep{he2023swinunetrv2}, used only to accelerate this stage, was trained on CAT08, ASOCA and the first batch of 500 ImageCAS images, and generated candidate ostium locations for the second batch. The same annotator (Annotator 4) then inspected every candidate against the corresponding CCTA volume and corrected it where necessary; the model thus only provided an initialization. This model-assisted correction is a lighter process than the independent, consensus-based annotation of the first stage, and the two stages are not methodologically equivalent; for this reason, the test set is drawn from the first batch only, and it was held out from the training of this model as well (Section~\ref{subsec:models-test-set}). The correction statistics of this second batch are reported in Section~\ref{subsec:res-annotations}.

This work uses only ECG-gated CCTA with an ostium reference: together with CAT08 and ASOCA, the extended ImageCAS forms the combined dataset on which the models of this work are trained and tested (Section~\ref{subsec:models-test-set}). Only the new annotations and centerlines are released: the images, masks and reference annotations of the three datasets are not redistributed and must be obtained from their original sources (Section~\ref{sec:data-availability}). As independent CCTA cases, acquired with different scanners and protocols and annotated under different operational conventions, the two smaller datasets both add training and test data and allow the robustness of the models to be assessed across datasets; their image characteristics are summarized in~\ref{subsec:app-datasets}\ofsupplementary.

\begin{figure*}[t]
  \centering
  \begin{minipage}{0.9\textwidth}
    \centering
    \includegraphics[width=\linewidth]{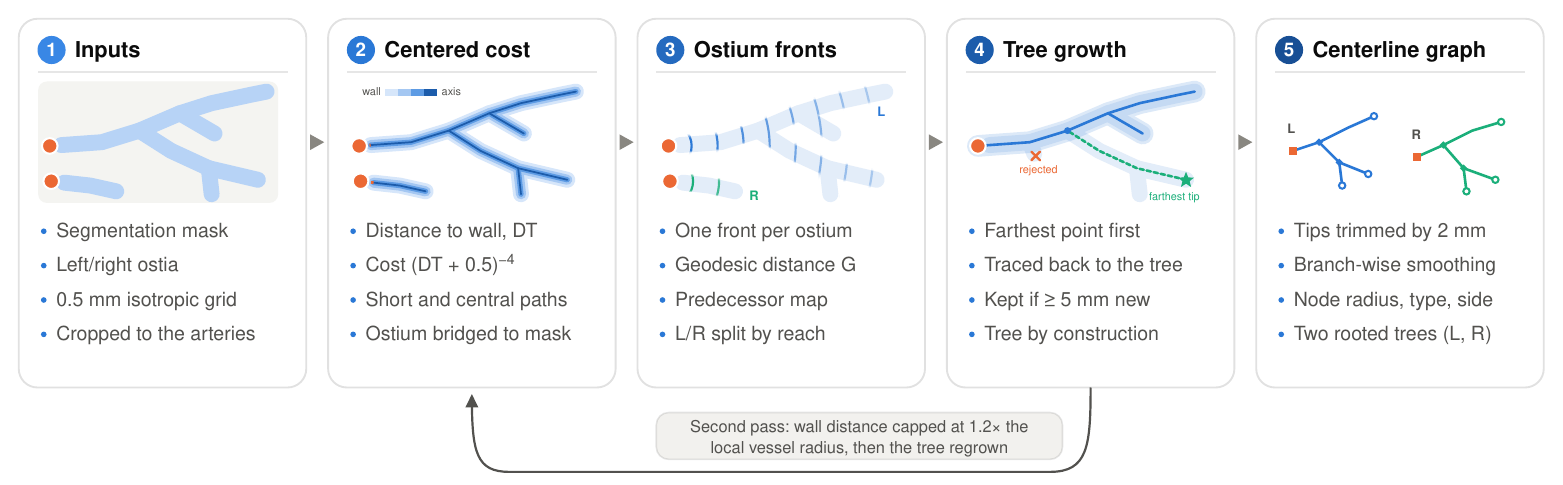}
    \caption{Centerline extraction pipeline. (1) Inputs: the coronary segmentation mask and the annotated left and right ostia (orange dots), resampled to a 0.5~mm isotropic grid and cropped to the arteries. (2) Centered cost: the cost of Eq.~\eqref{eq:centred-cost} decreases steeply with the distance DT from the vessel wall, so that minimal paths are both short and central; each ostium is bridged to the mask. (3) Ostium fronts: a minimal-cost front propagated from each ostium gives the geodesic distance $G$ and the minimal path (predecessor map) of every voxel, and splits the mask between the left (L) and right (R) trees according to the ostium that reaches each voxel; mask gaps of up to 10~mm are bridged. (4) Tree growth: the uncovered voxel farthest from the ostium (star) is traced back to the tree along its minimal path, which is kept as a new branch only if it adds at least 5~mm outside the region already covered (cross: rejected path), so that the result is a tree by construction. The growth is then repeated in a second pass with the wall distance capped at 1.2 times the local vessel radius (bottom arrow), which keeps the paths on the vessel axis at bulky bifurcations. (5) Centerline graph: the tips are trimmed by 2~mm and the trees smoothed branch by branch and resampled every 0.5~mm, yielding one tree per side, rooted at its ostium (square), with endpoints (circles), bifurcations, and per-node position, lumen radius, topological type and side, stored as one graph per patient. Full details are given in~\ref{subsec:app-centerlines}.}
    \label{fig:centerline-pipeline}
  \end{minipage}
\end{figure*}

\subsection{Extended ImageCAS dataset with centerline graphs}
\label{subsec:centerline-methods}

The coronary centerline graphs are extracted from the ImageCAS segmentation masks~\citep{imagecas} and rooted at the annotated ostia (Section~\ref{subsec:ostia-dataset}), yielding a left and a right arterial tree per patient. Instead of skeletonizing the mask by 3D thinning~\citep{lee1994skeleton}, which yields spurious branches and loops wherever the mask is irregular, we grow each tree from its ostium along minimal paths~\citep{cohen1997minimalpath,liu2014ostia} with the farthest-point strategy of TEASAR~\citep{sato2000teasar}, so that the result is a tree by construction. The procedure (Figure~\ref{fig:centerline-pipeline}) is deterministic, uses only the mask and the two ostia, and is detailed in~\ref{subsec:app-centerlines}.

The mask is resampled to 0.5~mm isotropic voxels, and the cost of traversing a voxel $\mathbf{x}$ of the mask is
\begin{equation}
c(\mathbf{x}) = \left(D(\mathbf{x}) + 0.5\right)^{-p}, \qquad p = 4,
\label{eq:centred-cost}
\end{equation}
where $D(\mathbf{x})$ is the distance of $\mathbf{x}$ from the mask boundary, in voxels, so that the cumulative cost of a path penalizes both its length and its distance from the vessel axis~\citep{deschamps2001minimalpaths}. Minimal-cost fronts propagated from each ostium~\citep{dijkstra1959note} give the geodesic distance and the minimal path of every voxel, and split the mask between the two trees; mask gaps of up to 10~mm, as at severe stenoses, are bridged.

Each tree is grown by repeatedly tracing the minimal path to the uncovered voxel farthest from the ostium and marking the mask around it as covered. A path becomes a new branch only if it runs for at least 5~mm outside the covered region, a single criterion that separates side branches from irregularities of the mask without heuristic pruning; as all the paths derive from the same minimal-path map, their union is a tree. A second growth pass, with $D$ capped near the local vessel radius, keeps the paths on the vessel axis at bulky bifurcations. Finally, the tips are trimmed and the trees smoothed and resampled every 0.5~mm; each node stores its position, lumen radius (that of the largest sphere inscribed in the mask), topological type and side, in one GML graph per patient.

The cases flagged by an automatic quality control, and a random sample of 50 unflagged ones, were reviewed visually. The procedure was validated on the 40 annotated volumes of ASOCA, whose reference centerlines derive from the same masks~\citep{asoca2}, which isolates the extraction error from the segmentation one, with the measures of the CAT08 evaluation framework~\citep{cat08}.

\subsection{Benchmark models for ostium localization}
\label{subsec:benchmark-models}
To compare architectural paradigms within a single localization framework, the benchmark includes transformer-based, residual convolutional, U-Net-based and ConvNeXt-inspired models, all formulated as 3D heatmap-based landmark localization networks. Each model receives a single-channel CCTA patch of $128\times128\times128$ voxels and produces three raw logit maps in the same spatial reference as the input, for the background, the right and the left coronary ostium. Loss, preprocessing, training strategy, postprocessing and evaluation are common to all the models and are described in the following sections, and the benchmark therefore isolates the differences between the architectures.

The primary architecture is a 3D \swinv2~\citep{he2023swinunetrv2}, an extension of SwinUNETR~\citep{swin-unetr} that combines a hierarchical shifted-window Transformer encoder with a U-Net-like decoder, modeling local anatomical information while progressively incorporating a broader spatial context; it is configured with a feature size of 24 and the residual convolutional blocks of the v2 configuration.
The ResNet50 and ResNet101 models use the 3D residual encoders~\citep{he2016resnet,hara2018resnet3d} of the MONAI implementation~\citep{monai}, trained from scratch, followed by a lightweight heatmap head (a $3\times3\times3$ convolution, group normalization and a ReLU activation, followed by a $1\times1\times1$ convolution mapping to the three output channels), whose output is trilinearly upsampled back to the input resolution.
\nnresunet is the residual-encoder U-Net architecture introduced for nnU-Net~\citep{nnunet}, with a five-stage encoder--decoder built from residual blocks, instance normalization and leaky ReLU activations, and per-stage feature widths of 32, 64, 128, 256 and 320.
MedNeXt is a ConvNeXt-inspired architecture developed for 3D medical image segmentation~\citep{mednext}, of which we trained the small (MedNeXt-S) and the base (MedNeXt-B) variants, with $3\times3\times3$ kernels and without deep supervision. The specifications of all the models are listed in Table~\ref{tab:models_specs}.

\subsection{Loss}
The models are trained with an ostium-localization loss that supervises the heatmaps through the coordinates they encode, following~\citet{gajowczyk2022ostia}. The loss relies on the Differentiable Spatial to Numerical Transform (DSNT)~\citep{nibali2018numericalcoordinateregressionconvolutional}, a differentiable operation that converts a spatial probability distribution into continuous landmark coordinates, which allows the network to be optimized directly on the coordinate error while still predicting dense heatmaps; a Jensen--Shannon divergence (JSD) term regularizes the shape of the heatmap toward a Gaussian distribution centered on the target landmark. For each ostium channel $c$, the predicted logits $z_c(v)$ over the patch grid $\Omega$ are converted into a spatial probability distribution through a temperature-scaled spatial softmax:
\begin{equation}
\hat{p}_c(v) =
\frac{\exp\big(z_c(v)/T\big)}
{\sum_{u \in \Omega}\exp\big(z_c(u)/T\big)} ,
\end{equation}
with temperature $T = 0.3$; since $T < 1$, the distribution is sharper than with a standard softmax, which limits the contribution of diffuse background activations to the expected coordinate. The DSNT then maps this distribution to continuous normalized coordinates in $[-1,1]^3$ as \begin{equation}
\hat{\mathbf{y}}_c =
\sum_{v \in \Omega} \hat{p}_c(v)\,\mathbf{g}(v),
\end{equation}
where $\mathbf{g}(v)$ denotes the normalized coordinate grid. 

For each side $c\in\{R,L\}$, the per-channel loss is defined as
\begin{equation}
\mathcal{L}^{\mathrm{ostium}}_c =
\mathcal{L}_{E,c}
+ \lambda_{JSD}\mathcal{L}_{JSD,c}
+ \lambda_{bg}\mathcal{L}_{BCE,c}.
\end{equation}
Here, $\mathcal{L}_{E,c}$ is the Euclidean distance between the DSNT-predicted coordinate $\hat{\mathbf{y}}_c$ and the target coordinate in normalized space:
\begin{equation}
\mathcal{L}_{E,c} =
\left\|\hat{\mathbf{y}}_c-\mathbf{y}_c\right\|_2 .
\end{equation}
The term $\mathcal{L}_{JSD,c}$ is the Jensen--Shannon divergence between the predicted probability distribution $\hat{p}_c$ and a normalized Gaussian target distribution $q_c$ centered on the ground-truth coordinate:
\begin{equation}
\mathcal{L}_{JSD,c}
=
\frac{1}{2}D_{KL}(\hat{p}_c \| m_c)
+
\frac{1}{2}D_{KL}(q_c \| m_c),
\qquad
m_c=\frac{1}{2}(\hat{p}_c+q_c).
\end{equation}
The term $\mathcal{L}_{BCE,c}$ is a voxel-wise binary cross-entropy computed from the raw logits and an unnormalized Gaussian target, included to suppress diffuse background activations outside the ostium region.

We use $\lambda_{JSD}=1.0$, $\lambda_{bg}=0.8$, and a Gaussian target standard deviation of $\sigma_R=\sigma_L=2.0$ voxels. Crops in which a given ostium is absent are masked out for that channel: the loss of each channel is computed only on the crops of the batch that contain its ostium, so that a randomly placed crop containing neither ostium contributes nothing to the loss, and a crop centered on one ostium supervises the other channel only if it contains that ostium too. What is not an ostium is learned from the surroundings of the ostium in the crops that contain it, such as the aortic wall, the neighboring vessels and the cardiac chambers, whose voxels are pushed toward zero by the background term and by the divergence from the Gaussian target.

The total loss combines the right and left ostium losses as
\begin{equation}
\mathcal{L} =
w_R\mathcal{L}^{\mathrm{ostium}}_R
+
w_L\mathcal{L}^{\mathrm{ostium}}_L.
\end{equation}
The weights start at $w_R=w_L=0.5$ and are adjusted during training within the interval $[0.25,0.75]$.

\subsection{Preprocessing}
\label{subsec:preprocessing}

For the benchmark models, all images are loaded as 3D volumes and resampled to an isotropic spacing of 0.5~mm before patch extraction or sliding-window inference. Intensities are clipped between -200 and 800 Hounsfield Units and linearly normalized to the $[-1,1]$ range. The same preprocessing is applied during training, validation, and inference so that model inputs are defined in a common spatial and intensity domain across ImageCAS, CAT08, and ASOCA.

\subsection{Postprocessing}
\label{subsec:postprocessing}

Full-volume prediction relies on 3D sliding-window inference, with windows of $128^3$ voxels, the patch size used in training, and an overlap of 50\% (64 voxels, i.e., a stride of 64 voxels). The windows cover the whole volume, with additional windows at the borders where needed and padding of any volume smaller than a window; the logits of the overlapping windows are averaged without any window weighting, the output activation of the model is applied to the averaged logits to obtain the class probabilities, and the result is mapped back to the native image grid.

The final 3D ostium coordinates are extracted independently for the right and left ostium channels. For each channel, the voxel with maximum heatmap response is first identified, and a local $11^3$ neighborhood (5.5~mm side at 0.5~mm isotropic spacing) centered on this peak is used to compute a probability-weighted centroid, yielding a sub-voxel coordinate estimate that is less sensitive to isolated noisy maxima than a pure argmax while preserving more than 96\% of the Gaussian energy. The resulting voxel coordinates are mapped back to physical space using the image-specific inverted affine matrix.

\subsection{Training strategy}
\label{subsec:training}
All the models are trained with the same recipe: the AdamW optimizer~\citep{AdamW} with mixed precision, on $128^3$ patches centered on either ostium or placed at random, with a light augmentation that omits flips and rotations by multiples of $90^\circ$ to preserve the standard right-anterior-superior (RAS) orientation of the images and the anatomical meaning of the right and left channels; the checkpoint with the lowest validation error is retained. The method was implemented with PyTorch~\citep{pytorch_neurips} and MONAI~\citep{monai} and trained on a single NVIDIA A100 GPU with 40~GB of memory; the complete training settings, including the software versions, are given in~\ref{subsec:app-training}\ofsupplementary.

\subsection{Evaluation metrics}
\label{subsec:evaluation-metrics}

Evaluation is performed at patient level on full volumes. For each case, the Euclidean distance in millimeters is computed between the predicted and reference coordinates for each coronary ostium $c\in\{R,L\}$ (right and left):
\[
    d_c = \sqrt{\sum_{i\in\{z,y,x\}}\left((\hat{p}_{c,i}-p_{c,i})s_i\right)^2},
\]
where $\hat{p}_{c,i}$ and $p_{c,i}$ are the predicted and ground-truth voxel coordinates for side $c$, and $s_i$ is the voxel spacing in millimeters. Results are reported over the whole test set (All) and separately for each of its datasets (ASOCA, CAT08 and ImageCAS; see Section~\ref{subsec:models-test-set}).
For each ostium, the distribution of $d_c$ over the test cases is summarized by its 25th percentile (p25), median (Med.), mean, 75th percentile (p75) and maximum (Max).
The combined statistics (Both) refer to $d_{\mathrm{Both}}=(d_R+d_L)/2$, the average of the left and right distances of each patient, so that they are computed over one value per patient rather than over the pooled left and right errors; the two definitions yield the same mean, but may differ in the other statistics (e.g., the median). 
Each localization is further graded into one of four accuracy buckets, \emph{Optimal}, \emph{Good}, \emph{Acceptable} and \emph{Failed}, delimited by three ostium-specific distance thresholds (Table~\ref{tab:thresholds}): an \emph{anatomical} threshold, a population-level criterion derived from the internal ostial diameter reported by \citet{nasr2018coronaryostia}, so that an error below it makes it likely, but not certain, that the point lies within the ostial opening of a given patient; an \emph{interoperator} threshold, equal to the 75th percentile of the distance of each individual annotator from the mean of the other two annotators of the same image (Table~\ref{tab:annotator-stats}); and a \emph{failure} threshold of 5~mm, the largest used in the literature \citep{zheng2011ostia,mostafa2021ostia}. The percentage of localizations in each bucket is computed for each ostium, and at patient level a case takes the worse bucket of its two ostia. We propose this grading as a common reporting standard for coronary ostium localization; its full definition is given in~\ref{subsec:app-buckets}.
\begin{table}[tb]
\centering
\caption{Distance thresholds (mm) delimiting the accuracy buckets.}
\label{tab:thresholds}
\begin{tabular}{l r r r}
\toprule
 & Anatomical & Interoperator & Failure \\
\midrule
Right ostium & 1.08 & 1.85 & 5.0 \\
Left ostium & 1.45 & 2.59 & 5.0 \\
\bottomrule
\end{tabular}
\end{table}
All the statistics are computed over all the cases, including the failed localizations. Finally, the computational cost is measured as the average time per patient (Time, s), obtained by dividing the total time needed to process the whole test set by the number of volumes.

\subsection{Models and test set}
\label{subsec:models-test-set}

The test set comprises 58 volumes, drawn from all three datasets: four from ASOCA (10\% of the total), four from CAT08 (12.5\%) and 50 from ImageCAS (5\%). The test set was drawn at random before any training and was held out from every model trained in this work, including the one used in the second annotation stage. The ImageCAS test cases were selected exclusively from the independently, manually annotated first batch (Section~\ref{subsec:ostia-dataset}), to base the evaluation on the most thoroughly verified labels, whereas every other ImageCAS image, including the whole model-assisted second batch, is used for training and validation. The training, validation and test volumes are 880, 70 and 50 for ImageCAS, 32, 4 and 4 for ASOCA, and 25, 3 and 4 for CAT08. Before being held out, the selected cases underwent a further verification pass by all the annotators, which confirmed every marker position without any modification.

Unless otherwise stated, every localization result refers to this test set, processed with the same preprocessing and postprocessing, with a single trained network per model (the checkpoint with the lowest validation error) and without test-time augmentation or ensembling; the ablation studies (\ref{sec:app-ablations}\ofsupplementary) change one setting of the training or of the inference at a time. Because the three datasets contribute very different numbers of volumes, each results table reports the whole test set in a black row, followed by one gray row for each dataset (ASOCA, CAT08 and ImageCAS). Inference was run on a single NVIDIA A100 GPU, with 20 of its 40~GB of memory available.

\begin{table}[h]
    \centering
    \renewcommand{\arraystretch}{1.2}
    \caption{Specifications of the models evaluated on the test dataset.}
    \scriptsize
    \begin{tabular}{lllc}
        \toprule
        \textbf{Model} & \textbf{Developer} & \textbf{Architecture} & \textbf{Parameters} \\
        \midrule
        \swinv2 & NVIDIA & Swin ViT + Conv. & 18.35M \\
        ResNet50 & Microsoft & Conv. & 49.70M \\
        ResNet101 & Microsoft & Conv. & 92.28M \\
        \nnresunet & ACVL & Conv. & 32.69M \\
        MedNeXt-S & MIC-DKFZ & ConvNeXt & 5.55M \\
        MedNeXt-B & MIC-DKFZ & ConvNeXt & 10.53M \\
        \bottomrule
    \end{tabular}
    \label{tab:models_specs}
\end{table}

\section{Results}
\label{sec:results}

\subsection{Ostia annotations of the ImageCAS CCTAs}
\label{subsec:res-annotations}
In the first, manually annotated batch of 500 ImageCAS images, the mean distance between each individual annotator marker and the mean of the other two markers placed on the same image (leave-one-out) was 1.45 $\pm$ 0.82~mm (mean $\pm$ SD) for the right ostium and 1.95 $\pm$ 1.04~mm for the left ostium. Table~\ref{tab:annotator-stats} reports the statistics of each annotator and of all the annotators pooled.

\begin{table*}[tb]
\centering
\caption{Per-annotator localization error (Euclidean distance, in mm, between
each annotator's manually placed point and the mean of the points placed on the same image by the other two annotators, i.e.\ leave-one-out)
for the left and right coronary ostia, individually and combined,
together with the pooled statistics across all six annotators. For each group
we report the mean $\pm$ SD and the median with interquartile range
(25th--75th percentile). The statistics are computed over 3000 placements by the six annotators (1500 landmarks per side, 250 per annotator per ostium, three annotators per volume).}
\label{tab:annotator-stats}
\small
\setlength{\tabcolsep}{3.5pt}
\begin{tabular}{l cc cc cc}
\toprule
\multirow{2}{*}{Annotator}
  & \multicolumn{2}{c}{Left ostium}
  & \multicolumn{2}{c}{Right ostium}
  & \multicolumn{2}{c}{Combined (Left+Right)} \\
\cmidrule(lr){2-3} \cmidrule(lr){4-5} \cmidrule(lr){6-7}
  & Mean $\pm$ SD & Median [IQR]
  & Mean $\pm$ SD & Median [IQR]
  & Mean $\pm$ SD & Median [IQR] \\
\midrule
Annotator 1 & 2.89 $\pm$ 0.95 & 2.73 [2.35--3.40] & 1.98 $\pm$ 0.78 & 1.90 [1.47--2.40] & 2.44 $\pm$ 0.98 & 2.37 [1.79--2.92] \\
Annotator 2 & 2.06 $\pm$ 0.95 & 1.99 [1.39--2.48] & 1.78 $\pm$ 0.95 & 1.63 [1.19--2.22] & 1.92 $\pm$ 0.96 & 1.75 [1.25--2.37] \\
Annotator 3 & 1.73 $\pm$ 0.88 & 1.54 [1.10--2.22] & 1.29 $\pm$ 0.71 & 1.19 [0.78--1.61] & 1.51 $\pm$ 0.83 & 1.34 [0.94--1.95] \\
Annotator 4 & 1.39 $\pm$ 0.90 & 1.19 [0.75--1.80] & 1.13 $\pm$ 0.61 & 1.04 [0.71--1.42] & 1.26 $\pm$ 0.78 & 1.09 [0.72--1.57] \\
Annotator 5 & 1.44 $\pm$ 0.83 & 1.29 [0.81--1.94] & 1.19 $\pm$ 0.67 & 1.10 [0.69--1.55] & 1.31 $\pm$ 0.76 & 1.20 [0.76--1.69] \\
Annotator 6 & 2.18 $\pm$ 0.90 & 2.15 [1.56--2.85] & 1.35 $\pm$ 0.76 & 1.11 [0.83--1.65] & 1.76 $\pm$ 0.93 & 1.60 [1.00--2.40] \\
\midrule
\textbf{All annotators (pooled)} & 1.95 $\pm$ 1.04 & 1.83 [1.16--2.60] & 1.45 $\pm$ 0.82 & 1.31 [0.86--1.85] & 1.70 $\pm$ 0.97 & 1.53 [0.97--2.27] \\
\bottomrule
\end{tabular}

\vspace{1mm}
\footnotesize
\textit{Notes.} All values are in mm. (1) The reference of each annotator, per image and per ostium,
is the mean of the points of the other two annotators who saw that image (annotator groups of
three rotate every 25 images, 6 annotators total, 250 images each by design); the distances from the
released consensus, which also includes the annotator's own point, are given in
Table~\ref{tab:app-annotator-consensus}. (2) The
\emph{Combined (Left+Right)} column for a given annotator pools that
annotator's L and R point-wise errors together before computing each
statistic -- it is \emph{not} an average of the L and R columns, since means,
SDs, medians and percentiles are not in general recoverable from the
per-ostium values once the samples are pooled. (3) \emph{All annotators
(pooled)} likewise pools every individual point-wise error across all six
annotators -- it is \emph{not} an average of the six per-annotator statistics.
(4) Standard deviations are population SD (ddof=0). (5) Percentiles are
computed with linear interpolation between order statistics.
\end{table*}

In the second, model-assisted batch (Table~\ref{tab:correction-stats}), whose candidates were generated by a \swinv2 model trained on the first batch together with CAT08 and ASOCA (its test results are reported in~\ref{sec:app-stage2}\ofsupplementary), the reviewing annotator refined the candidate position in 88 of the 500 images (17.6\%), and the model did not detect an ostium in 10 images (2.0\%), which were annotated manually. The review required about 5 person-hours, against approximately 144 person-hours for the fully manual annotation of the first batch, in which every volume was annotated three times (six annotators, about 24 hours each).

\begin{table}[tb]
\small
\centering
\caption{Semi-automated correction statistics for the second batch of 500
ImageCAS images (patients 501--1000), annotated using \swinv2
predictions followed by single-annotator visual review.}
\label{tab:correction-stats}
\setlength{\tabcolsep}{4.5pt}
\begin{tabular}{l c}
\toprule
\textbf{Metric} & \textbf{Value} \\
\midrule
Images requiring manual refinement$^{a}$      & 88 / 500 (17.6\%) \\
Images with an undetected ostium (outlier)    & 10 / 500 (2.0\%) \\
\quad Right ostium                            & 8 \\
\quad Left ostium                             & 2 \\
\quad Both ostia                              & 0 \\
\midrule
Annotation time, first batch$^{b}$  & $\approx$24~h (144 person-h) \\
Annotation time, this batch$^{c}$   & $\approx$12~h (5 person-h) \\
\bottomrule
\end{tabular}

\vspace{1mm}
\raggedright\footnotesize
$^{a}$Ostium position adjusted by 1~mm or less relative to the model
prediction.
$^{b}$Fully manual annotation of the same number of images (500) by the
first batch's six-annotator team (rotating in groups of three per image):
6 annotators $\times$ 8~h/day $\times$ 3~days.
$^{c}$\swinv2 inference (7~h, unattended) followed by single-annotator
visual review and correction (5~h); only the review time counts as human
annotation effort.
\end{table}

\subsection{Consistency of the ostium definition across datasets}
\label{subsec:res-ostia-consistency}
The ostium annotations of CAT08, ASOCA and ImageCAS derive from different operational definitions (Section~\ref{subsec:ostia-dataset}). Re-expressed relative to the right-left midpoint of the same patient, the right and left ostium point clouds of the three datasets are essentially superimposed, and the two ostia lie, on average, 33.0~mm apart (SD 5.9~mm). Four permutation tests, Holm-Bonferroni corrected, compared the datasets on the position of each ostium (two tests that are necessarily equivalent, since each patient's left ostium mirrors its right one about the midpoint), on the inter-ostial distance and on the direction of the left-to-right axis; only the last one detects a difference between the datasets ($p=0.0028$), driven by an offset of about 5--6$^\circ$ of the ImageCAS axis with respect to ASOCA and CAT08. The point clouds, the design of the tests and their full results are reported in~\ref{sec:app-consistency}\ofsupplementary.

\subsection{Training on the complete ImageCAS dataset}
\begin{table*}[t!]
\centering
\fontsize{7}{8.6}\selectfont
\setlength{\tabcolsep}{2.6pt}
\newcommand{\bucketgapAll}{-2pt}
\newcommand{\blockgapAll}{5pt}
\caption{Euclidean distance $d_c$ (mm) of the benchmark models trained on the complete ImageCAS dataset (the 950 CCTAs not in the test set) and tested on 58 CCTAs, for the right ostium, the left ostium, and both (per-patient average of the right and left distances). The line below each row gives the percentage of cases in each accuracy bucket (for both, at patient level): O = Optimal, G = Good, A = Acceptable, F = Failed. See Section~\ref{subsec:evaluation-metrics}. In each distance column, the best value is in bold and the second best is underlined, comparing the rows of the same kind (overall and, separately, each dataset).}
\label{tab:res-1000}
\begin{tabular}{l *{15}{r} r}
\toprule
 & \multicolumn{5}{c}{Right ostium} & \multicolumn{5}{c}{Left ostium} & \multicolumn{5}{c}{Both} & \\
\cmidrule(lr){2-6} \cmidrule(lr){7-11} \cmidrule(lr){12-16}
\makecell[l]{Model} & p25 & Med. & Mean & p75 & Max & p25 & Med. & Mean & p75 & Max & p25 & Med. & Mean & p75 & Max & \makecell{Time\\(s)} \\
\midrule
\swinv2 & \underline{0.53} & \underline{0.74} & \textbf{0.83} & \textbf{1.02} & \underline{2.07} & \textbf{0.61} & \textbf{0.85} & \underline{0.98} & \underline{1.32} & \underline{2.51} & \textbf{0.68} & \underline{0.86} & \underline{0.91} & \underline{1.10} & \textbf{1.94} & 55.6 \\[\bucketgapAll]
 & \multicolumn{5}{c}{\bucketrow{77.59}{20.69}{1.72}{0.00}} & \multicolumn{5}{c}{\bucketrow{84.48}{15.52}{0.00}{0.00}} & \multicolumn{5}{c}{\bucketrow{68.97}{29.31}{1.72}{0.00}} & \\
\multicolumn{1}{r}{\subds{ASOCA}} & \subds{\textbf{0.53}} & \subds{\underline{0.95}} & \subds{\underline{0.95}} & \subds{\underline{1.37}} & \subds{\underline{1.45}} & \subds{\textbf{0.54}} & \subds{\textbf{0.60}} & \subds{\textbf{0.58}} & \subds{\textbf{0.64}} & \subds{\textbf{0.71}} & \subds{\underline{0.58}} & \subds{\underline{0.81}} & \subds{\underline{0.77}} & \subds{\underline{0.99}} & \subds{\textbf{1.02}} &  \\[\bucketgapAll]
 & \multicolumn{5}{c}{\subds{\bucketrow{50.00}{50.00}{0.00}{0.00}}} & \multicolumn{5}{c}{\subds{\bucketrow{100.00}{0.00}{0.00}{0.00}}} & \multicolumn{5}{c}{\subds{\bucketrow{50.00}{50.00}{0.00}{0.00}}} & \\
\multicolumn{1}{r}{\subds{CAT08}} & \subds{\textbf{0.60}} & \subds{\textbf{1.02}} & \subds{\textbf{1.04}} & \subds{\textbf{1.46}} & \subds{\underline{1.75}} & \subds{\textbf{0.59}} & \subds{\textbf{1.04}} & \subds{\underline{1.20}} & \subds{\underline{1.64}} & \subds{\underline{2.25}} & \subds{\underline{0.98}} & \subds{\textbf{1.05}} & \subds{\underline{1.12}} & \subds{\textbf{1.20}} & \subds{\textbf{1.46}} &  \\[\bucketgapAll]
 & \multicolumn{5}{c}{\subds{\bucketrow{50.00}{50.00}{0.00}{0.00}}} & \multicolumn{5}{c}{\subds{\bucketrow{75.00}{25.00}{0.00}{0.00}}} & \multicolumn{5}{c}{\subds{\bucketrow{25.00}{75.00}{0.00}{0.00}}} & \\
\multicolumn{1}{r}{\subds{ImageCAS}} & \subds{\underline{0.54}} & \subds{\underline{0.74}} & \subds{\underline{0.81}} & \subds{\textbf{0.98}} & \subds{\underline{2.07}} & \subds{\textbf{0.62}} & \subds{\underline{0.89}} & \subds{\underline{1.00}} & \subds{\underline{1.32}} & \subds{\underline{2.51}} & \subds{\textbf{0.68}} & \subds{\underline{0.84}} & \subds{\underline{0.90}} & \subds{\underline{1.10}} & \subds{\textbf{1.94}} &  \\[\bucketgapAll]
 & \multicolumn{5}{c}{\subds{\bucketrow{82.00}{16.00}{2.00}{0.00}}} & \multicolumn{5}{c}{\subds{\bucketrow{84.00}{16.00}{0.00}{0.00}}} & \multicolumn{5}{c}{\subds{\bucketrow{74.00}{24.00}{2.00}{0.00}}} & \\
\addlinespace[\blockgapAll]
ResNet50 & 1.75 & 2.54 & \underline{2.62} & 3.34 & 4.75 & 1.81 & 2.49 & 2.47 & 3.21 & 4.53 & \underline{1.93} & 2.65 & 2.54 & 3.16 & \underline{4.46} & 36.0 \\[\bucketgapAll]
 & \multicolumn{5}{c}{\bucketrow{8.62}{18.97}{72.41}{0.00}} & \multicolumn{5}{c}{\bucketrow{17.24}{37.93}{44.83}{0.00}} & \multicolumn{5}{c}{\bucketrow{1.72}{15.52}{82.76}{0.00}} & \\
\multicolumn{1}{r}{\subds{ASOCA}} & \subds{0.78} & \subds{1.20} & \subds{1.32} & \subds{1.75} & \subds{2.33} & \subds{0.97} & \subds{1.34} & \subds{1.65} & \subds{2.01} & \subds{3.16} & \subds{1.14} & \subds{1.42} & \subds{1.48} & \subds{1.76} & \subds{2.00} &  \\[\bucketgapAll]
 & \multicolumn{5}{c}{\subds{\bucketrow{50.00}{25.00}{25.00}{0.00}}} & \multicolumn{5}{c}{\subds{\bucketrow{50.00}{25.00}{25.00}{0.00}}} & \multicolumn{5}{c}{\subds{\bucketrow{0.00}{50.00}{50.00}{0.00}}} & \\
\multicolumn{1}{r}{\subds{CAT08}} & \subds{1.88} & \subds{2.57} & \subds{2.46} & \subds{3.15} & \subds{3.22} & \subds{2.28} & \subds{2.88} & \subds{2.79} & \subds{3.39} & \subds{3.93} & \subds{2.54} & \subds{2.93} & \subds{2.63} & \subds{3.02} & \subds{3.17} &  \\[\bucketgapAll]
 & \multicolumn{5}{c}{\subds{\bucketrow{0.00}{25.00}{75.00}{0.00}}} & \multicolumn{5}{c}{\subds{\bucketrow{0.00}{50.00}{50.00}{0.00}}} & \multicolumn{5}{c}{\subds{\bucketrow{0.00}{25.00}{75.00}{0.00}}} & \\
\multicolumn{1}{r}{\subds{ImageCAS}} & \subds{1.99} & \subds{2.61} & \subds{2.73} & \subds{3.65} & \subds{4.75} & \subds{1.93} & \subds{2.50} & \subds{2.51} & \subds{3.21} & \subds{4.53} & \subds{\underline{2.10}} & \subds{2.68} & \subds{2.62} & \subds{3.19} & \subds{\underline{4.46}} &  \\[\bucketgapAll]
 & \multicolumn{5}{c}{\subds{\bucketrow{6.00}{18.00}{76.00}{0.00}}} & \multicolumn{5}{c}{\subds{\bucketrow{16.00}{38.00}{46.00}{0.00}}} & \multicolumn{5}{c}{\subds{\bucketrow{2.00}{12.00}{86.00}{0.00}}} & \\
\addlinespace[\blockgapAll]
ResNet101 & 1.96 & 2.63 & 2.68 & 3.50 & 4.73 & 1.81 & 2.71 & 2.70 & 3.72 & 5.36 & \underline{1.93} & 2.70 & 2.69 & 3.33 & 4.73 & 37.3 \\[\bucketgapAll]
 & \multicolumn{5}{c}{\bucketrow{6.90}{17.24}{75.86}{0.00}} & \multicolumn{5}{c}{\bucketrow{17.24}{31.03}{50.00}{1.72}} & \multicolumn{5}{c}{\bucketrow{0.00}{15.52}{82.76}{1.72}} & \\
\multicolumn{1}{r}{\subds{ASOCA}} & \subds{0.77} & \subds{1.18} & \subds{1.31} & \subds{1.71} & \subds{2.31} & \subds{1.06} & \subds{1.36} & \subds{1.73} & \subds{2.04} & \subds{3.15} & \subds{1.24} & \subds{1.48} & \subds{1.52} & \subds{1.76} & \subds{2.00} &  \\[\bucketgapAll]
 & \multicolumn{5}{c}{\subds{\bucketrow{50.00}{25.00}{25.00}{0.00}}} & \multicolumn{5}{c}{\subds{\bucketrow{50.00}{25.00}{25.00}{0.00}}} & \multicolumn{5}{c}{\subds{\bucketrow{0.00}{50.00}{50.00}{0.00}}} & \\
\multicolumn{1}{r}{\subds{CAT08}} & \subds{1.85} & \subds{2.53} & \subds{2.46} & \subds{3.13} & \subds{3.23} & \subds{2.26} & \subds{2.87} & \subds{2.78} & \subds{3.38} & \subds{3.92} & \subds{2.54} & \subds{2.91} & \subds{2.62} & \subds{2.99} & \subds{\underline{3.15}} &  \\[\bucketgapAll]
 & \multicolumn{5}{c}{\subds{\bucketrow{0.00}{25.00}{75.00}{0.00}}} & \multicolumn{5}{c}{\subds{\bucketrow{0.00}{50.00}{50.00}{0.00}}} & \multicolumn{5}{c}{\subds{\bucketrow{0.00}{25.00}{75.00}{0.00}}} & \\
\multicolumn{1}{r}{\subds{ImageCAS}} & \subds{2.11} & \subds{2.68} & \subds{2.81} & \subds{3.55} & \subds{4.73} & \subds{1.94} & \subds{2.76} & \subds{2.78} & \subds{3.77} & \subds{5.36} & \subds{2.17} & \subds{2.78} & \subds{2.79} & \subds{3.36} & \subds{4.73} &  \\[\bucketgapAll]
 & \multicolumn{5}{c}{\subds{\bucketrow{4.00}{16.00}{80.00}{0.00}}} & \multicolumn{5}{c}{\subds{\bucketrow{16.00}{30.00}{52.00}{2.00}}} & \multicolumn{5}{c}{\subds{\bucketrow{0.00}{12.00}{86.00}{2.00}}} & \\
\addlinespace[\blockgapAll]
\nnresunet & \textbf{0.52} & \textbf{0.70} & \textbf{0.83} & \underline{1.13} & \textbf{2.03} & \underline{0.68} & \underline{0.87} & \textbf{0.96} & \textbf{1.21} & \textbf{2.32} & \textbf{0.68} & \textbf{0.81} & \textbf{0.89} & \textbf{1.01} & \textbf{1.94} & 41.1 \\[\bucketgapAll]
 & \multicolumn{5}{c}{\bucketrow{72.41}{24.14}{3.45}{0.00}} & \multicolumn{5}{c}{\bucketrow{89.66}{10.34}{0.00}{0.00}} & \multicolumn{5}{c}{\bucketrow{67.24}{29.31}{3.45}{0.00}} & \\
\multicolumn{1}{r}{\subds{ASOCA}} & \subds{\underline{0.61}} & \subds{\textbf{0.67}} & \subds{\textbf{0.77}} & \subds{\textbf{0.83}} & \subds{\textbf{1.20}} & \subds{\underline{0.58}} & \subds{\underline{0.74}} & \subds{\underline{0.67}} & \subds{\underline{0.83}} & \subds{\underline{1.06}} & \subds{\textbf{0.57}} & \subds{\textbf{0.68}} & \subds{\textbf{0.72}} & \subds{\textbf{0.83}} & \subds{\underline{1.13}} &  \\[\bucketgapAll]
 & \multicolumn{5}{c}{\subds{\bucketrow{75.00}{25.00}{0.00}{0.00}}} & \multicolumn{5}{c}{\subds{\bucketrow{100.00}{0.00}{0.00}{0.00}}} & \multicolumn{5}{c}{\subds{\bucketrow{75.00}{25.00}{0.00}{0.00}}} & \\
\multicolumn{1}{r}{\subds{CAT08}} & \subds{\underline{0.68}} & \subds{\underline{1.22}} & \subds{\underline{1.17}} & \subds{\underline{1.71}} & \subds{\textbf{1.74}} & \subds{\underline{1.03}} & \subds{\underline{1.17}} & \subds{\textbf{1.05}} & \subds{\textbf{1.20}} & \subds{\textbf{1.23}} & \subds{\textbf{0.93}} & \subds{\underline{1.08}} & \subds{\textbf{1.11}} & \subds{\underline{1.26}} & \subds{\textbf{1.46}} &  \\[\bucketgapAll]
 & \multicolumn{5}{c}{\subds{\bucketrow{50.00}{50.00}{0.00}{0.00}}} & \multicolumn{5}{c}{\subds{\bucketrow{100.00}{0.00}{0.00}{0.00}}} & \multicolumn{5}{c}{\subds{\bucketrow{50.00}{50.00}{0.00}{0.00}}} & \\
\multicolumn{1}{r}{\subds{ImageCAS}} & \subds{\textbf{0.48}} & \subds{\textbf{0.70}} & \subds{\textbf{0.80}} & \subds{\underline{1.07}} & \subds{\textbf{2.03}} & \subds{\underline{0.68}} & \subds{\textbf{0.87}} & \subds{\textbf{0.97}} & \subds{\textbf{1.23}} & \subds{\textbf{2.32}} & \subds{\textbf{0.68}} & \subds{\textbf{0.80}} & \subds{\textbf{0.89}} & \subds{\textbf{0.97}} & \subds{\textbf{1.94}} &  \\[\bucketgapAll]
 & \multicolumn{5}{c}{\subds{\bucketrow{74.00}{22.00}{4.00}{0.00}}} & \multicolumn{5}{c}{\subds{\bucketrow{88.00}{12.00}{0.00}{0.00}}} & \multicolumn{5}{c}{\subds{\bucketrow{68.00}{28.00}{4.00}{0.00}}} & \\
\addlinespace[\blockgapAll]
MedNeXt-S & 1.72 & 2.45 & 12.21 & 3.91 & 181.91 & 2.82 & 6.56 & 52.17 & 110.68 & 162.70 & 3.31 & 7.28 & 32.19 & 59.61 & 168.67 & 49.4 \\[\bucketgapAll]
 & \multicolumn{5}{c}{\bucketrow{8.62}{22.41}{46.55}{22.41}} & \multicolumn{5}{c}{\bucketrow{3.45}{15.52}{24.14}{56.90}} & \multicolumn{5}{c}{\bucketrow{0.00}{5.17}{27.59}{67.24}} & \\
\multicolumn{1}{r}{\subds{ASOCA}} & \subds{1.95} & \subds{2.92} & \subds{4.74} & \subds{5.72} & \subds{11.84} & \subds{2.04} & \subds{2.57} & \subds{31.07} & \subds{31.60} & \subds{117.52} & \subds{2.92} & \subds{5.03} & \subds{17.91} & \subds{20.01} & \subds{59.85} &  \\[\bucketgapAll]
 & \multicolumn{5}{c}{\subds{\bucketrow{0.00}{25.00}{50.00}{25.00}}} & \multicolumn{5}{c}{\subds{\bucketrow{0.00}{50.00}{25.00}{25.00}}} & \multicolumn{5}{c}{\subds{\bucketrow{0.00}{25.00}{25.00}{50.00}}} & \\
\multicolumn{1}{r}{\subds{CAT08}} & \subds{1.54} & \subds{1.67} & \subds{28.45} & \subds{28.58} & \subds{108.95} & \subds{66.05} & \subds{98.54} & \subds{85.88} & \subds{118.37} & \subds{144.11} & \subds{33.81} & \subds{58.62} & \subds{57.17} & \subds{81.97} & \subds{109.37} &  \\[\bucketgapAll]
 & \multicolumn{5}{c}{\subds{\bucketrow{0.00}{75.00}{0.00}{25.00}}} & \multicolumn{5}{c}{\subds{\bucketrow{0.00}{25.00}{0.00}{75.00}}} & \multicolumn{5}{c}{\subds{\bucketrow{0.00}{25.00}{0.00}{75.00}}} & \\
\multicolumn{1}{r}{\subds{ImageCAS}} & \subds{1.75} & \subds{2.63} & \subds{11.51} & \subds{3.90} & \subds{181.91} & \subds{3.17} & \subds{6.56} & \subds{51.16} & \subds{110.04} & \subds{162.70} & \subds{3.32} & \subds{6.93} & \subds{31.33} & \subds{58.24} & \subds{168.67} &  \\[\bucketgapAll]
 & \multicolumn{5}{c}{\subds{\bucketrow{10.00}{18.00}{50.00}{22.00}}} & \multicolumn{5}{c}{\subds{\bucketrow{4.00}{12.00}{26.00}{58.00}}} & \multicolumn{5}{c}{\subds{\bucketrow{0.00}{2.00}{30.00}{68.00}}} & \\
\addlinespace[\blockgapAll]
MedNeXt-B & 4.26 & 6.30 & 43.45 & 86.69 & 189.12 & 3.08 & 5.37 & 31.39 & 9.04 & 163.07 & 4.66 & 8.82 & 37.42 & 67.14 & 165.62 & 50.4 \\[\bucketgapAll]
 & \multicolumn{5}{c}{\bucketrow{0.00}{1.72}{37.93}{60.34}} & \multicolumn{5}{c}{\bucketrow{1.72}{18.97}{22.41}{56.90}} & \multicolumn{5}{c}{\bucketrow{0.00}{1.72}{20.69}{77.59}} & \\
\multicolumn{1}{r}{\subds{ASOCA}} & \subds{4.20} & \subds{6.85} & \subds{46.49} & \subds{49.15} & \subds{168.35} & \subds{5.63} & \subds{45.92} & \subds{47.81} & \subds{88.10} & \subds{96.44} & \subds{35.60} & \subds{47.49} & \subds{47.15} & \subds{59.05} & \subds{85.66} &  \\[\bucketgapAll]
 & \multicolumn{5}{c}{\subds{\bucketrow{0.00}{0.00}{50.00}{50.00}}} & \multicolumn{5}{c}{\subds{\bucketrow{0.00}{0.00}{25.00}{75.00}}} & \multicolumn{5}{c}{\subds{\bucketrow{0.00}{0.00}{0.00}{100.00}}} & \\
\multicolumn{1}{r}{\subds{CAT08}} & \subds{101.13} & \subds{125.24} & \subds{114.31} & \subds{138.42} & \subds{154.80} & \subds{3.14} & \subds{4.74} & \subds{37.27} & \subds{38.86} & \subds{137.64} & \subds{52.04} & \subds{69.46} & \subds{75.79} & \subds{93.20} & \subds{135.30} &  \\[\bucketgapAll]
 & \multicolumn{5}{c}{\subds{\bucketrow{0.00}{0.00}{0.00}{100.00}}} & \multicolumn{5}{c}{\subds{\bucketrow{0.00}{25.00}{25.00}{50.00}}} & \multicolumn{5}{c}{\subds{\bucketrow{0.00}{0.00}{0.00}{100.00}}} & \\
\multicolumn{1}{r}{\subds{ImageCAS}} & \subds{4.20} & \subds{5.84} & \subds{37.54} & \subds{59.21} & \subds{189.12} & \subds{3.04} & \subds{5.23} & \subds{29.61} & \subds{8.32} & \subds{163.07} & \subds{4.08} & \subds{6.88} & \subds{33.57} & \subds{65.21} & \subds{165.62} &  \\[\bucketgapAll]
 & \multicolumn{5}{c}{\subds{\bucketrow{0.00}{2.00}{40.00}{58.00}}} & \multicolumn{5}{c}{\subds{\bucketrow{2.00}{20.00}{22.00}{56.00}}} & \multicolumn{5}{c}{\subds{\bucketrow{0.00}{2.00}{24.00}{74.00}}} & \\
\bottomrule
\end{tabular}
\end{table*}

Table~\ref{tab:res-1000} reports the results of the benchmark models, overall and stratified by dataset, after training on the full development set, i.e., the 950 ImageCAS volumes not in the test set, together with the training volumes of CAT08 and ASOCA (labeled 1000 in the tables).
\swinv2 and \nnresunet localized both ostia in all the test volumes, with no failed localization, median errors of 0.74 and 0.85~mm (\swinv2) and of 0.70 and 0.87~mm (\nnresunet) for the right and left ostium, and 69.0\% and 67.2\% of the cases Optimal at patient level; their processing times were 55.6 and 41.1~s per patient. Compared with the model of the second annotation stage (\ref{sec:app-stage2}\ofsupplementary), which was trained with half of the ImageCAS volumes and on images not yet reoriented to the common RAS orientation, \swinv2 had no failed localization, against six volumes with at least one failed ostium, and a larger share of Optimal cases at patient level (69.0\% against 53.4\%).
The two ResNet models were markedly less precise, with median errors between 2.49 and 2.71~mm and most of their localizations only Acceptable, although only one left ostium failed (ResNet101); they were the fastest models. The two MedNeXt models were unreliable: at patient level, 67.2\% (MedNeXt-S) and 77.6\% (MedNeXt-B) of the cases failed, with failures in all three datasets, and none was Optimal.
Overall, \swinv2 and \nnresunet were the best-performing models, followed by the ResNets and the two MedNeXt models.

\subsection{Ablation studies}
\label{subsec:res-ablations}
The training and inference pipeline of the benchmark models was examined with five ablation studies on the same test set: the number of ImageCAS images used for training, the voxel spacing used for training, a mismatch between the training and the inference spacing, the size of the peak window, and the overlap of the sliding windows. They were restricted to the two best-performing models, the only ones whose use the ablations could inform: all of them were run on \swinv2, and those on the training voxel spacing and on the sliding-window overlap, the two settings that most directly determine the accuracy and the processing time experienced by the user, were repeated on \nnresunet. They were carried out after the benchmark models had been trained and evaluated, and none of them influenced the configuration of those models. Their tables, plots and description are reported in~\ref{sec:app-ablations}\ofsupplementary, and their findings are discussed in Section~\ref{sec:discussion-ablations}. In brief, \swinv2 trained without ImageCAS images failed on 34.5\% of the right and 53.4\% of the left ostia, whereas with 250 ImageCAS images it had no failure; trained and tested at 1.0~mm instead of 0.5~mm, it had no failure either, with a processing time of 21.9 instead of 55.6~s per patient, whereas testing at 1.0~mm a model trained at 0.5~mm made 22.4\% of the left ostia fail.

\subsection{Comparison between trained models and inter-annotator variability}
\begin{figure}[t]
    \centering
    \begin{tikzpicture}
    \begin{axis}[
        boxplot/draw direction=y,
        ylabel={Euclidean distance (mm, log scale)},
        xtick={1,2,3,4,5,6,7},
        xticklabels={{Annotator\\variability}, \swinv2, ResNet50, ResNet101, \nnresunet, MedNeXt-S, MedNeXt-B},
        x tick label style={align=center, font=\scriptsize, rotate=45, anchor=north east},
        ymode=log, ymin=0.04, ymax=250,
        log basis y=10,
        xmin=0.5, xmax=7.5,
        enlarge x limits=0.08,
        ymajorgrids=true,
        grid style=dashed,
        width=\linewidth,
        height=4cm,
    ]

    \addplot+[
        boxplot prepared={
            draw position=1,
            median=1.63,
            upper quartile=2.48,
            lower quartile=1.03,
            upper whisker=5.79,
            lower whisker=0.20
        },
        fill=blue!20, draw=blue, thick
    ] coordinates {};

    \addplot+[
        boxplot prepared={
            draw position=2,
            median=0.86,
            upper quartile=1.10,
            lower quartile=0.68,
            upper whisker=1.94,
            lower whisker=0.32
        },
        fill=red!20, draw=red, thick
    ] coordinates {};

    \addplot+[
        boxplot prepared={
            draw position=3,
            median=2.65,
            upper quartile=3.16,
            lower quartile=1.93,
            upper whisker=4.46,
            lower whisker=0.73
        },
        fill=red!20, draw=red, thick
    ] coordinates {};

    \addplot+[
        boxplot prepared={
            draw position=4,
            median=2.70,
            upper quartile=3.33,
            lower quartile=1.93,
            upper whisker=4.73,
            lower whisker=1.12
        },
        fill=red!20, draw=red, thick
    ] coordinates {};

    \addplot+[
        boxplot prepared={
            draw position=5,
            median=0.81,
            upper quartile=1.01,
            lower quartile=0.68,
            upper whisker=1.94,
            lower whisker=0.38
        },
        fill=red!20, draw=red, thick
    ] coordinates {};

    \addplot+[
        boxplot prepared={
            draw position=6,
            median=7.28,
            upper quartile=59.61,
            lower quartile=3.31,
            upper whisker=168.67,
            lower whisker=1.45
        },
        fill=red!20, draw=red, thick
    ] coordinates {};

    \addplot+[
        boxplot prepared={
            draw position=7,
            median=8.82,
            upper quartile=67.14,
            lower quartile=4.66,
            upper whisker=165.62,
            lower whisker=1.50
        },
        fill=red!20, draw=red, thick
    ] coordinates {};

    \end{axis}
    \end{tikzpicture}
    \caption{Comparison of inter-annotator variability (distance of each annotator's marker from
    the mean of the other two annotators of the same image, as in Table~\ref{tab:annotator-stats}, on the 50 ImageCAS test volumes) against the benchmark models' localization
    error on the test set (Table~\ref{tab:res-1000}), both ostia combined.}
    \label{fig:boxplot_variabilita}
\end{figure}
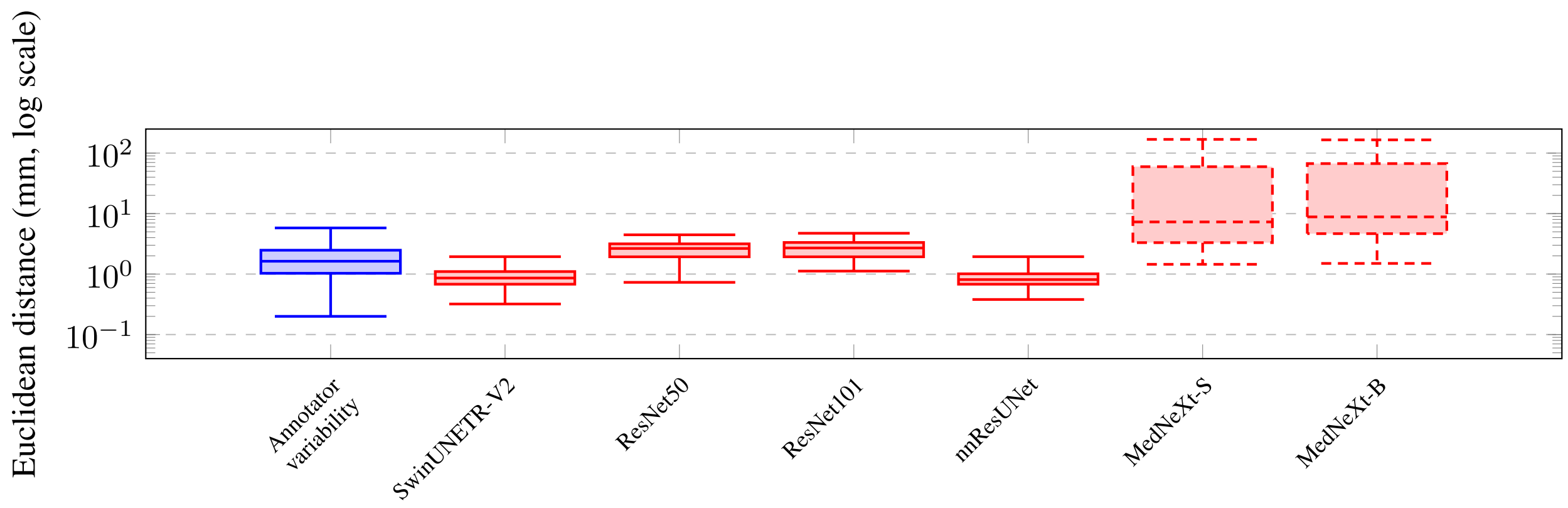

To assess whether the localization errors of the benchmark models are compatible with the variability between human readers, Figure~\ref{fig:boxplot_variabilita} compares the models' errors on the 58 test volumes (Table~\ref{tab:res-1000}) with the distance between each annotator's marker and the mean of the other two markers of the same volume (Table~\ref{tab:annotator-stats}), for both ostia combined. The annotators worked only on ImageCAS, so the annotator distances are those of the 50 ImageCAS test volumes, which are among the volumes on which the models are tested.

On these volumes, the annotators placed their markers at a median distance of 1.48~mm (interquartile range 0.95--2.33~mm) from the mean of the other two annotators for the right ostium, of 1.81~mm (1.16--2.62~mm) for the left ostium, and of 1.63~mm (1.03--2.48~mm) for the two ostia pooled. The errors of \swinv2 and \nnresunet were lower: their median distance over the two ostia was 0.86~mm (interquartile range 0.68--1.10~mm) and 0.81~mm (0.68--1.01~mm), their per-side medians of 0.70--0.87~mm were below the annotators' medians of both sides, and the largest per-patient error was 1.94~mm for both models, whereas the median errors of the ResNet and MedNeXt models lay above the annotators' upper quartile. Since the models are trained to reproduce the consensus, this indicates an agreement with the reference at the level of a human reader, not a superiority over one (\ref{subsec:app-buckets}\ofsupplementary).

The two distributions are not computed on exactly the same samples: the annotator distances are measured on the 50 ImageCAS test volumes, and the model errors on all the 58 test volumes, the other 8 being from ASOCA and CAT08, which were not annotated by the annotators. They are measured in the same way, however: each annotator against the mean of two markers to which it did not contribute, and each model against a consensus to which it did not contribute, on volumes not used for its training; the two-marker reference of the annotators is noisier than the three-marker consensus used for the models, a difference that favors the models and is one more reason to read the comparison as agreement with the reference, not as superiority. Moreover, the combined annotator distribution pools the right and left distances, whereas the combined model errors are the per-patient means of the two ostia (Section~\ref{subsec:evaluation-metrics}), whose averaging narrows their spread; the comparison of the medians is less affected by this difference than that of the extremes.

\subsection{Coronary centerline extraction}
\label{subsec:res-centerlines}
The centerline extraction of Section~\ref{subsec:centerline-methods} completed on 999 of the 1000 ImageCAS volumes. In the remaining patient, no tree could be grown from the right ostium, since the mask does not reach it. The extracted coronary graphs have a median cumulative length of 693~mm per patient (interquartile range 592--797~mm), 404~mm for the left and 278~mm for the right tree, with a median of five endpoints per tree; flattened views of the trees of 20 patients are shown in~\ref{subsec:app-flattened}\ofsupplementary. In 17 patients, 19 interruptions of the mask, from 1.0 to 8.7~mm long, were bridged. The other statistics of the extraction (distance of the ostia from the mask, activity of the capped second pass, re-attached branches, effect of the smoothing) are reported in~\ref{subsec:app-centerlines-imagecas}\ofsupplementary.

The quality control found no anomaly in 756 patients. In 129 patients, it raised a point for inspection only (\ref{subsec:app-centerlines}\ofsupplementary), such as a bridged gap or the left and right trees touching, which occurred in 25 patients overall. 111 patients were flagged as possibly missing a branch, and 4 as failed: the patient in which the extraction did not complete, and three patients in which a mask region of at least 20~mm was left unattached or a tree was shorter than 20~mm. Across the cohort, 662 accepted branches, in 459 patients, add between 5 and 10~mm of new vessel to their tree.

The visual review of the flagged patients confirmed a missed branch in 29 of the 111 patients flagged as possibly missing one, that is, in 2.9\% of the 999 extracted patients; all of these branches are small and distal, and the count includes a few very short terminal branches whose omission is borderline. In the other flagged patients, no relevant branch was missing, and none was missing in the random sample of 50 unflagged patients. Of the 25 patients whose left and right trees touch in the mask, 12 had a distal vessel segment assigned to the wrong tree, or one that could not be attributed to either tree with confidence. The other flagged situations (ostia far from the mask, paths moved by the capped pass, bridged gaps) were confirmed correct, with a single bridge left unconfirmed because of an ambiguous mask (\ref{subsec:app-centerlines-imagecas}\ofsupplementary).

On the 40 annotated ASOCA volumes, the extraction completed in every case, and the agreement with the reference centerlines is reported in Table~\ref{tab:centerlines-asoca}. The extracted trees cover a median of 99.5\% of the reference length (recall), and no reference branch or distal portion of at least 10~mm was missed or truncated in any patient. A median of 98.0\% of the extracted length is matched by the reference (precision); the unmatched part consists of 13 extra branches, from 10 to 45~mm long, in 10 patients, all of them vessels segmented in the ASOCA masks but not traced in the reference centerlines, and of shorter side branches. Where both centerlines are present, the extracted one lies at a mean distance of 0.22~mm from the reference (median over patients), the 95th percentile of the symmetric distance is 0.56~mm, and the lumen radius differs from the reference one by a median of $-0.05$~mm. On average, 96.3\% of the reference endpoints and 97.5\% of the reference bifurcations have an extracted counterpart within 10~mm and 5~mm, respectively. The left and the right tree perform similarly, with a slightly lower precision for the left tree, where most of the extra branches lie.

\begin{table}[t]
\centering
\caption{Agreement between the centerlines extracted from the 40 annotated ASOCA masks and the ASOCA reference centerlines, for both trees together and for each tree, as median [interquartile range] over the patients. Recall: fraction of the reference length matched (coverage); precision: fraction of the extracted length matched; OV, OT, OF: overlap, overlap of vessels at least 1.5~mm wide, and overlap until first error (first 40~mm of every path ignored); AI: average inside, the mean distance of the matched reference points from the extracted centerline; HD95: 95th-percentile symmetric Hausdorff distance; length ratio: extracted over reference length; radius error: extracted minus reference radius on matched points (signed mean) and its mean absolute value; endpoints: one-to-one pairing within 10~mm; bifurcations: fraction of reference bifurcations paired within 5~mm.}
\label{tab:centerlines-asoca}
\fontsize{6.4}{7.5}\selectfont
\setlength{\tabcolsep}{3pt}
\begin{tabular}{lccc}
\toprule
 & Both trees & Left tree & Right tree \\
\midrule
Recall (\%)             & 99.5 [98.9--99.8] & 99.4 [98.7--99.8] & 99.7 [99.0--99.9] \\
Precision (\%)          & 98.0 [96.5--99.6] & 98.1 [95.7--99.8] & 99.4 [97.7--100] \\
OV (\%)                 & 98.6 [98.2--99.0] & 98.5 [97.6--99.1] & 99.1 [98.4--99.5] \\
OT (\%)                 & 99.1 [98.8--99.9] & 99.0 [98.4--100] & 100 [99.0--100] \\
OF (\%)                 & 99.1 [96.8--99.5] & 99.0 [96.3--99.6] & 99.5 [98.6--99.9] \\
AI (mm)                 & 0.22 [0.21--0.24] & 0.22 [0.21--0.25] & 0.22 [0.20--0.24] \\
HD95 (mm)               & 0.56 [0.49--0.62] & 0.57 [0.52--0.73] & 0.49 [0.44--0.59] \\
Length ratio            & 0.99 [0.97--1.01] & 0.99 [0.95--1.02] & 0.98 [0.96--1.00] \\
Radius error (mm)       & $-$0.05 [$-$0.07--$-$0.03] & $-$0.04 [$-$0.07--$-$0.02] & $-$0.06 [$-$0.08--$-$0.04] \\
Radius abs.\ error (mm) & 0.10 [0.09--0.11] & 0.10 [0.08--0.11] & 0.11 [0.09--0.12] \\
Endpoint recall (\%)    & 100 [90.9--100] & 100 [100--100] & 100 [100--100] \\
Endpoint precision (\%) & 88.9 [83.3--100] & 85.7 [80.0--100] & 100 [95.8--100] \\
Endpoint distance (mm)  & 1.75 [1.49--2.28] & 1.71 [1.38--2.19] & 1.82 [1.18--2.54] \\
Bifurcations $\leq$5~mm (\%) & 100 [100--100] & 100 [100--100] & 100 [100--100] \\
\bottomrule
\end{tabular}
\end{table}

ImageCAS-X~\citep{imagecasx2026} provides curated masks for 800 of the 1000 scans and centerlines derived from them. A difference between its centerlines and ours could come from the extraction method or from the masks, so we ran our extraction on both sets of masks of the 800 shared scans and compared the results in three ways (Table~\ref{tab:imagecasx-differences}): on the ImageCAS-X masks, to check that the method alone does not alter the paths; at the start of the trees, to compare the ImageCAS-X start points with our annotated ostia; and across the two sets of masks, to see what depends on the masks. On the ImageCAS-X masks our extraction reproduces its centerlines (median recall 100\%, precision 99.9\%, mean distance 0.23~mm; \ref{sec:app-centerlines-imagecasx}\ofsupplementary). Its start points lie at a median of 2.45~mm (left) and 1.07~mm (right) from the annotated ostia, and its trees are stored as separate segments that must be rejoined in 15.6\% of the trees. The masks differ in both directions: a median of 8.8\% of the ImageCAS-X centerline length lies outside the official masks and 18.7\% of ours outside the ImageCAS-X masks, and our right trees are longer and branch more on the official masks than on the ImageCAS-X masks (median length 251 against 198~mm, 3 against 1 bifurcations). In short, the paths agree where the masks are the same, and the differences lie in the masks, in the start point and in how the trees are stored; what our centerlines add is one connected graph per side, rooted at an annotated ostium and carrying the lumen radius at every node, for 999 of the 1000 scans of the official release, including the 200 that ImageCAS-X does not contain.

\begin{table}[tb]
\centering
\caption{Our centerlines compared with those released with ImageCAS-X, on the 800 scans the two datasets share. Median with its interquartile range below, or percentage of the trees or of the pooled length. \emph{Detached length}: part of a tree that cannot be reached from the start point when segments closer than 1~mm are joined; \emph{bridged by the mask}: share of it that the ImageCAS-X mask connects (a gap of one voxel allowed). \emph{Start point}: ImageCAS-X start point against our annotated ostium (interoperator threshold: Table~\ref{tab:thresholds}). \emph{Inside the other masks}: share of the centerline length lying within one voxel of the other set's masks. \emph{Tree size}: our extraction on the official and on the ImageCAS-X masks of the same patients (all differences significant, paired Wilcoxon signed-rank test, $p<0.01$).}
\label{tab:imagecasx-differences}
\fontsize{6.4}{7.5}\selectfont
\setlength{\tabcolsep}{3pt}
\newcommand{\medIQR}[2]{\makecell{#1\\[-2.5pt]{\fontsize{5}{5.5}\selectfont[#2]}}}
\begin{tabular}{@{}p{3.3cm}ccc@{}}
\toprule
 & Both trees & Left tree & Right tree \\
\midrule
\multicolumn{4}{l}{\textit{ImageCAS-X trees: connectivity}} \\
Trees with detached segments (\%)        & 15.6 & 28.6 & 2.5 \\
Detached length, in those trees (\%)     & \medIQR{47.2}{36.1--58.5} & \medIQR{47.7}{38.2--59.2} & \medIQR{24.9}{21.1--36.9} \\
Detached length, all trees (\%)          & 9.7 & 14.6 & 1.0 \\
Detached length bridged by the mask (\%) & 97.7 & 97.6 & 100 \\
\midrule
\multicolumn{4}{l}{\textit{ImageCAS-X start points against our ostia}} \\
Distance (mm)                            & -- & \medIQR{2.45}{1.84--3.06} & \medIQR{1.07}{0.71--1.60} \\
Within the interoperator threshold (\%)  & -- & 55.5 & 83.5 \\
Beyond 5~mm (\%)                         & -- & 1.0 & 0.4 \\
\midrule
\multicolumn{4}{l}{\textit{Centerline length inside the other masks (\%)}} \\
ImageCAS-X, in the official masks        & \medIQR{91.2}{86.1--94.7} & \medIQR{91.3}{84.7--95.8} & \medIQR{93.0}{85.7--98.4} \\
Ours, in the ImageCAS-X masks            & \medIQR{81.3}{75.3--86.3} & \medIQR{89.6}{81.6--94.7} & \medIQR{71.4}{62.9--80.8} \\
\midrule
\multicolumn{4}{l}{\textit{Tree size (ours, on each set of masks)}} \\
Length (mm), official masks              & -- & \medIQR{364}{304--423} & \medIQR{251}{198--315} \\
Length (mm), ImageCAS-X masks            & -- & \medIQR{353}{308--408} & \medIQR{198}{167--232} \\
Bifurcations, official masks             & -- & \medIQR{5}{4--6} & \medIQR{3}{2--4} \\
Bifurcations, ImageCAS-X masks           & -- & \medIQR{4}{4--5} & \medIQR{1}{1--2} \\
\bottomrule
\end{tabular}
\end{table}

\FloatBarrier
\section{Discussion}
\label{sec:discussion}

\subsection{Implications and reuse}
\label{sec:discussion-implications}

Progress on methods for coronary ostium localization has been difficult to measure. They were developed on private datasets ranging from 118 to 1360 scans \citep{zheng2011ostia, wolterink2018landmarks, mostafa2021ostia,gajowczyk2022ostia}. The public data were limited to the 72 scans of CAT08 and ASOCA, in which the ostium is only implicitly encoded and of which only ASOCA is still openly accessible, and to the 77 non-gated aortic CT angiography scans of ImageTBAD \citep{yao2021imagetbad} annotated by \citet{gajowczyk2022ostia}. These scans, acquired to study Type-B aortic dissection and not dedicated to imaging the coronary ostia, were not used in this work. This work addresses ECG-gated CCTA, in which the ostia are imaged at diagnostic quality, whereas on these non-gated scans the only model evaluated so far had median errors of 3.48~mm (left) and 2.97~mm (right), against 1.14 and 0.98~mm on gated CCTA \citep{gajowczyk2022ostia}, a gap its authors attribute to the lower resolution and the artefacts of the images (Section~\ref{subsec:rw-ostia}). CAT08 and ASOCA were included to add other centers, scanners and operational definitions of the ostium, whereas ImageTBAD was acquired at the same hospital as ImageCAS. The publication describes no definition of the ostium and no measure of its variability, so that they would have needed a harmonization of their own (\ref{sec:app-consistency}\ofsupplementary), and the training-set-size ablation shows that, beyond a few hundred volumes, more training data change the results little (Section~\ref{sec:discussion-datasize}). Their use as an external test on non-gated scans is left to future work. Each work adopted its own definition of the ostium, from the start point of reference centerlines \citep{cat08} to protocols designed for valve planning \citep{wolterink2020landmarks}, measured the variability of manual placement only on small subsets \citep{wolterink2020landmarks,jang2021ostia}, if at all, and summarized its results with its own statistics. This work addresses these issues together. It provides annotations of both ostia for 1000 volumes, 500 by consensus and 500 by model-assisted correction, harmonized with the implicit definitions of CAT08 and ASOCA into a single landmark; it measures the annotation variability on 1500 annotations of each ostium, three per volume in 500 volumes; and it releases the weights and the test split, with results reported both as distributional statistics and as threshold-based grades.

Beyond the benchmark, the resources are meant to be reused as components of larger systems. The training-set-size ablation (Section~\ref{sec:discussion-datasize}) indicates that the new annotations are needed to learn the task reliably, and the best released models localize both ostia with median errors below 1~mm, and should therefore be accurate enough to provide automatic seeds for coronary segmentation and centerline tracking, and the ostium positions required by the planning of surgical and transcatheter interventions and by the assessment of anomalous coronary origins \citep{aortic_valve_01,aortic_valve_02,ImageALCAPA}. The centerline graphs, rooted at the same ostia and always connected, can be used directly for curved planar reformation, branch labeling, stenosis grading or hemodynamic modeling. The second annotation stage, finally, shows how such resources can be extended at a low cost: correcting the candidates of a model trained on the first batch took about 5 person-hours, against about 144 for the first batch, annotated three times (Section~\ref{subsec:res-annotations}), and the released models can likewise assist the annotation of further datasets.

\subsection{Consistency of the ostium definition across datasets}
\label{sec:discussion-consistency}
The single ostium definition reconciles three different operational procedures: the native centerline start point of CAT08, the first centerline point of ASOCA, and the consensus of the ImageCAS annotators. The comparison of Section~\ref{subsec:res-ostia-consistency} found no gross definitional inconsistency: the only difference, an offset of about 5--6$^\circ$ in the left-to-right axis of ImageCAS, displaces each ostium by about 1.5~mm, within the variability between annotators, and may equally reflect differences in population, scanners or patient positioning rather than in the landmark (\ref{sec:app-consistency}\ofsupplementary).

\subsection{Model choice}
None of the network architectures of the benchmark had been evaluated as a dedicated ostium localizer with per-ostium results. Residual U-Nets had been used by \citet{gajowczyk2022ostia}, but not in the nnU-Net residual-encoder configuration adopted here; SwinUNETR only as the student network of a federated, multi-task model for transcatheter valve planning, which did not evaluate the ostia separately \citep{tolle2025federated}; and ResNet-based heatmap regressors and MedNeXt not at all. In the absence of prior evidence, \swinv2 was selected for the first experiments, including the model that generated the candidates of the second annotation stage (\ref{sec:app-stage2}\ofsupplementary), as a widely adopted general-purpose 3D architecture, without a preliminary comparison.

\subsection{Benchmark results}
\label{sec:discussion-models}
The results (Table~\ref{tab:res-1000}) show that neither the architecture type nor the size of the models (Table~\ref{tab:models_specs}) determines the outcome alone. \swinv2 (hybrid Swin transformer and convolutional decoder, 18.35M parameters) and \nnresunet (purely convolutional residual U-Net, 32.69M) reached nearly the same accuracy, with differences too small to be resolved by the test set: neither model failed any localization, and 98.3\% and 96.6\% of the cases, respectively, were Optimal or Good on both sides. Almost every predicted point therefore lies within the ostial radius expected for the population or within the spread of a human annotator, although this does not guarantee it for an individual patient with a smaller ostium. The convergence of a model built on self-attention and a purely convolutional network on the same error below the interoperator thresholds suggests that on this task the limit is set by the reference annotations and by the training recipe rather than by the type of architecture; the choice between the two can then rest on secondary considerations, such as the processing time (41.1~s per CCTA for \nnresunet against 55.6~s for \swinv2).

For the two ResNet models, the number of parameters does not seem to determine the accuracy: ResNet101 has almost twice the parameters of ResNet50 and no better accuracy, and both are larger than \swinv2 and \nnresunet yet about three times less precise. Their errors are systematic: failures are almost absent, yet 82.8\% of the cases of ResNet50 were only acceptable, so the networks find the right region but place the point imprecisely, a difference that a binary failure criterion would not reveal. A possible explanation lies in the output stage: the ResNets predict the heatmap with a light head and recover the input resolution by trilinear upsampling, without a decoder with skip connections, and fine spatial detail is lost regardless of the depth of the encoder. Being the fastest models, they offer a lower-cost but clearly less precise alternative.

The weakness of the two MedNeXt models is their reliability: many ostia were localized more than 5~mm away, with errors up to about 190~mm, i.e.\ on a different structure. Since MedNeXt-B, with almost twice the parameters of MedNeXt-S, is not better, the cause more likely lies in the training of these networks on this task, or in how well their design suits it, than in their capacity; as the training recipe was the same for all the models, the result says little about the MedNeXt design as such. In this setting, therefore, the models that restore the spatial resolution through a proper decoder are the accurate ones, whereas neither depth alone (ResNet101) nor a larger version of a different design (MedNeXt-B) compensates for it.

A comparison with the published methods remains indirect, since most of them were developed on private datasets \citep{zheng2011ostia,wolterink2020landmarks,mostafa2021ostia}. On the CAT08 cases, \swinv2 and \nnresunet reached mean errors between 1.04 and 1.20\,mm, against 1.94\,mm (right) and 2.22\,mm (left) reported by \citet{jang2021ostia}, although on four rather than eight test cases from a different split, and all their predictions fell within 2.6\,mm, whereas previous works reported errors above 5\,mm in 0.8\% to 7.3\% of the ostia \citep{zheng2011ostia,mostafa2021ostia}. Since a distance in millimeters depends on the cohort, on the image resolution and on the definition of the ostium, these values, collected in~\ref{sec:app-literature}\ofsupplementary, are only indicative; for this reason, we release the test split and the model weights.

All these considerations come from a single training run per model, on the same recipe, and the processing time includes volume loading and the export of the results, so small differences, including those between \swinv2 and \nnresunet, should not be read as a ranking.

\subsection{Ablation studies}
\label{sec:discussion-ablations}

\subsubsection{Training set size}
\label{sec:discussion-datasize}
Trained on the volumes of CAT08 and ASOCA alone, \swinv2 fails on about a third of the right and half of the left ostia, including those of CAT08 and ASOCA themselves (Table~\ref{tab:app-datasize}): \nnresunet, trained in the same way, is more robust, with median errors of 1.33 and 1.03~mm, but still fails on 6.9\% of the right and 3.4\% of the left ostia, in five ImageCAS volumes, whereas the same model trained on the three datasets never failed on this test set (\ref{subsec:app-datasize}\ofsupplementary): the public resources that existed before are not sufficient to train a reliable ostium localizer. Adding 250 ImageCAS volumes, with the same architecture, loss and inference pipeline, removes all the failures and brings the median errors below 1~mm. Adding these volumes therefore turned a model that could not be used into one that can; as the architecture, loss and inference pipeline were the same, the gain is attributable to the added data, although, with one training per arm, their number and their annotations cannot be told apart. Beyond 250 volumes, the differences are small and not monotonic, and the models trained with 500 and 750 volumes failed again on one and three ostia, mostly of ASOCA, although the model trained on the complete dataset obtains the best value in six of the ten distance columns, more than any smaller training set. The ablation thus supports the premise of this work, that the public data available before were not sufficient, and suggests that a few hundred volumes annotated with this protocol are enough to reach an accuracy close to that of the full set, which is relevant for whoever needs to extend the annotation to another dataset. The occasional failures at intermediate sizes, and the reasons why the benefit of the larger training sets may be underestimated by this test set, are discussed in~\ref{subsec:app-datasize}\ofsupplementary.

\paragraph{External validation} When trained on ImageCAS alone, \swinv2 and \nnresunet still localized the ostia of the CAT08 and ASOCA volumes, which they had never seen, with one failure and two failures, respectively, in 72 volumes; the median errors were somewhat higher than on ImageCAS (0.77--1.18~mm against 0.66--0.91~mm for the two models) and the share of Optimal localizations was lower, but most localizations were Optimal or Good (Table~\ref{tab:app-imagecas-only}). On the eight volumes that no model had seen, the errors of both models were within a few tenths of a millimeter of those of the models trained on the three datasets, higher for the right ostium and similar or lower for the left one. The failures were isolated and of the order of a hundred millimeters, which suggests that a different structure was located rather than the ostium imprecisely placed, and the two architectures behaved alike. The models therefore transfer, at a slightly lower accuracy, to other scanners, centers and operational definitions of the ostium, and the accuracy of the benchmark models on CAT08 and ASOCA is not merely the result of having trained on them. Read together with the previous paragraph, the two experiments are the mirror image of each other. Trained on ImageCAS, the models work on the volumes of CAT08 and ASOCA, which they had never seen. Trained on these two datasets, the models do not work on ImageCAS: \swinv2 fails on a third to half of the ostia, also on the CAT08 and ASOCA volumes it was not trained on, and \nnresunet, although more robust, fails on 8\% of the right and 4\% of the left ostia of the ImageCAS test volumes and has a median error of 1.37 and 1.09~mm, against 0.70 and 0.87~mm without failures when trained on the three datasets. Although these experiments differ in more than the amount of data, with about 950 annotated volumes against 64, the architecture, loss and inference pipeline being the same, the asymmetry follows the number of annotated volumes more than the similarity between the images: a large and consistently annotated training set carries the model across scanners and centers, whereas a small one yields a model that is less accurate and less reliable, to an extent that depends on the architecture (more for \swinv2 than for \nnresunet). This is the gap that the annotation of ImageCAS fills, and the reason why the size of the annotated dataset, more than the choice of the architecture, is what makes the training of a reliable ostium localizer possible. The conclusion rests on 72 external volumes and on a single training per model, so that isolated failures cannot be attributed to either architecture (\ref{subsec:app-imagecas-only}\ofsupplementary).

\subsubsection{Spacing, peak window, and overlap}
\label{sec:discussion-respacing-optimal}
The other four ablations (\ref{sec:app-ablations}\ofsupplementary) translate into practical rules for the use of the released models. Training and testing at a coarser spacing retains most of the accuracy: neither \swinv2 nor \nnresunet fails at any of the tested spacings, up to 1.5~mm; at 1.0~mm, the share of Optimal cases at patient level is the same or higher than at 0.5~mm, and the processing time per patient falls to less than half for \swinv2 (from 55.6 to 21.9~s) and to about half for \nnresunet (from 41.1 to 21.3~s), whereas at 1.5~mm the errors grow and the time decreases by at most 3~s more; 1.0~mm therefore appears to be the coarsest spacing worth using (\ref{subsec:app-respacing-optimal}\ofsupplementary). The models, however, tolerate only a small mismatch between the training and the inference spacing: \swinv2 trained at 0.5~mm starts to fail from 0.7~mm and fails on every ostium at 1.5~mm, while the model trained at 1.0~mm does not fail at all, so the failures are caused by the change in the apparent scale of the anatomy, not by the loss of resolution (\ref{subsec:app-respacing-mismatch}). The size of the peak window has little influence up to 21 voxels, so the default of 11 voxels, fixed beforehand, lies in a flat region (\ref{subsec:app-window}); the sliding-window overlap matters little from 32 voxels upwards, and an overlap of 32 voxels makes the processing about 40\% (\swinv2) and 24\% (\nnresunet) faster than the default of 64 voxels, whereas removing it lowers the accuracy (\ref{subsec:app-overlap}). In short, the volumes must be resampled to the spacing at which the model was trained, a faster processing is obtained by training at a coarser spacing rather than by resampling at inference, and an overlap of 32 voxels is a good compromise between accuracy and speed.

\subsection{Centerline extraction}
\label{sec:discussion-centerlines}

On ASOCA, where the reference and the extracted centerlines derive from the same masks, the comparison measures the extraction alone, and its error is well below the resolution of the data (Table~\ref{tab:centerlines-asoca}): the matched centerlines lie at a mean distance from the reference smaller than half of the 0.5~mm grid on which the trees are grown, and no reference branch of 10~mm or more is missed. The differences lie mostly in the small side branches that the extraction follows down to 5~mm of new length and the reference does not trace; whether such branches should be part of a reference tree is a matter of convention. Since the reference centerlines of ASOCA were also computed as minimal paths, by the Vascular Modeling Toolkit~\citep{antiga2003geometry}, part of the close agreement may reflect the similarity of the two methods rather than the accuracy of either.

Growing the trees along minimal paths from the ostia, instead of thinning the mask, removes the two classical artifacts of skeletonization at the source: the result is always a tree, so no loop has to be detected and cut, and spurious side branches are discarded by a single length criterion, chosen from the lengths of the rejected candidates, without heuristic pruning; the accepted branches shorter than 10~mm are listed in the release, so that a user who prefers a more conservative tree can prune them. The capped second pass rarely changes the result (\ref{subsec:app-centerlines-imagecas}\ofsupplementary), but it leaves every regular vessel segment unchanged by construction and corrects the path at the bulky bifurcations, where a centerline displaced from the lumen would compromise branch labeling, curved planar reformation or the measurement of bifurcation angles; it is therefore kept as a safety net.

On ImageCAS, the quality of the centerlines is bounded by that of the masks, and the quality control is designed to expose the mask defects. Where the left and right masks touch, every shared voxel is assigned to the nearer ostium along the mask, but the boundary between the two trees is then arbitrary; where an interruption is bridged, an interrupted vessel cannot be told apart from a nearby fragment, which is why the bridge is limited to 10~mm. Both situations are flagged, and the patients flagged as failed all trace back to defects of the masks, not to the extraction (\ref{subsec:app-centerlines-imagecas}\ofsupplementary). The missed branches confirmed by the visual review are all small and distal vessels whose absence does not change the structure of the tree, and none was found in the random sample of unflagged patients, which supports the quality control as a screen for this kind of error. The main weakness of the official masks lies elsewhere: irregular protrusions, spurious stubs and implausible fragments, which the extraction keeps out of the trees, and, where the left and right masks touch, distal segments assigned to the wrong tree, mostly in the inferior part of the heart, where the distal vessels are thin, run close to each other and are imaged with less contrast.

ImageCAS-X~\citep{imagecasx2026} re-annotates the masks of 800 ImageCAS scans and derives centerlines from them automatically by thinning the new masks; unlike this work, it provides no ostium landmarks, and its centerlines are not rooted at them. In contrast, our extraction, run on its masks, reproduces the paths of its centerlines (\ref{sec:app-centerlines-imagecasx}\ofsupplementary). The two centerline sets differ in the masks they follow, in their coverage, and in their content (Table~\ref{tab:imagecasx-differences}). A median of 8.8\% of the ImageCAS-X centerline length lies outside the original masks, so our graphs provide the centerlines of the official masks, and they cover all the scans of the official four-fold cross-validation of ImageCAS~\citep{imagecas}, 200 of which are excluded from ImageCAS-X in favor of a split of its own; on the official masks, our trees also contain proximal side branches, mostly on the right. Each of our trees is a single connected graph that carries the lumen radius at every node, as needed for stenosis grading and hemodynamic modeling, and is rooted at an annotated ostium, the same landmark localized by the benchmark models, so that ostium localization and centerline extraction can be chained and evaluated against one shared reference. The ImageCAS-X centerlines carry no radius, start at skeleton endpoints near an automatically segmented aorta, and are stored as separate segments, which in 15.6\% of the trees must be rejoined across gaps of a few millimeters. Since the extraction needs only a mask and the two ostia, it was also applied to the ImageCAS-X masks: the official set is the default, and the graphs of the ImageCAS-X masks are released as an alternative set, for users of those masks and where the official mask is defective (\ref{sec:app-data-availability}\ofsupplementary).

\subsection{Limitations}
This work has some limitations. First, the CCTAs of ImageCAS were acquired at a single institution, while CAT08 and ASOCA contribute only 32 and 40 volumes, respectively, so the per-dataset statistics are fragile, and differences of a few tenths of a millimeter between models or settings should be interpreted with caution. Gender was not considered in the design of the study or in the analysis: the ImageCAS cohort includes 586 male and 414 female patients~\citep{imagecas}, but the gender of the individual patients is not distributed with the images, so neither the annotations nor the benchmark results could be stratified by gender.

Second, the reference positions have an intrinsic uncertainty of the same order of magnitude as the errors of the models (Table~\ref{tab:annotator-stats}). Additionally, the annotators were biomedical engineers (five highly trained master's students and one PhD student) rather than clinicians, and the three datasets define the ostium in different operational ways. Moreover, the labels of the second half of the ImageCAS dataset were obtained by a single annotator, the one with the lowest mean distance from the other two annotators in the first stage (Annotator 4), who reviewed and corrected the candidates of a model. Additionally, the labels have not been verified independently. For this reason, the test set was drawn only from the first, independently annotated half. Since the model only proposed the candidate positions, each of which the annotator inspected on the CCTA volume and corrected where needed, the labels should not differ materially from manual ones, but the accepted candidates could still favor the \swinv2 architecture; that \nnresunet reaches the same accuracy suggests that the effect, if any, is small.

Third, the annotation protocol assumes exactly one ostium per side: anatomical variants, such as separate ostia of the left anterior descending and circumflex arteries, a single coronary ostium or anomalous origins, are not represented by the labels, and the behavior of the models on them has not been quantified. Fourth, the centerline graphs inherit the defects of the ImageCAS segmentation masks, which do not affect the ostium annotations, placed on the images; their validation against reference centerlines covers the 40 ASOCA volumes only, and on ImageCAS the visual review covered the flagged patients and a random sample of 50 of the others; a small number of branches missed in unflagged patients therefore cannot be excluded.

Finally, the benchmark covers a limited set of architectures and a single training recipe, each model is evaluated from a single training run, and the processing time was measured over the whole test-set run on a single GPU limited to 20 of its 40~GB of memory, including image loading and export, without a separate measurement of the network inference time or of CPU-only execution. The models are research resources: they were evaluated retrospectively, on public data, and have not been validated clinically or prospectively, so their use within a decision-support system would require a dedicated clinical evaluation.

\section{Conclusions}
\label{sec:conclusions}

This work extends ImageCAS, the largest public CCTA dataset for coronary artery segmentation, with coronary ostium landmarks for all its 1000 CCTAs and with coronary centerline graphs rooted at those landmarks, and establishes on them an open-weight benchmark for ostium localization. The ostia of the first 500 volumes were placed by three annotators each and consolidated by consensus; those of the remaining 500 were obtained by correcting the candidates of a deep learning model trained on the first batch. 
On the benchmark, \swinv2 and \nnresunet localized every ostium of the test set without failures, with median errors between 0.70~mm and 0.87~mm, below the interoperator thresholds, whereas the ResNets were markedly less precise and the MedNeXt models unreliable. The training-set-size ablation shows that the previously available public data alone do not suffice to train a reliable localizer, whereas 250 ImageCAS volumes removed all the failures.
The centerlines were extracted in 999 of the 1000 patients and, on ASOCA, cover a median of 99.5\% of the reference length, with a median per-patient mean distance of 0.22~mm between matched reference points and centerlines, without missing any reference branch of 10~mm or longer.
Future work includes the independent validation of the second batch, the evaluation on anomalous coronary origins and on non-gated CT angiography, such as the annotated ImageTBAD volumes, and the use of the ostia as seeds for coronary tracking and segmentation. The annotations, the centerlines, the code and the weights of the models are available, together with an evaluation scheme based on the ostial anatomy and on the variability between annotators, as a common basis on which future methods and decision-support pipelines for CCTA analysis can be built and compared.

\section{Data and Code Availability}
\label{sec:data-availability}

The CCTA volumes and segmentation masks belong to the public ImageCAS~\citep{imagecas}, CAT08~\citep{cat08,cat08_2}, ASOCA~\citep{asoca1} and ImageCAS-X~\citep{imagecasx2026} datasets and are not redistributed: they must be obtained from their original sources. The ostium annotations of all the 1000 ImageCAS volumes and the centerline graphs are released on Zenodo~\citep{awakening_dataset}, \url{https://doi.org/10.5281/zenodo.23243093}, and the weights of the \swinv2, ResNet50, ResNet101 and \nnresunet models in a separate record~\citep{awakening_weights}, \url{https://doi.org/10.5281/zenodo.23399623}, both under the CC BY-NC 4.0 license. The code for training, inference and evaluation, with the lists of the training, validation and test volumes, is on GitHub~\citep{awakening_code}, \url{https://github.com/AAMIASoftwares-research/HCATOstiaLocalizationRelease}, under the MIT license. All of them are linked from \url{https://hal9000-lab.github.io/}. The contents, formats and licenses of the released material, and the terms of the third-party data, are detailed in \ref{sec:app-data-availability}\ofsupplementary.

\section*{CRediT Authorship Contribution Statement}

\textbf{Matteo Leccardi}: Conceptualization, Methodology, Software, Validation, Formal analysis, Investigation, Data curation, Visualization, Writing -- original draft, Writing -- review \& editing.
\textbf{Andrea Moglia}: Conceptualization, Methodology, Validation, Supervision, Project administration, Writing -- review \& editing. Closely followed the project throughout its development and acted as the main reviewer of the methodology, the results and the manuscript.
\textbf{Marco Morandini}, \textbf{Michele Brembilla}, \textbf{Alessandro Nicosia} and \textbf{Andrea Penserini}: Conceptualization, Methodology, Software, Validation, Formal analysis, Investigation, Data curation, Visualization, Writing -- original draft, Writing -- review \& editing. All four authors contributed equally to the design and implementation of the annotation protocol, to the manual and semi-automated annotation of the coronary ostia, to the development, training and evaluation of the benchmark models, and to the preparation of the manuscript.
\textbf{Christian Brandini}: Investigation, Data curation. Contributed to the manual annotation of the coronary ostia in the first annotation phase.
\textbf{Ender Konukoglu}: Supervision, Resources, Writing -- review \& editing.
\textbf{Marco Marcon}, \textbf{Luca Mainardi} and \textbf{Pietro Cerveri}: Conceptualization, Supervision, Project administration, Resources, Funding acquisition, Writing -- review \& editing.

\section*{Declaration of Competing Interest}

The authors declare that they have no known competing financial interests or personal relationships that could have appeared to influence the work reported in this paper.

\section*{Acknowledgments}

This work was partially supported by P.E. PE00000013-FUTURE ARTIFICIAL
INTELLIGENCE RESEARCH (FAIR), Italian Ministry of Research and University, and
Fondo di Beneficenza IntesaSanPaolo (IT) (grant: B/2023/0201). AM is partially supported by the European Union project SMASH-HCM under grant
101137115.

\section*{Declaration of Generative AI and AI-Assisted Technologies in the Writing Process}

During the preparation of this work the authors used the chat assistants Gemini (Google) and Claude (Anthropic) in order to reword sentences and to check grammar and spelling. After using these tools, the authors reviewed and edited the content as needed and take full responsibility for the content of the publication.

\bibliographystyle{unsrtnat}
\bibliography{Bibliography/landmark_detection,Bibliography/datasets,Bibliography/others}

\clearpage
\FloatBarrier
\appendix
\counterwithin*{figure}{section}
\counterwithin*{table}{section}
\section{Additional Details on the Data and the Methods}
\label{sec:app-methods}
This appendix completes the description of the data and the methods of the article: the characteristics of the three datasets (\ref{subsec:app-datasets}) and the protocol of the manual ostium annotation (\ref{subsec:app-annotation}); the centerline extraction (\ref{subsec:app-centerlines}), its run statistics and flagged patients on ImageCAS (\ref{subsec:app-centerlines-imagecas}), a flattened view of the extracted trees (\ref{subsec:app-flattened}) and the release format of the centerlines (\ref{subsec:app-centerline-release}); the training settings (\ref{subsec:app-training}); and the full definition of the accuracy buckets (\ref{subsec:app-buckets}).

\subsection{Datasets}
\label{subsec:app-datasets}
Table~\ref{table:dataset-comparison} summarizes the raw image characteristics of the three datasets used in this work (Section~\ref{subsec:ostia-dataset}). CAT08 provides the start points of its reference centerlines in separate files (\texttt{pointS.txt}), which are used as ostium references. ASOCA does not provide explicit ostium annotations: for each patient and each side, the first points of the centerline segments of that side were averaged into a single ostium position. All the ostium references of both datasets were then inspected visually.

ASOCA~\citep{asoca1} was collected at Mercy Angiography, Auckland (New Zealand), using a LightSpeed 64 Slice scanner by General Electric Healthcare (Chicago, United States), and its segmentations were made with 3D Slicer~\citep{asoca2}. CAT08~\citep{cat08} was acquired at the Erasmus MC, University Medical Center Rotterdam (The Netherlands), 20 volumes with a 64-slice scanner (Sensation 64) and 12 with a dual-source scanner (Somatom Definition), both by Siemens Medical Solutions (Forchheim, Germany), with an image size of $512\times512\times(229-351)$; its image quality was graded as poor (six volumes), moderate (11) and good (15), based on the presence of artifacts and confidence evaluation, and its calcification as low (12 volumes), moderate (16) and severe (4)~\citep{cat08}. ImageCAS~\citep{imagecas} was acquired with a 128-slice dual-source scanner by Siemens Healthineers, with an image size of $512\times512\times(206-275)$.

\begin{table}[tb]
\centering
\caption{Comparison of the raw image characteristics of the CAT08, ASOCA and ImageCAS datasets. Spacing: in-plane pixel spacing for ImageCAS and CAT08; for ASOCA, the slice spacing, which is its largest spacing (in-plane 0.3--0.4~mm).}
\scriptsize
\begin{tabular}{l l l l l} 
  \hline \afterHLineSpacer
    & Spacing (mm) & \makecell[l]{Size\\{\tiny(with labels)}} & Resolution & N\textdegree{} of slices \vertspace\\ 
  \hline \afterHLineSpacer
  \makecell[l]{ImageCAS \\ \citep{imagecas}}
   & 0.25 - 0.45 & 1000 & 512 $\times$ 512 & 206 - 275 \vertspace\\ 
  \makecell[l]{CAT08 \\ \citep{cat08_2}} & 0.262 - 0.408 & 32 & 512 $\times$ 512 & 229 - 351 \vertspace\\ 
  \makecell[l]{ASOCA \\ \citep{asoca1}} & 0.625 & 40 & 512 $\times$ 512 & 200 \vertspace\\ 
  \hline \afterHLineSpacer
\end{tabular}
\label{table:dataset-comparison}
\end{table}

\subsection{Ostium annotation protocol}
\label{subsec:app-annotation}
Five of the six annotators of the first stage (Section~\ref{subsec:ostia-dataset}) were biomedical engineering master's students, highly trained for this task by the first author, who was the sixth annotator and had worked with CCTA images for four and a half years as the core of a doctoral project. The training combined a written definition of the coronary ostium, given to all the annotators and summarized in Figure~\ref{fig:overview-pipeline}, with the ostium references of CAT08 and ASOCA (\ref{subsec:app-datasets}), which served as worked examples of the landmark to place. Each ostium was placed in 3D Slicer by navigating the axial, coronal and sagittal views together, and every marker was checked in all three planes before being saved. The rotation of the annotator groups every 25 images, the consensus and the final visual inspection are described in Section~\ref{subsec:ostia-dataset}.

\subsection{Centerline extraction}
\label{subsec:app-centerlines}
This section describes in full the centerline extraction summarized in Section~\ref{subsec:centerline-methods} and in Figure~\ref{fig:centerline-pipeline}.

\begin{figure}[!b]
  \centering
  \includegraphics[width=0.86\linewidth]{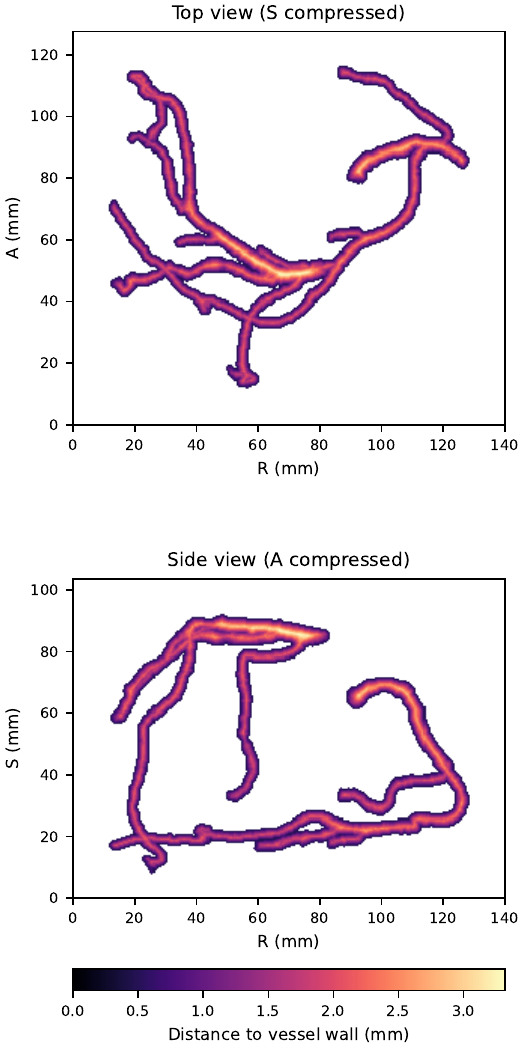}
  \caption{Distance from the vessel wall, $D$, on which the cost of the minimal paths is built, for one ImageCAS patient: maximum projections along the superior--inferior (top) and anterior--posterior (bottom) axes, in millimeters from the corner of the volume cropped around the arteries. $D$ peaks along the axis of every vessel, which the minimal paths therefore follow; it is largest in the proximal vessels and at their junctions, which motivates the cap applied in the second growth pass.}
  \label{fig:distance-map}
\end{figure}

\paragraph{Grid and cost} The mask is resampled by nearest-neighbor interpolation to an isotropic grid of 0.5~mm in Right-Anterior-Superior (RAS) orientation and cropped to the bounding box of the mask and of both ostia. To repair connections broken by the resampling, the mask is dilated by one voxel; the added voxels can be traversed, but at a cost 1000 times higher than the mask itself, so that paths use them only where the mask is actually interrupted. The distance $D$ of Eq.~\eqref{eq:centred-cost} is measured from the boundary of the dilated mask (Figure~\ref{fig:distance-map}). Moving off the axis by $\delta$ in a vessel of radius $R$ multiplies the local cost by $\left(R/(R-\delta)\right)^{p}$, and the high exponent $p=4$ keeps the paths centered through bends instead of cutting their corners. An annotated ostium lies on the aortic wall, at the edge of the coronary mask or just outside it; each ostium is therefore connected to the mask through the smallest ball around it that reaches the mask, traversed at the same high cost as the dilation voxels, and a case is rejected if the mask lies more than 25~mm from either ostium.

\paragraph{Fronts, tree domains and gap bridging} From each ostium, two minimal-cost fronts are propagated over the 26-connected grid with Dijkstra's algorithm~\citep{dijkstra1959note}: one with unit cost, which gives the geodesic distance $G$ from the ostium through the mask, and one with the centered cost $c$, whose predecessor map links every reachable voxel to the ostium along its minimal path. The domain of each tree is the set of voxels reached by its front; where the two trees touch and both fronts reach a voxel, the voxel is assigned to the ostium with the smaller geodesic distance. A segmentation can be interrupted for a few millimeters, for instance at a severe stenosis, leaving distal portions of a vessel disconnected from the ostium. Before the fronts are propagated, every mask piece separated from the portions reached from the ostia by a gap of at most 10~mm is therefore joined to them by a straight segment between their closest voxels, traversed at the same high cost as the dilation voxels, so that it is used only to cross the gap; pieces farther away are not bridged.

\paragraph{Tree growth} Each tree is grown by repeating the following steps until every voxel of its domain is covered. The uncovered voxel with the largest geodesic distance $G$ is selected, and its minimal path is read from the predecessor map up to the point where it joins the tree. Every mask voxel within $1.5\,r + 1$~mm of the new path, $r$ being the local lumen radius, and whose straight segment to the nearest point of the path lies entirely inside the mask, is marked as covered; the latter condition prevents a distinct vessel running alongside the path, but separated from it by background, from being absorbed, so that it is traced in turn. A path that does not run for at least 5~mm outside the region covered before it was traced is discarded, and the region it covered prevents it from being selected again; the paths of irregularities of the mask surface leave the covered region for a few millimeters only, so that no explicit search for endpoints is needed.

\paragraph{Capped second pass} The centered cost rewards depth without limit, so where the mask is locally bulky, as at some bifurcations, the deepest voxels can pull a path away from the axis of the vessels that meet there. Each tree is therefore grown a second time with $D$ capped, at every voxel, at 1.2 times the local vessel radius, estimated as the median of $D$ over the first-pass tree within a 5~mm window along it. Beyond one radius of depth the cost then becomes flat, and the path follows the shortest route that remains a full radius deep; wherever the distance does not exceed the cap, which is the case along regular vessel segments, the cost is unchanged (see Section~\ref{sec:discussion-centerlines}).

\paragraph{Trimming, re-attachment and node attributes} Each tip of the tree reaches the very end of the mask, where the lumen tapers and the path is poorly constrained, so 2~mm of centerline are removed from every endpoint, never beyond a bifurcation. A bifurcation at the ostium itself, or within a few millimeters of it, is anatomically implausible, yet it can arise when the annotated ostium lies where the mask already splits; when the first bifurcation of a tree is closer than 2~mm to its ostium, the branches leaving before that distance are therefore re-attached to the main vessel at the first point of its centerline at least 2~mm away from the ostium, which becomes the first bifurcation. Every node is then assigned its position in physical coordinates, the ostium nodes keeping the exact annotated coordinate; a lumen radius, equal to the Euclidean distance transform of the mask at the node, i.e.\ the radius of the largest sphere centered at the node and contained in the mask, the ostium nodes, which lie on the aortic wall, taking the radius of their only neighbor; a topological type --- ostium, segment, bifurcation or endpoint --- assigned by node degree; and the side of the tree it belongs to. This radius is the maximum inscribed sphere radius that also comes with the ASOCA reference centerlines~\citep{asoca2,antiga2003geometry}, whereas the radius of the CAT08 reference centerlines is that of the circle with the same area as the lumen cross-section~\citep{cat08}, which is never smaller, and is larger where the lumen is not circular.

\paragraph{Smoothing and output} A path on a voxel grid lies within a fraction of a voxel of the vessel axis, but its direction changes by up to $45^\circ$ at every step, and any use of the local direction --- curved planar reformation, cross-sectional analysis, curvature measurements --- inherits this jitter. The trees are therefore smoothed branch by branch. Each tree is decomposed into branches by following, from the ostium and at every bifurcation, the child with the longest path to an endpoint, every other child starting a side branch at that bifurcation; each branch is processed after the branch it originates from. The positions and radii of the nodes of a branch are replaced by their Gaussian-weighted average over arc length, with $\sigma = 2$~mm and point-reflection padding at both ends. This leaves the ostium, the endpoints and the junction of every side branch with its already smoothed parent exactly in place, and lets main vessels pass through their bifurcations without kinks. No node is displaced by more than 1~mm, and a node lying inside the mask is never moved out of it: where its smoothed position would fall outside the mask, the displacement is halved until it lies inside, and, failing that, limited to 0.4 times the local radius. The graphs are finally resampled to one node every 0.5~mm. The output is one graph per patient, holding the left and right trees as two disjoint components rooted at the annotated ostia, and stored in GML format.

\paragraph{Quality control} The mask regions that the final tree does not explain under the same coverage rule, and those unreachable from both ostia, are measured. A case is flagged as failed when an unattached mask region extends over at least 20~mm, or when either tree is shorter than 20~mm; as possibly missing a branch when such a region extends over 10 to 20~mm, or when a region of the domain of a tree not explained by it exceeds 10~mm; and for inspection when an ostium lies more than 3~mm from the mask, when a gap had to be bridged, when the domains of the left and right trees touch, or when the length of a tree or its number of endpoints falls below the 5th percentile of the cohort. Every flagged case was then reviewed visually against its mask, together with a random sample of 50 unflagged cases, which checks for missed branches among the latter; every accepted branch shorter than 10~mm is listed in the released table for the same purpose.

\paragraph{Validation against ASOCA} The trees were extracted from the 40 annotated ASOCA masks, rooted at the ASOCA ostia defined in Section~\ref{subsec:ostia-dataset}, with exactly the same settings as for ImageCAS, and compared with the expert reference centerlines and lumen radii of ASOCA~\citep{asoca2}. Both centerlines were resampled every 0.25~mm, each sample weighted by the centerline length it represents, and compared with the measures of the CAT08 evaluation framework~\citep{cat08}. A reference point is matched when the extracted centerline passes within the reference lumen radius at that point, with a minimum tolerance of 1~mm, and a point of the extracted centerline is matched when it lies within the radius of its nearest reference point. We report the recall (fraction of the reference length that is matched, i.e.\ the coverage of the reference tree), the precision (fraction of the extracted length that is matched), the overlap OV (matched length of both centerlines over their total length), the overlap OT restricted to reference vessels at least 1.5~mm in diameter, and the overlap until first error OF, i.e.\ the fraction of each reference ostium-to-endpoint path matched before the first unmatched point, averaged over the paths. Since the ostium definitions of the reference and of this work can differ at the very origin of the vessels, OF ignores the first 40~mm of every path, instead of the 5~mm of the original framework, and paths not longer than 40~mm are left out. The spatial accuracy is measured by the average inside AI, the mean distance from the matched reference points to the extracted centerline, and by the 95th-percentile symmetric Hausdorff distance (HD95). We also report the length ratio between the extracted and the reference tree, the error of the lumen radius on matched points, the endpoints of the two trees paired one-to-one within 10~mm (recall, precision and mean distance), and the fraction of reference bifurcations paired one-to-one with an extracted bifurcation within 5~mm. Finally, every reference vessel portion at least 10~mm long that the extracted tree misses is classified as a missed branch (when it leaves the tree) or a truncation (when it extends a tip), and every extracted portion at least 10~mm long that is absent from the reference as an extra branch. All the measures are computed per patient, for each tree and for both trees together, and summarized over the patients.

\subsection{Centerline extraction on ImageCAS: run statistics and flagged patients}
\label{subsec:app-centerlines-imagecas}
This section completes the results of Section~\ref{subsec:res-centerlines} and the discussion of Section~\ref{sec:discussion-centerlines}.

\paragraph{Run statistics} The extracted coronary graphs are made of a median of 1189 nodes. The ostia lay at a median of 0.33~mm from the mask (95th percentile 1.96~mm); 26 of the 1998 ostia were more than 3~mm away from it, with a maximum of 14.6~mm. Besides the 19 bridged interruptions, in 2 patients a mask region extending over more than 10~mm remained unreachable. The second growth pass, with the capped distance, moved the path by more than 1~mm in 12 patients, and capped at most 0.54\% of the voxels of the tree domains. In 192 of the 1998 trees, branches leaving within 2~mm of the ostium were re-attached further along the main vessel. Smoothing reduced the median change of direction between consecutive 0.5~mm steps from $19.6^\circ$ to $1.7^\circ$, and its 95th percentile, per patient, from a median of $46^\circ$ to $5.9^\circ$; the largest displacement of a node had a median of 0.81~mm per patient, and no node lying inside the mask was moved out of it.

\paragraph{Visual review} Of the 25 patients whose left and right trees touch in the mask, 12 had a distal vessel segment assigned to the wrong tree, or one that could not be attributed to either tree with confidence; in 2, the trees met only at their terminal branches, and in the others they were separated correctly, although the mask was often irregular. The ostia lying more than 3~mm from the mask and the paths moved by the capped distance were all correct, the former reflecting masks that stop short of the ostium. Of the 17 patients with bridged gaps, one had a mask too ambiguous around the gap to confirm the bridge.

\paragraph{Failed patients} The few cases in which the extraction did not produce a complete tree all trace back to the segmentation masks, not to the extraction itself. On the ImageCAS masks, four of the 1000 patients were flagged as failed, each for a different defect of the mask. In patient 501, whose right coronary artery arises unusually close to the left one through a small orifice, the mask does not reach the annotated right ostium and extends far from it instead, so that no right tree can be grown from the ostium; the ImageCAS-X mask of the same patient reaches the ostium, and the extraction on it is complete. In patient 556, the right coronary artery is segmented only over its first few centimeters, so that the right tree is necessarily truncated. In patient 865, a segment in the middle of the right coronary artery is missing from the mask, and the distal part lies beyond the largest gap that the extraction is allowed to bridge. In patient 66, a disconnected piece of the mask is an erroneous segmentation of a structure inside a cardiac cavity, which the extraction correctly leaves unattached. Three of these four patients (556, 865 and 66) are among the 200 scans that ImageCAS-X excluded because of image artifacts \citep{imagecasx2026}. On the 800 ImageCAS-X masks, a single patient (510) was flagged: the mask contains a short arterial segment that is anatomically plausible but disconnected from both trees, at a distance at which reattaching it would be a guess rather than a reconstruction, and as such it is left out by the extraction by design.

\subsection{Flattened view of the coronary trees}
\label{subsec:app-flattened}
\begin{figure*}[p]
  \centering
  \includegraphics[width=\textwidth]{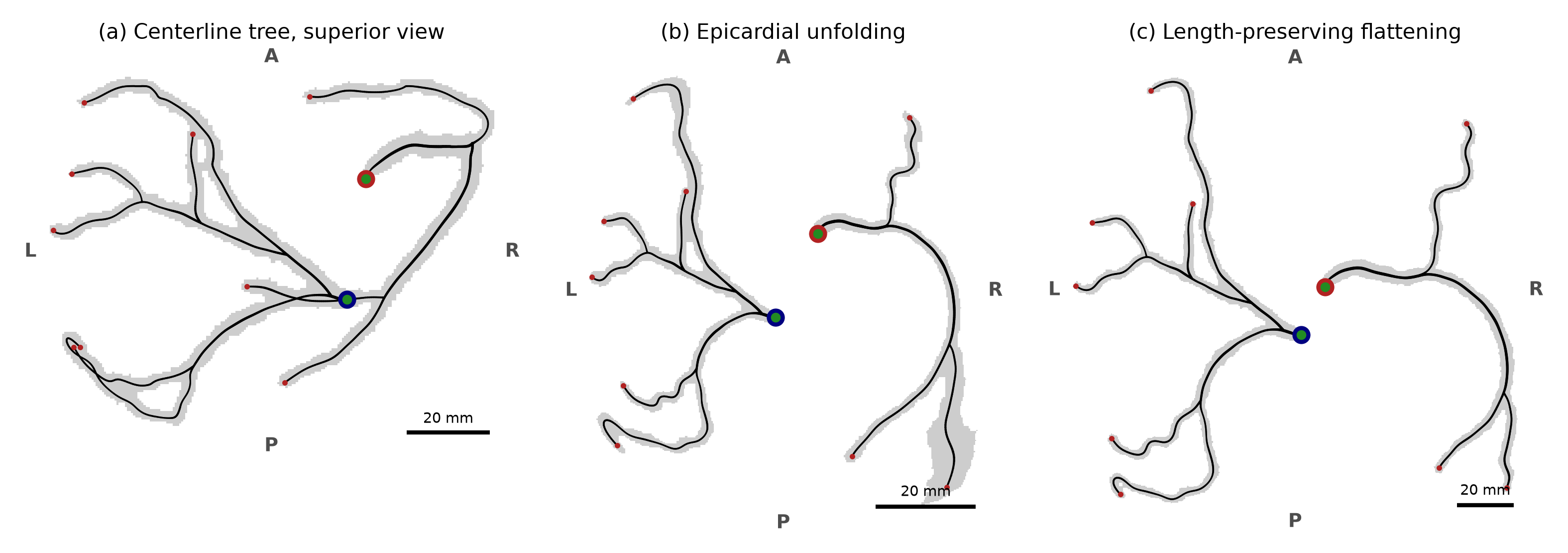}
  \caption{Flattening of the coronary trees of one ImageCAS patient (343), seen from superior (anterior up, patient's right on the right). (a) Centerline graph projected on the axial plane: vessels running toward the apex are foreshortened and branches overlap. (b) Epicardial unfolding: the trees are mapped from a spheroid fitted around the heart onto a plane centered on the aortic root; vessels no longer overlap, but distances across the meridians are stretched toward the apex (bottom of the right tree). (c) Length-preserving flattening: each tree is rebuilt from its ostium with the directions of (b) and the true length of every segment. Black lines: centerline graph; green dots: ostia (right with a red, left with a blue outline); red dots: endpoints. The vessels (gray) are drawn as tubes of the lumen radius stored in the graph, with a slightly irregular outline added for display, not from the segmentation masks. Scale bar: 20~mm. Flattened views of this kind were also used in the visual review of the released centerlines, to assess the quality of the extraction and the defects of the segmentation masks. See~\ref{subsec:app-flattened}.}
  \label{fig:app-flattened-pipeline}

  \vspace{0.8\baselineskip}
  \includegraphics[width=0.80\textwidth]{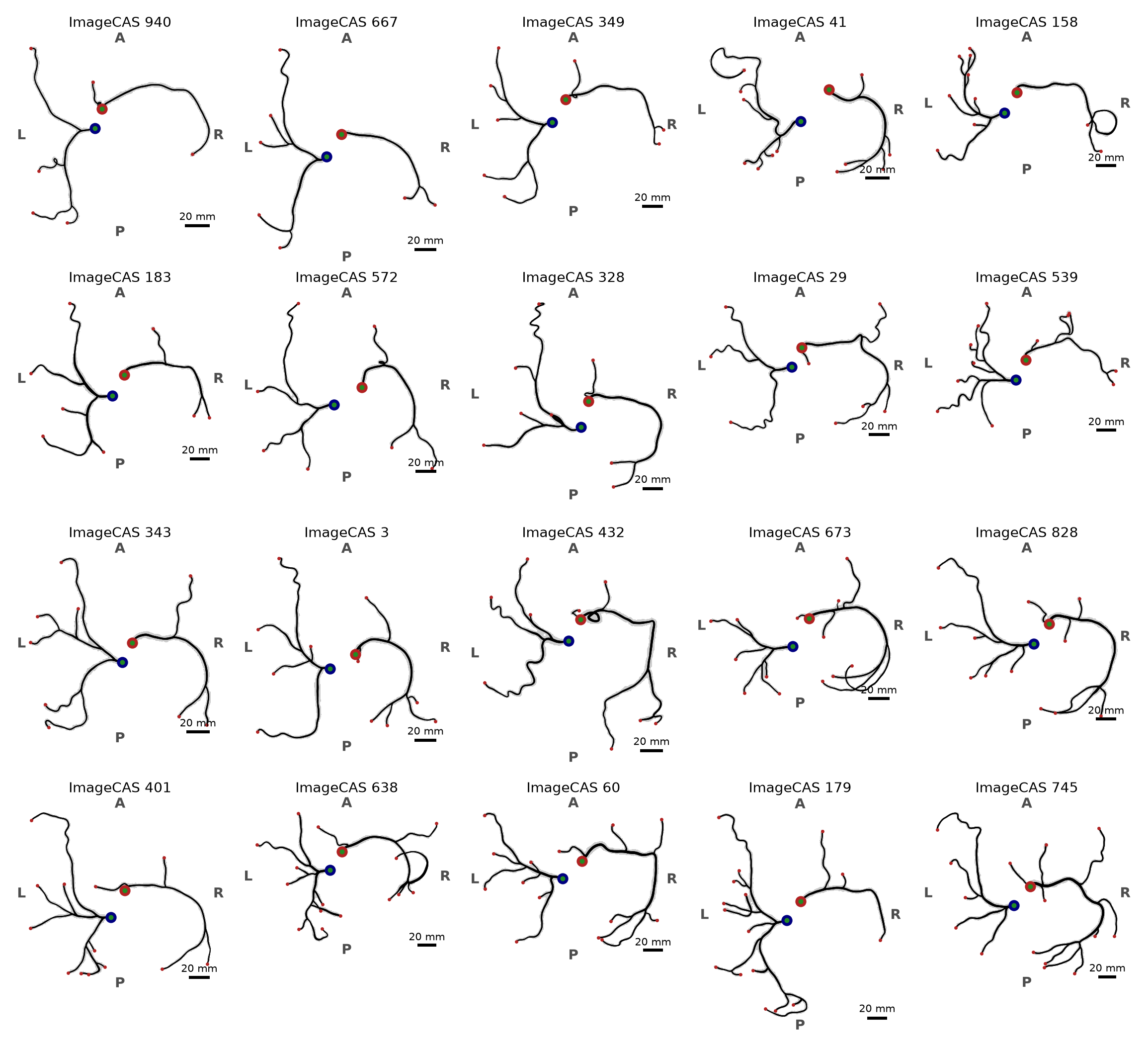}
  \caption{Length-preserving flattened coronary trees of 20 ImageCAS patients from the released centerline set, as in Figure~\ref{fig:app-flattened-pipeline}c: one patient for each of 20 strata of total tree length, from 497 to 1096~mm, drawn at random among the 756 patients in which the quality control found no anomaly. Same elements and orientation as in Figure~\ref{fig:app-flattened-pipeline}; each panel has its own scale (bar: 20~mm).}
  \label{fig:app-flattened-grid}
\end{figure*}

A coronary tree wraps around the heart, so any projection of it on a plane foreshortens the vessels that run toward the apex and makes branches overlap (Figure~\ref{fig:app-flattened-pipeline}a). To show a whole tree in one image, the centerline graphs were unfolded onto a plane in two steps. First, a spheroid whose axis passes through the midpoint of the two ostia, i.e.\ through the aortic root, is fitted by least squares to the nodes of both trees and scaled about its center until no node lies inside it; every point is then placed at its longitude around the axis and at its arc length along the meridian from the aortic root, which gives a map centered on the aortic root with the apex toward the rim (Figure~\ref{fig:app-flattened-pipeline}b). Lengths along the meridians are preserved, but distances across them are stretched more and more toward the apex, where the map is singular; the axis is therefore tilted, by the smallest angle that keeps every node at least 10~mm from its half opposite the aortic root. Second, each tree is rebuilt from its ostium, segment by segment, keeping the direction that each segment has on the map but its true length, so that the lengths of the vessels and the distances along the trees are those of the graph (Figure~\ref{fig:app-flattened-pipeline}c). Figure~\ref{fig:app-flattened-grid} shows the result for 20 patients of the released set, which span the range of tree lengths. Since they show a whole tree at once, without overlaps, flattened views of this kind were also used in the visual review of the centerline dataset (Section~\ref{subsec:res-centerlines} and~\ref{subsec:app-centerlines-imagecas}), both to assess the quality of the extraction, such as missed or spurious branches, and to spot the quality issues and defects of the segmentation masks, such as interruptions, disconnected fragments or touching left and right trees.

\subsection{Release format of the centerlines}
\label{subsec:app-centerline-release}
The centerlines are released as follows. Each patient has one graph in GML format, which can be read with the generic \texttt{networkx} library or with HCATNetwork, a library dedicated to coronary centerline graphs, which restores the typed node attributes and adds coronary-specific operations. Every node carries its position in millimeters in the world (RAS) coordinates of the ImageCAS image, its lumen radius, its topological type (ostium, segment, bifurcation or endpoint) and its side, and nodes are spaced 0.5~mm apart. The same trees are also provided in VTK format, readable by common visualization and vessel-analysis software, both as an undirected graph, with one line per edge, and as one polyline per ostium-to-endpoint path. For direct visualization, each patient also has a set of 3D Slicer markups, with one open curve for every ostium-to-endpoint path, which can be used for curved planar reformation, and one list of the ostia, bifurcations and endpoints, each named after its side, its type, its order from the ostium and the identifier of the corresponding graph node. A table lists, for every patient, the outcome of the quality control with its details, the length of each tree, any comment from the visual review of the flagged cases, and which graph to use. The official set, extracted from the ImageCAS masks, is recommended by default; the alternative set extracted from the ImageCAS-X masks is recommended only where the official graph is missing, failed, may miss a branch or has touching left and right trees, while the alternative one is free of these problems. This is the case for 96 patients, since the visual review confirmed two official graphs whose trees touch only at their terminal branches. Independently of this recommendation, the set to use follows from the masks: users working with the ImageCAS masks, the only ones covering all 1000 patients, should use the official set, and users working with the ImageCAS-X masks the alternative one, so that trees and masks agree.

\subsection{Training details}
\label{subsec:app-training}
This section gives the complete training settings summarized in Section~\ref{subsec:training}.

\paragraph{Optimization and computational setup} Models are trained with a batch size of 1 and gradient accumulation over four iterations. Mixed-precision training is used, and gradients are clipped with maximum norm 1.0. Validation is run every five epochs using the full-volume sliding-window procedure. Checkpoints are saved during training, with the best-model checkpoint locked during the first 10 warmup epochs to avoid selecting unstable early models. The AdamW optimizer~\citep{AdamW} is used with learning rate $5\cdot10^{-5}$ and weight decay $10^{-5}$. The learning-rate schedule uses a custom linear warmup followed by decay over the remaining training steps. During training, validation uses the MONAI sliding-window inferer with Gaussian window blending, whereas the test-time implementation described in Section~\ref{subsec:postprocessing} averages the overlapping windows without weighting. The method was implemented in Python 3.11.15 using PyTorch 2.6.0+cu124~\citep{pytorch_neurips} and MONAI 1.5.2~\citep{monai}. Training was performed on a single NVIDIA A100-PCIE GPU with 40~GB of memory, using 8 CPU cores and 128~GB of allocated RAM. No final activation function is applied during training, since the loss operates directly on the raw heatmap logits, and gradient checkpointing is employed to reduce the GPU memory consumption of \swinv2.

\paragraph{Data augmentation} Training uses lightweight 3D data augmentation to improve robustness while preserving the anatomical meaning of the left and right ostium channels. Augmentations include random zoom in range of $[0.9;1.1]$, small random rotations of $\pm~15^\circ$, smooth elastic deformations (maximum displacement of 2 voxels, i.e.\ 1~mm), additive Gaussian noise and intensity shifts. Each augmentation is applied to a training patch with its own probability: 0.8 for rotation and zoom, 0.3 for the elastic deformation, and 0.5 for the noise and for the intensity shift. Standard random flips along the spatial axes and rotations by multiples of $90^\circ$ were omitted to take advantage of the standard RAS orientation of medical images. Rotation, zoom and elastic deformation are composed in a single resampling, and the geometric augmentations are applied consistently to the image and the target heatmaps.

\paragraph{Patch extraction strategy} Training patches are extracted from the resampled volumes with a class-aware probabilistic crop strategy. For each patch, one of three modes is sampled: random background crop with probability 0.32, right-ostium-centered crop with probability 0.34, or left-ostium-centered crop with probability 0.34. When the selected foreground target is present, the crop is placed so that the corresponding ostium is contained in the $128^3$ field of view; otherwise, a random crop is used. This strategy balances the extreme class imbalance between the small ostium targets and the surrounding CCTA volume while still exposing the model to background-only anatomy.

\subsection{Accuracy buckets}
\label{subsec:app-buckets}
The accuracy buckets of Section~\ref{subsec:evaluation-metrics} grade each localization error against three distance thresholds, defined separately for the right and the left ostium (Table~\ref{tab:thresholds}). They are given here in full, as a reporting standard for coronary ostium localization: the definition comes first, followed by the rationale of each threshold, the rules for reporting the buckets, their use on other test sets, and the reasons for a graded scale.

\paragraph{Definition} Let $d_c$ be the localization error of ostium $c\in\{R,L\}$, i.e.\ the Euclidean distance in millimeters between the predicted and the reference point (Section~\ref{subsec:evaluation-metrics}), and let $\tau^{\mathrm{A}}_c$, $\tau^{\mathrm{I}}_c$ and $\tau^{\mathrm{F}}$ be the anatomical, interoperator and failure thresholds of Table~\ref{tab:thresholds}, with $\tau^{\mathrm{A}}_c<\tau^{\mathrm{I}}_c<\tau^{\mathrm{F}}$. Each localization is assigned to one of four buckets:
\begin{equation}
B(d_c) =
\begin{cases}
\textit{Optimal} & \text{if } d_c \le \tau^{\mathrm{A}}_c,\\
\textit{Good} & \text{if } \tau^{\mathrm{A}}_c < d_c \le \tau^{\mathrm{I}}_c,\\
\textit{Acceptable} & \text{if } \tau^{\mathrm{I}}_c < d_c \le \tau^{\mathrm{F}},\\
\textit{Failed} & \text{if } d_c > \tau^{\mathrm{F}} \text{ or no point is predicted.}
\end{cases}
\label{eq:buckets}
\end{equation}
At patient level, a case takes the worse bucket of its two ostia: a patient is Optimal only when both ostia are Optimal, and Failed when at least one ostium is Failed. Each of the three thresholds answers a different question, discussed in turn below.

\paragraph{Anatomical threshold} The anatomical threshold asks whether the predicted point lies inside the ostium itself. The reference points are placed at the center of the ostial opening, so a prediction lies within the ostium when its error is smaller than the ostial radius. The threshold is half the lower bound $\mu-2\sigma$ of the internal ostial diameter reported for female hearts by \citet{nasr2018coronaryostia}, where $\mu$ and $\sigma$ are the mean and the standard deviation of that diameter: assuming an approximately normal distribution, an error below it places the point within the ostial opening in about 98\% of women, and in an even larger share of men, whose coronary ostia are larger. It is deliberately the most stringent of the three criteria, and it is a property of the anatomy, independent of the dataset and of the annotators. The reference values come from a single adult cohort, whose population differs from those of the datasets used here; however, as the bound lies two standard deviations below the mean of the smaller, female ostia, it is expected to hold for adults of any population, whereas it may not apply to pediatric patients, whose ostia are smaller. The threshold is also close to the resolution of the data: on the 0.5~mm isotropic grid on which the benchmark models operate (Section~\ref{subsec:preprocessing}), it corresponds to about two voxels for the right and three voxels for the left ostium, and at the native in-plane spacing of CAT08 and ImageCAS (0.25--0.45~mm, Table~\ref{table:dataset-comparison}) to between about two and six pixels. An Optimal localization is therefore attainable, but only with sub-voxel accuracy in the extraction of the coordinates (Section~\ref{subsec:postprocessing}).

\paragraph{Interoperator threshold} This threshold asks whether the prediction is roughly comparable with the placement of a human reader. Since the reference positions are themselves manual, an error smaller than the typical disagreement between annotators cannot be distinguished from annotation noise. The threshold is the 75th percentile of the distance between the point placed by an individual annotator and the mean of the points placed on the same image by the other two annotators (leave-one-out), pooled over the six annotators of the first ImageCAS batch (about 1500 placements per ostium) and computed separately for each ostium (Table~\ref{tab:annotator-stats}); it coincides with the upper end of the interquartile range reported there. Three quarters of the individual human placements lie within this distance of the other two annotators' mean, so a prediction at or below it falls within the ordinary spread of the annotators. The threshold was chosen so as not to be too tight: the median would have excluded from the same grade half of the human placements themselves, whereas a bound based on the mean and the standard deviation would have been driven by the annotators who deviated most from the others (Table~\ref{tab:annotator-stats}). The threshold is larger for the left ostium (2.59~mm against 1.85~mm), as the annotators agree less on it: the left ostium is wider than the right one~\citep{nasr2018coronaryostia}, so the point where the aorta ends and the artery begins is less sharply defined.

Two caveats apply to this threshold. First, each annotator is measured against the mean of the other two annotators, not against the released consensus, which also includes the annotator's own point: the annotator is thus compared with a reference to which they did not contribute, just as the models are compared with a consensus to which they did not contribute. On an image annotated by three annotators, the distance from the consensus is exactly two thirds of the leave-one-out distance, since
\begin{equation}
\mathbf{p}_1-\tfrac{1}{3}\left(\mathbf{p}_1+\mathbf{p}_2+\mathbf{p}_3\right)=\tfrac{2}{3}\left(\mathbf{p}_1-\tfrac{1}{2}\left(\mathbf{p}_2+\mathbf{p}_3\right)\right),
\end{equation}
where $\mathbf{p}_1$ is the point of the annotator and $\mathbf{p}_2$, $\mathbf{p}_3$ those of the other two (Table~\ref{tab:app-annotator-consensus}). A threshold computed against the consensus would therefore have been 1.5 times smaller, as it would partly reproduce the annotator's own point. The leave-one-out reference, on the other hand, averages two points instead of three, so it is slightly noisier than the consensus against which the models are evaluated: if the errors of the annotators are independent, unbiased and equally dispersed, the expected leave-one-out distance is about 6\% larger than the distance from the mean of three other annotators, so the threshold is, if anything, slightly lenient.
Second, the models are trained to reproduce the consensus, so their errors and the annotators' deviations are not directly comparable in absolute terms, and a prediction below this threshold indicates human-level agreement with the reference, not superiority over a human reader.

\begin{table*}[tb]
\centering
\caption{Per-annotator distance (Euclidean distance, in mm) between each annotator's manually placed point
and the released consensus of that image, i.e.\ the mean of the points of all its annotators, the
annotator's own point included, for the left and right coronary ostia, individually and combined,
together with the pooled statistics across all six annotators, computed over the same 3000 placements (1500 landmarks per side, 250 per annotator per ostium, three annotators per volume, two ostia per volume - left and right). Same layout and statistics as
Table~\ref{tab:annotator-stats}, which reports the leave-one-out distances; on an image annotated by three
annotators each distance here is exactly two thirds of the corresponding leave-one-out distance.}
\label{tab:app-annotator-consensus}
\small
\setlength{\tabcolsep}{3.5pt}
\begin{tabular}{l cc cc cc}
\toprule
\multirow{2}{*}{Annotator}
  & \multicolumn{2}{c}{Left ostium}
  & \multicolumn{2}{c}{Right ostium}
  & \multicolumn{2}{c}{Combined (Left+Right)} \\
\cmidrule(lr){2-3} \cmidrule(lr){4-5} \cmidrule(lr){6-7}
  & Mean $\pm$ SD & Median [IQR]
  & Mean $\pm$ SD & Median [IQR]
  & Mean $\pm$ SD & Median [IQR] \\
\midrule
Annotator 1 & 1.93 $\pm$ 0.64 & 1.82 [1.56--2.28] & 1.32 $\pm$ 0.52 & 1.27 [0.98--1.60] & 1.63 $\pm$ 0.66 & 1.58 [1.19--1.95] \\
Annotator 2 & 1.37 $\pm$ 0.64 & 1.33 [0.93--1.65] & 1.19 $\pm$ 0.63 & 1.09 [0.79--1.49] & 1.28 $\pm$ 0.64 & 1.16 [0.83--1.58] \\
Annotator 3 & 1.15 $\pm$ 0.59 & 1.03 [0.73--1.47] & 0.86 $\pm$ 0.47 & 0.79 [0.52--1.08] & 1.00 $\pm$ 0.55 & 0.90 [0.63--1.30] \\
Annotator 4 & 0.93 $\pm$ 0.60 & 0.79 [0.50--1.20] & 0.75 $\pm$ 0.41 & 0.70 [0.48--0.95] & 0.84 $\pm$ 0.52 & 0.73 [0.48--1.05] \\
Annotator 5 & 0.96 $\pm$ 0.56 & 0.86 [0.54--1.29] & 0.79 $\pm$ 0.45 & 0.74 [0.46--1.04] & 0.88 $\pm$ 0.51 & 0.80 [0.50--1.13] \\
Annotator 6 & 1.45 $\pm$ 0.60 & 1.42 [1.04--1.90] & 0.90 $\pm$ 0.51 & 0.74 [0.55--1.10] & 1.17 $\pm$ 0.62 & 1.07 [0.67--1.60] \\
\midrule
\textbf{All annotators (pooled)} & 1.30 $\pm$ 0.69 & 1.22 [0.77--1.73] & 0.97 $\pm$ 0.55 & 0.88 [0.57--1.23] & 1.13 $\pm$ 0.65 & 1.02 [0.65--1.51] \\
\bottomrule
\end{tabular}

\vspace{1mm}
\footnotesize
\textit{Notes.} All values are in mm. Only the annotators who placed exactly one point on a side of an image
enter the consensus of that side, and the statistics. Standard deviations are population SD (ddof=0);
percentiles are computed with linear interpolation between order statistics.
\end{table*}

\paragraph{Failure threshold} The failure threshold of 5~mm, the same for both ostia, marks the point beyond which a localization can no longer be considered usable, for instance as the starting point of vessel tracking. The thresholds adopted in the literature vary between works, from 2~mm \citep{mostafa2021ostia} and 3.5~mm \citep{gajowczyk2022ostia} to 5~mm \citep{zheng2011ostia,mostafa2021ostia}. We adopt the largest of them, which makes the failure rate comparable with the published ones, and which is large compared with both the ostial dimensions and the annotator variability, so that a failed case reflects a genuine miss, not a borderline placement.

\paragraph{Reporting rules} To make results comparable across works, the buckets should be computed and reported as follows.
\begin{itemize}
\item The error $d_c$ is measured in millimeters in physical space, between the predicted and the reference point.
\item The bucket boundaries are inclusive, as in Eq.~\eqref{eq:buckets}: an error equal to a threshold belongs to the better bucket.
\item An ostium with no predicted point counts as Failed, and is never excluded from the test set.
\item The percentage of localizations in each of the four buckets is reported for each ostium separately, over all the test cases, and at patient level; the four percentages of each row sum to 100\%.
\item The distance statistics reported with the buckets, such as the median, the mean and the maximum, are computed over all the cases, the Failed ones included.
\end{itemize}

\paragraph{Use on other test sets} The thresholds used so far in the literature were chosen case by case, with values from 2 to 5~mm and without a stated rationale, so that the reported success and failure rates cannot be compared across works. The scheme proposed here derives each threshold from a stated criterion: the ostial anatomy for the anatomical threshold, the measured variability of human readers for the interoperator threshold, and the most permissive value used in the literature for the failure threshold. The anatomical and failure thresholds do not depend on any dataset. The interoperator threshold rests on six annotators and 1500 placements per ostium, a sample large enough to characterize the variability of human readers on this task, and is released together with the annotations. The values of Table~\ref{tab:thresholds} can therefore be used as default thresholds on other test sets, which keeps the results obtained on different data comparable with each other and with this benchmark; the interoperator values, however, derive from a single annotation team and protocol.

\paragraph{Why a graded scale} Grading the errors, instead of counting failures only, avoids two drawbacks of a binary criterion. First, a model could have no failures and still be systematically less precise than the annotators, which a failure rate alone would not reveal. Second, a fixed threshold at the anatomical level would label as failures errors that are still within the variability of the reference, whereas one at the failure level would not discriminate between the models that are accurate enough for the task.

\FloatBarrier
\section{Model of the Second Annotation Stage}
\label{sec:app-stage2}
The candidates of the second, model-assisted annotation stage (Section~\ref{subsec:ostia-dataset}) were generated by a \swinv2 model trained on CAT08, ASOCA and the first, manually annotated batch of 500 ImageCAS images, all without the 58 test volumes, which were held out from every training of this work. The localization errors of \swinv2 on the test set are reported in Table~\ref{tab:res-500}. The model correctly localized both ostia in 52 of the 58 test volumes; 5 right (8.6\%) and 2 left (3.4\%) ostia failed, in 6 volumes overall. All four right ostia of CAT08 failed, together with one left ostium of CAT08, one left ostium of ASOCA, and one right ostium of ImageCAS. Because of these failures, the mean errors (10.3~mm for the right and 5.17~mm for the left ostium) are much higher than the median errors (0.82~mm and 0.94~mm, respectively); on ImageCAS alone, the median errors were 0.73~mm (right) and 0.92~mm (left). In terms of accuracy buckets (Section~\ref{subsec:evaluation-metrics}), 65.5\% of the right and 77.6\% of the left ostia were Optimal, 24.1\% and 19.0\% Good, 1.7\% (one right ostium) and none Acceptable, and 8.6\% and 3.4\% Failed; at patient level, 53.4\% of the cases were Optimal, 34.5\% Good, 1.7\% Acceptable and 10.3\% Failed. The average processing time was 50.7~s per patient.

\begin{table}[tb]
\centering
\fontsize{6.3}{8}\selectfont
\setlength{\tabcolsep}{1.5pt}
\newcommand{\bucketgapHalf}{-2pt}
\newcommand{\blockgapHalf}{5pt}
\caption{Euclidean distance $d_c$ (mm) of \swinv2 trained with the first 500 annotated ImageCAS images. Means and percentiles include the failed localizations. The line below each row gives the percentage of cases in each accuracy bucket for the right and left ostium: O = Optimal, G = Good, A = Acceptable, F = Failed. See Section~\ref{subsec:evaluation-metrics}.}
\label{tab:res-500}
\begin{tabular}{l *{10}{r} r}
\toprule
 & \multicolumn{5}{c}{Right ostium} & \multicolumn{5}{c}{Left ostium} & \\
\cmidrule(lr){2-6} \cmidrule(lr){7-11}
\makecell[l]{Model} & p25 & Med. & Mean & p75 & Max & p25 & Med. & Mean & p75 & Max & \makecell{Time\\(s)} \\
\midrule
\swinv2 & 0.63 & 0.82 & 10.3 & 1.21 & 150.0 & 0.65 & 0.94 & 5.17 & 1.31 & 131.6 & 50.7 \\[\bucketgapHalf]
 & \multicolumn{5}{c}{\bucketrow{65.52}{24.14}{1.72}{8.62}} & \multicolumn{5}{c}{\bucketrow{77.59}{18.97}{0.00}{3.45}} & \\
\multicolumn{1}{r}{\subds{ASOCA}} & \subds{0.91} & \subds{1.05} & \subds{1.18} & \subds{1.32} & \subds{1.88} & \subds{1.29} & \subds{1.72} & \subds{29.2} & \subds{29.7} & \subds{112.8} & \\[\bucketgapHalf]
 & \multicolumn{5}{c}{\subds{\bucketrow{50.00}{25.00}{25.00}{0.00}}} & \multicolumn{5}{c}{\subds{\bucketrow{25.00}{50.00}{0.00}{25.00}}} & \\
\multicolumn{1}{r}{\subds{CAT08}} & \subds{98.3} & \subds{118.6} & \subds{121.2} & \subds{141.5} & \subds{150.0} & \subds{0.82} & \subds{1.50} & \subds{33.8} & \subds{34.5} & \subds{131.6} & \\[\bucketgapHalf]
 & \multicolumn{5}{c}{\subds{\bucketrow{0.00}{0.00}{0.00}{100.00}}} & \multicolumn{5}{c}{\subds{\bucketrow{50.00}{25.00}{0.00}{25.00}}} & \\
\multicolumn{1}{r}{\subds{ImageCAS}} & \subds{0.53} & \subds{0.73} & \subds{2.20} & \subds{1.11} & \subds{70.4} & \subds{0.64} & \subds{0.92} & \subds{0.95} & \subds{1.18} & \subds{2.56} & \\[\bucketgapHalf]
 & \multicolumn{5}{c}{\subds{\bucketrow{72.00}{26.00}{0.00}{2.00}}} & \multicolumn{5}{c}{\subds{\bucketrow{84.00}{16.00}{0.00}{0.00}}} & \\
\bottomrule
\end{tabular}
\end{table}

\paragraph{Comparison with the final model} The accuracy of this model was lower than that of the model trained on the complete dataset (Table~\ref{tab:res-1000}): median errors of 0.82~mm and 0.94~mm for the right and left ostium against 0.74~mm and 0.85~mm, failures in six volumes against none, and 53.4\% against 69.0\% of Optimal cases at patient level. Two factors differed between the two models and cannot be separated: the first model was trained on half as many ImageCAS volumes, and on volumes that had not been reoriented to the common RAS orientation adopted for all subsequent experiments. Its accuracy also bounds the quality of the candidates of the second stage, all of which were reviewed by an annotator (Table~\ref{tab:correction-stats}).

\FloatBarrier
\section{Consistency of the Ostium Definition across Datasets}
\label{sec:app-consistency}
This appendix reports in full the comparison summarized in Section~\ref{subsec:res-ostia-consistency} and discussed in Section~\ref{sec:discussion-consistency}.

\begin{figure}[tb]
  \centering
  \includegraphics[width=0.85\linewidth]{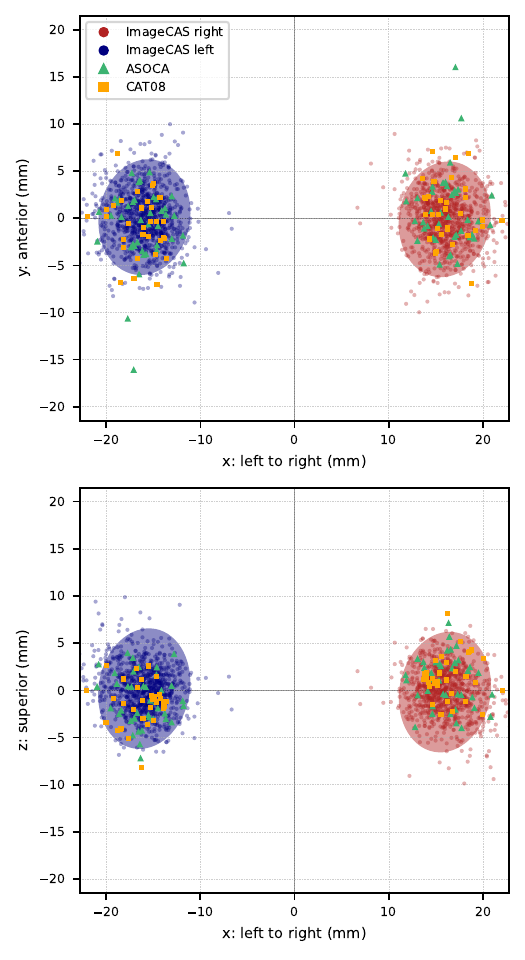}
  \caption{Right and left coronary ostium positions of all three datasets, after removing each
  patient's own absolute position: every ostium is expressed relative to that patient's own
  right-left midpoint, then rotated into a common frame whose $x$ axis is the population-average
  left-to-right direction and whose $y$/$z$ axes are the anatomical anterior/superior directions
  made orthogonal to it. Top: coronal-like projection ($x$-$y$). Bottom: sagittal-like projection
  ($x$-$z$). ASOCA (triangles) and CAT08 (squares) are shown as raw points; ImageCAS (circles,
  semi-transparent) additionally shows, for each side and projection, a 2$\sigma$ confidence
  ellipse computed from the ImageCAS points only. Its axes are the principal axes (principal
  component analysis) of the covariance of the projected points, and its semi-axes are twice
  the standard deviation along them, so it is an ellipse, not a circle, because the spread of
  the points differs between directions; it encloses about 86\% of a bivariate normal
  distribution. Right ostium in firebrick, left ostium in navy.}
  \label{fig:ostia-scatter}
\end{figure}

The ostium annotations of CAT08, ASOCA and ImageCAS derive from different operational definitions (Section~\ref{subsec:ostia-dataset}). To check how the three landmarks relate to each other, Figure~\ref{fig:ostia-scatter} re-expresses every ostium relative to the right-left midpoint of the same patient and compares the resulting point clouds of the three datasets. The right and left ostium point clouds of ASOCA, CAT08 and ImageCAS are essentially superimposed: the population-average left-to-right direction is shared by individual patients with a mean resultant length of 0.97 (1 would indicate perfect agreement), and the two ostia lie, on average, 33.0~mm apart (SD 5.9~mm) across the combined datasets.

The pooled statistics above do not exclude that one dataset sits at a systematically different position, distance or orientation than the other two, so this was tested directly with four non-parametric, permutation-based tests (10\,000 permutations of the dataset labels each, Holm-Bonferroni corrected across the four), one per quantity of interest: the right-ostium position, the left-ostium position, the inter-ostial distance, and the left-to-right axis direction, all expressed in the patient-centered frame of Figure~\ref{fig:ostia-scatter}. For each quantity, the null hypothesis $H_0$ is that ASOCA, CAT08 and ImageCAS share the same mean value (no systematic difference between datasets), against the alternative $H_1$ that at least one dataset's mean differs from the other two; the test statistic is the between-dataset sum of squared distances from the grand mean (a PERMANOVA-style statistic), and its null distribution is obtained by repeatedly reassigning the dataset label of each patient at random, keeping the three dataset sizes fixed.

Of the four tests, only the axis-direction one rejects $H_0$ at $\alpha=0.05$ after correction (statistic $0.56$, $p=0.0007$ raw, $p=0.0028$ Holm-corrected): the population-average left-to-right direction differs between datasets. The right- and left-ostium position tests give the same statistic ($192.46$, since the left position is the exact mirror image of the right one for every patient) and do not survive correction ($p=0.038$ raw, $p=0.114$ Holm-corrected), and neither does the inter-ostial distance (statistic $11.97$, $p=0.374$). Neither the position of each ostium relative to its own patient's midpoint nor the distance between the two ostia, therefore, differs detectably between datasets; only their orientation does. The difference is driven by ImageCAS: its mean axis direction differs from ASOCA's and from CAT08's by about 5--6$^\circ$ (computed from the reported mean directions), while ASOCA's and CAT08's own mean directions differ from each other by under 1$^\circ$.

\paragraph{Interpretation} The comparison was meant as a screen for gross definitional inconsistencies, for instance, one source systematically placing the point a few millimeters further into the aorta than the others, and it revealed none. Formally establishing the equivalence of the three definitions would require an equivalence test against a prespecified margin, which ASOCA and CAT08 are too small to support for small offsets. Such a test is, however, not needed here, for three reasons. First, the comparison cannot separate a definitional offset from genuine differences between the datasets: the cohorts differ in population, scanners and acquisition protocols, so differences in ostial anatomy or in patient positioning are expected, and would not indicate an inconsistent landmark. Second, the offset is small in absolute terms: with the two ostia about 33~mm apart, a rotation of the axis by 5--6$^\circ$ displaces each ostium by about 1.5~mm, which is of the same order as the distance of the individual annotators from the mean of the other two annotators (1.45~mm for the right and 1.95~mm for the left ostium, Table~\ref{tab:annotator-stats}) and below the interoperator thresholds adopted to grade the errors (1.85~mm and 2.59~mm), so it cannot be told apart from the variability of the manual annotation itself. Third, the use of the labels does not depend on the datasets being identical: the models are trained on the pooled data and localize the ostia of the three datasets with errors of the order of 1~mm (Table~\ref{tab:res-1000}), although ASOCA and CAT08 contribute only four test volumes each.

\FloatBarrier

\section{Ablation Studies}
\label{sec:app-ablations}
This appendix collects the ablation studies of the benchmark. Unless stated otherwise, they refer to \swinv2 and all the models are evaluated on the same test set of 58 volumes described in Section~\ref{subsec:models-test-set}. The errors are Euclidean distances in millimeters, and the line below each row of the tables gives the share of cases in each accuracy bucket (O = Optimal, G = Good, A = Acceptable, F = Failed; see Section~\ref{subsec:evaluation-metrics}). The implications of each experiment are summarized in the Discussion (Section~\ref{sec:discussion-ablations}), and its detailed interpretation follows the results here. All the ablations were run after the final benchmark models had been trained, selected and evaluated, and none of their outcomes was used to modify the final models, their hyperparameters or the inference settings reported in the main text (the voxel spacing of 0.5~mm, the peak window of 11 voxels and the sliding-window overlap of 64 voxels were all fixed beforehand); the test set therefore played no role in any design choice.

\subsection{Ablation: training set size}
\label{subsec:app-datasize}
\begin{figure}[H]
\centering
\definecolor{cRight}{HTML}{B22222}
\definecolor{cLeft}{HTML}{000080}
\begin{tikzpicture}
\begin{axis}[
    name=top,
    width=\linewidth, height=5.4cm,
    ylabel={Euclidean distance $d_c$ (mm)},
    label style={font=\scriptsize}, tick label style={font=\scriptsize},
    xmin=0.5, xmax=5.5, xtick={1,2,3,4,5}, xticklabels={,,,,},
    ymode=log, ymin=0.4, ymax=150, ytick={0.5,1,2,5,10,20,50,100}, yticklabels={0.5,1,2,5,10,20,50,100},
    ymajorgrids=true, grid style={dashed, gray!30},
    legend style={at={(0.5,1.02)}, anchor=south, legend columns=2, draw=none, font=\scriptsize, /tikz/every even column/.append style={column sep=2pt}},
    legend cell align=left,
]
\addplot[draw=none, fill=cRight, fill opacity=0.18, forget plot] coordinates {(1,53.8851) (2,1.0788) (3,1.0691) (4,1.0844) (5,1.0158) (5,0.5272) (4,0.5850) (3,0.5452) (2,0.6245) (1,1.1804)} \closedcycle;
\addplot[draw=none, fill=cLeft,  fill opacity=0.18, forget plot] coordinates {(1,99.0118) (2,1.1635) (3,1.2892) (4,1.2778) (5,1.3243) (5,0.6087) (4,0.6176) (3,0.6517) (2,0.6107) (1,2.0757)} \closedcycle;
\addplot[color=cRight, mark=square*, mark size=1.3pt, mark options={solid}, thick] coordinates {(1,1.9276) (2,0.8049) (3,0.7717) (4,0.7715) (5,0.7395)}; \addlegendentry{Right}
\addplot[color=cLeft,  mark=triangle*, mark size=1.6pt, mark options={solid}, thick] coordinates {(1,36.8858) (2,0.8793) (3,0.9449) (4,0.9037) (5,0.8520)}; \addlegendentry{Left}
\end{axis}
\begin{axis}[
    at={(top.south)}, anchor=north, yshift=-0.35cm,
    width=\linewidth, height=2.6cm,
    xlabel={ImageCAS images}, ylabel={Failed (\%)},
    label style={font=\scriptsize}, tick label style={font=\scriptsize},
    xmin=0.5, xmax=5.5, xtick={1,2,3,4,5}, xticklabels={0,250,500,750,1000},
    ymin=-6, ymax=62, ytick={0,25,50},
    ymajorgrids=true, grid style={dashed, gray!30},
]
\addplot[color=cRight, mark=square*, mark size=1.3pt, mark options={solid}, thick] coordinates {(1,34.4828) (2,0.0000) (3,1.7241) (4,0.0000) (5,0.0000)};
\addplot[color=cLeft,  mark=triangle*, mark size=1.6pt, mark options={solid}, thick] coordinates {(1,53.4483) (2,0.0000) (3,0.0000) (4,5.1724) (5,0.0000)};
\end{axis}
\end{tikzpicture}
\caption{Euclidean distance $d_c$ (mm) of \swinv2 vs. the number of ImageCAS images used for training (equidistant axis; 1000 denotes the complete dataset, i.e., the 950 images not in the test set; CAT08 and ASOCA always included). Top: medians (lines) and p25--p75 (bands) of the right and left ostium over the whole test set, on a logarithmic axis. Bottom: percentage of failed localizations (error above 5~mm). See Section~\ref{subsec:evaluation-metrics}.}
\label{fig:app-datasize-plot}
\end{figure}
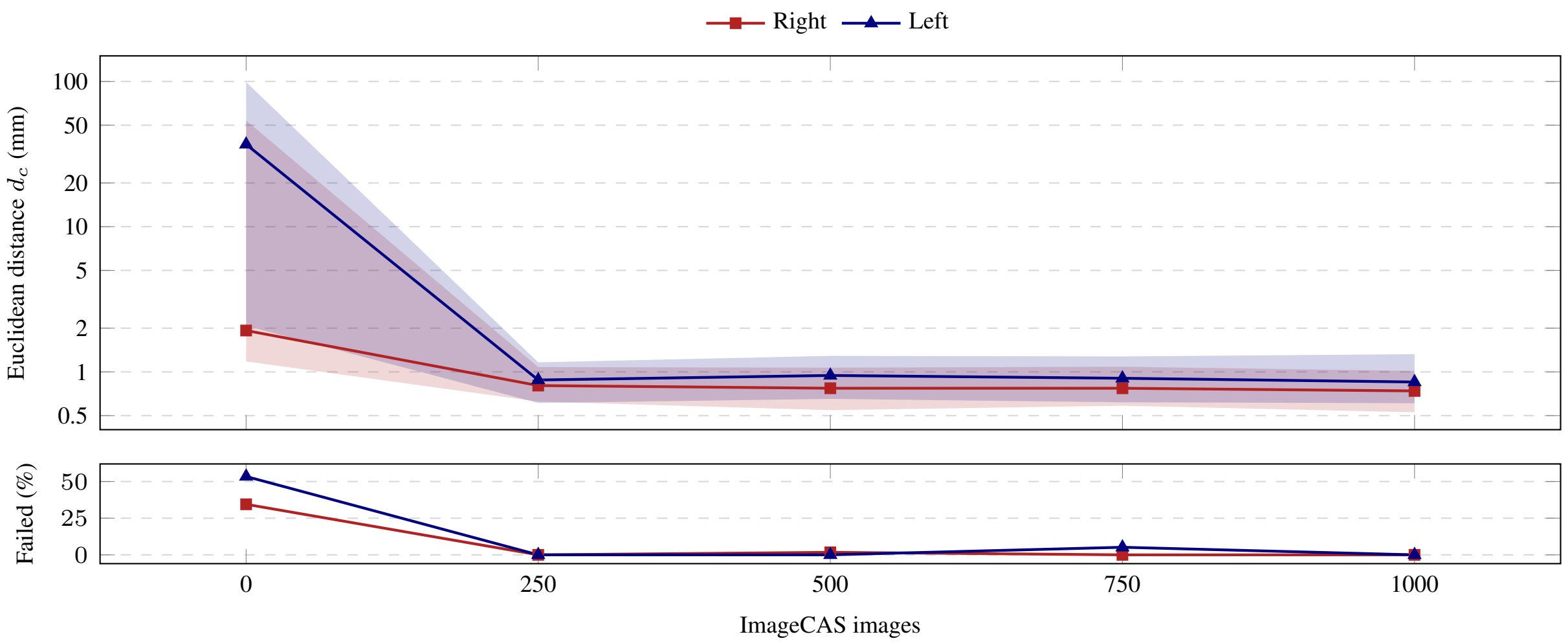

To quantify the contribution of the new annotations, we trained \swinv2 with an increasing number of ImageCAS images, namely 0, 250, 500, 750 and 1000, always together with all the volumes of CAT08 and ASOCA. The arm without ImageCAS images was also trained with \nnresunet. The ImageCAS images were drawn at random from those not belonging to the test set, using a fixed random seed (31018) so that the selection is reproducible, with the smaller subsets nested within the larger ones, and shared between the training and validation sets in the same proportion as in the complete dataset; the test set, which contains cases from the three datasets, was the same for all the models. The numbers of images refer to the training and validation sets combined (the label 1000 denotes the complete dataset, i.e., all the 950 ImageCAS images not in the test set), and, since the images are drawn at random from the whole dataset, each subset mixes manually and semi-automatically annotated cases. The results are reported in Table~\ref{tab:app-datasize} and Figure~\ref{fig:app-datasize-plot}.
Without any ImageCAS image, the model is accurate on the right ostium only in the central part of the distribution (median 1.93~mm) and fails on 34.48\% of the right and 53.45\% of the left ostia of the test set, with a mean error of 34.53~mm and 50.94~mm respectively; the failures concern the ImageCAS cases (34\% right, 58\% left) as well as those of ASOCA and CAT08 (25--50\% of the four cases of each dataset, depending on the ostium). With 250 ImageCAS images the errors drop to a median of 0.80~mm (right) and 0.88~mm (left), no failure is recorded for either ostium, and 74.14\% (right) and 86.21\% (left) of the predictions fall in the Optimal bucket. The complete dataset (labeled 1000) yields a median of 0.74~mm and 0.85~mm and 77.59\% and 84.48\% of Optimal predictions, i.e. values close to those obtained with 250 images. With 500 images the median errors are 0.77~mm (right) and 0.94~mm (left), 75.86\% and 82.76\% of the predictions are Optimal, and a single right ostium fails (1.72\%, an ASOCA volume, which raises the mean error of the right ostium to 3.05~mm), whereas no left ostium fails. With 750 images the median errors are 0.77~mm (right) and 0.90~mm (left), 74.14\% and 81.03\% of the predictions are Optimal, and no right ostium fails, but three left ostia do (5.17\%): two of the four ASOCA volumes and one ImageCAS volume, with errors of up to 141.24~mm, which raise the mean error of the left ostium to 7.47~mm (0.83~mm for the right one). Overall, the median errors of the models trained with 250 or more ImageCAS images span only 0.06~mm on the right and 0.09~mm on the left ostium, with no monotonic trend in the number of images, although the model trained on the complete dataset has the best value in six of the ten distance columns of Table~\ref{tab:app-datasize}, whereas the few failures (one right ostium with 500 images and three left ostia with 750 images) occur at intermediate sizes and mostly concern ASOCA volumes.

The arm without ImageCAS images was repeated with \nnresunet, trained on the same 64 volumes of CAT08 and ASOCA and tested on the same 58 volumes. Its median errors were 1.33~mm (right) and 1.03~mm (left), against 1.93~mm and 36.89~mm for \swinv2 trained in the same way, and 0.70~mm and 0.87~mm for \nnresunet trained on the three datasets. Of its right ostia, 32.8\% were Optimal, 51.7\% Good, 8.6\% Acceptable and 6.9\% Failed (four ostia, with errors of up to 74.73~mm); of the left ostia, 72.4\% were Optimal, 19.0\% Good, 5.2\% Acceptable and 3.4\% Failed (two ostia, up to 117.52~mm), and the mean error was 5.76~mm and 4.78~mm. All the failures were ImageCAS volumes (five of the 50 test volumes, one with both ostia failed), so that the median errors on ImageCAS were 1.37~mm and 1.09~mm and the shares of Optimal localizations 34\% and 68\%, against 74\% and 88\% for the model trained on the three datasets, which has no failure. On the four ASOCA and the four CAT08 volumes of the test set, which belong to the distribution of its training data, no ostium failed: the median errors were 1.09~mm (right) and 0.74~mm (left) for ASOCA and 1.35~mm and 0.76~mm for CAT08.

\paragraph{Interpretation} The model without ImageCAS images fails on a quarter to a half of the four ASOCA and the four CAT08 volumes too, which belong to the same distribution as its training data: the two small datasets are not sufficient to train a reliable \swinv2 even for their own distribution, and the contrast with the models trained with ImageCAS images cannot be explained only by a domain shift toward the ImageCAS test volumes. \nnresunet is less affected: it does not fail on the ASOCA and CAT08 test volumes and localizes most ImageCAS ostia, but with larger errors and with failures that the same model trained on the three datasets does not have, so that, for both architectures, the reliability on ImageCAS depends on the presence of its volumes in training, and the amount of data limits the model more than the architecture does. The occasional failures at intermediate sizes (one right ostium with 500 images, three left ostia with 750 images, mostly in ASOCA volumes) are compatible with run-to-run variability rather than with a systematic degradation, but they show that the robustness of a single training is not guaranteed by a larger training set, and that the mean errors, which these few cases dominate (7.47~mm for the left ostium at 750 images), are far less stable than the medians. ASOCA, which is the most different from ImageCAS in acquisition and contributes only four test volumes, appears to be the most sensitive dataset.

Two considerations suggest that the benefit of the volumes beyond 250 may be larger than these differences show. First, errors of 0.8--0.9~mm are already close to the level at which the annotations themselves are uncertain, so the metric has little room to discriminate between models, and any gain in robustness on rare and difficult cases would be hard to detect with this test set, whose size limits its sensitivity to infrequent failures. Second, the experiment with several numbers of images concerns a single architecture (only the arm without ImageCAS images was repeated with \nnresunet) and a fixed training recipe, and the benefit of the larger dataset may be different for models with more capacity or less tolerant of noise. The experiment shows that the public data available before were not sufficient for this task. It does not show that the whole dataset is needed to reach the accuracy of the final model, although the complete dataset gave the most accurate model overall. The implications of this ablation are summarized in Section~\ref{sec:discussion-datasize}.
\begin{table}[H]
\centering
\scriptsize
\setlength{\tabcolsep}{1.5pt}
\newcommand{\bucketgapDataSize}{-2pt}
\newcommand{\blockgapDataSize}{5pt}
\caption{Euclidean distance $d_c$ (mm) of \swinv2 trained with an increasing number of ImageCAS images (training and validation sets combined; 1000 denotes the complete dataset, i.e., the 950 images not in the test set), always together with the whole of CAT08 and ASOCA (excluding the held-out cases of the test set). All the models are evaluated on the same test set. The line below each row gives the percentage of cases in each accuracy bucket for the right and left ostium: O = Optimal, G = Good, A = Acceptable, F = Failed. See Section~\ref{subsec:evaluation-metrics}. In each distance column, the best value is in bold and the second best is underlined, comparing the rows of the same kind (overall and, separately, each dataset).}
\label{tab:app-datasize}
\begin{tabular}{l *{10}{r} r}
\toprule
 & \multicolumn{5}{c}{Right ostium} & \multicolumn{5}{c}{Left ostium} & \\
\cmidrule(lr){2-6} \cmidrule(lr){7-11}
\makecell[l]{ImageCAS\\images} & p25 & Med. & Mean & p75 & Max & p25 & Med. & Mean & p75 & Max & \makecell{Time\\(s)} \\
\midrule
0 & 1.18 & 1.93 & 34.53 & 53.89 & 170.58 & 2.08 & 36.89 & 50.94 & 99.01 & 139.27 & 48.7 \\[\bucketgapDataSize]
 & \multicolumn{5}{c}{\bucketrow{24.14}{22.41}{18.97}{34.48}} & \multicolumn{5}{c}{\bucketrow{18.97}{13.79}{13.79}{53.45}} & \\
\multicolumn{1}{r}{\subds{ASOCA}} & \subds{0.64} & \subds{1.17} & \subds{15.79} & \subds{16.31} & \subds{60.34} & \subds{2.94} & \subds{3.25} & \subds{21.07} & \subds{21.39} & \subds{75.31} & \\[\bucketgapDataSize]
 & \multicolumn{5}{c}{\subds{\bucketrow{50.00}{25.00}{0.00}{25.00}}} & \multicolumn{5}{c}{\subds{\bucketrow{0.00}{25.00}{50.00}{25.00}}} & \\
\multicolumn{1}{r}{\subds{CAT08}} & \subds{1.16} & \subds{16.31} & \subds{43.08} & \subds{58.24} & \subds{138.70} & \subds{0.94} & \subds{2.04} & \subds{32.58} & \subds{33.68} & \subds{125.74} & \\[\bucketgapDataSize]
 & \multicolumn{5}{c}{\subds{\bucketrow{25.00}{25.00}{0.00}{50.00}}} & \multicolumn{5}{c}{\subds{\bucketrow{50.00}{0.00}{25.00}{25.00}}} & \\
\multicolumn{1}{r}{\subds{ImageCAS}} & \subds{1.20} & \subds{1.94} & \subds{35.34} & \subds{53.89} & \subds{170.58} & \subds{2.08} & \subds{64.17} & \subds{54.79} & \subds{101.20} & \subds{139.27} & \\[\bucketgapDataSize]
 & \multicolumn{5}{c}{\subds{\bucketrow{22.00}{22.00}{22.00}{34.00}}} & \multicolumn{5}{c}{\subds{\bucketrow{18.00}{14.00}{10.00}{58.00}}} & \\
\addlinespace[\blockgapDataSize]
250 & 0.62 & 0.80 & \underline{0.91} & 1.08 & 2.21 & \textbf{0.61} & \underline{0.88} & \textbf{0.95} & \textbf{1.16} & \underline{2.36} & 48.1 \\[\bucketgapDataSize]
 & \multicolumn{5}{c}{\bucketrow{74.14}{20.69}{5.17}{0.00}} & \multicolumn{5}{c}{\bucketrow{86.21}{13.79}{0.00}{0.00}} & \\
\multicolumn{1}{r}{\subds{ASOCA}} & \subds{0.79} & \subds{0.87} & \subds{\underline{0.92}} & \subds{\textbf{1.00}} & \subds{\underline{1.21}} & \subds{0.56} & \subds{\underline{0.60}} & \subds{0.61} & \subds{\underline{0.64}} & \subds{0.72} & \\[\bucketgapDataSize]
 & \multicolumn{5}{c}{\subds{\bucketrow{75.00}{25.00}{0.00}{0.00}}} & \multicolumn{5}{c}{\subds{\bucketrow{100.00}{0.00}{0.00}{0.00}}} & \\
\multicolumn{1}{r}{\subds{CAT08}} & \subds{0.75} & \subds{1.27} & \subds{1.36} & \subds{1.87} & \subds{2.21} & \subds{\underline{0.79}} & \subds{\textbf{0.95}} & \subds{\textbf{0.90}} & \subds{\textbf{1.07}} & \subds{\underline{1.24}} & \\[\bucketgapDataSize]
 & \multicolumn{5}{c}{\subds{\bucketrow{50.00}{25.00}{25.00}{0.00}}} & \multicolumn{5}{c}{\subds{\bucketrow{100.00}{0.00}{0.00}{0.00}}} & \\
\multicolumn{1}{r}{\subds{ImageCAS}} & \subds{0.61} & \subds{0.80} & \subds{0.87} & \subds{\underline{1.04}} & \subds{2.17} & \subds{0.65} & \subds{\textbf{0.89}} & \subds{\textbf{0.98}} & \subds{\textbf{1.24}} & \subds{\underline{2.36}} & \\[\bucketgapDataSize]
 & \multicolumn{5}{c}{\subds{\bucketrow{76.00}{20.00}{4.00}{0.00}}} & \multicolumn{5}{c}{\subds{\bucketrow{84.00}{16.00}{0.00}{0.00}}} & \\
\addlinespace[\blockgapDataSize]
500 & \underline{0.55} & \underline{0.77} & 3.05 & \underline{1.07} & 128.48 & 0.65 & 0.94 & 1.01 & 1.29 & \textbf{2.34} & 46.1 \\[\bucketgapDataSize]
 & \multicolumn{5}{c}{\bucketrow{75.86}{17.24}{5.17}{1.72}} & \multicolumn{5}{c}{\bucketrow{82.76}{17.24}{0.00}{0.00}} & \\
\multicolumn{1}{r}{\subds{ASOCA}} & \subds{0.75} & \subds{\underline{0.82}} & \subds{32.69} & \subds{32.77} & \subds{128.48} & \subds{\textbf{0.40}} & \subds{\textbf{0.50}} & \subds{\textbf{0.46}} & \subds{\textbf{0.56}} & \subds{\textbf{0.69}} & \\[\bucketgapDataSize]
 & \multicolumn{5}{c}{\subds{\bucketrow{75.00}{0.00}{0.00}{25.00}}} & \multicolumn{5}{c}{\subds{\bucketrow{100.00}{0.00}{0.00}{0.00}}} & \\
\multicolumn{1}{r}{\subds{CAT08}} & \subds{\underline{0.69}} & \subds{1.17} & \subds{1.23} & \subds{1.71} & \subds{1.98} & \subds{1.00} & \subds{1.05} & \subds{1.04} & \subds{\underline{1.10}} & \subds{\underline{1.24}} & \\[\bucketgapDataSize]
 & \multicolumn{5}{c}{\subds{\bucketrow{50.00}{25.00}{25.00}{0.00}}} & \multicolumn{5}{c}{\subds{\bucketrow{100.00}{0.00}{0.00}{0.00}}} & \\
\multicolumn{1}{r}{\subds{ImageCAS}} & \subds{\textbf{0.51}} & \subds{\textbf{0.74}} & \subds{\underline{0.83}} & \subds{\underline{1.04}} & \subds{2.26} & \subds{0.70} & \subds{0.98} & \subds{1.05} & \subds{1.34} & \subds{\textbf{2.34}} & \\[\bucketgapDataSize]
 & \multicolumn{5}{c}{\subds{\bucketrow{78.00}{18.00}{4.00}{0.00}}} & \multicolumn{5}{c}{\subds{\bucketrow{80.00}{20.00}{0.00}{0.00}}} & \\
\addlinespace[\blockgapDataSize]
750 & 0.59 & \underline{0.77} & \textbf{0.83} & 1.08 & \textbf{1.95} & \underline{0.62} & 0.90 & 7.47 & \underline{1.28} & 141.24 & 47.8 \\[\bucketgapDataSize]
 & \multicolumn{5}{c}{\bucketrow{74.14}{24.14}{1.72}{0.00}} & \multicolumn{5}{c}{\bucketrow{81.03}{13.79}{0.00}{5.17}} & \\
\multicolumn{1}{r}{\subds{ASOCA}} & \subds{\textbf{0.44}} & \subds{\textbf{0.79}} & \subds{\textbf{0.78}} & \subds{\underline{1.13}} & \subds{\textbf{1.16}} & \subds{0.67} & \subds{57.55} & \subds{60.23} & \subds{117.11} & \subds{125.21} &  \\[\bucketgapDataSize]
 & \multicolumn{5}{c}{\subds{\bucketrow{50.00}{50.00}{0.00}{0.00}}} & \multicolumn{5}{c}{\subds{\bucketrow{50.00}{0.00}{0.00}{50.00}}} & \\
\multicolumn{1}{r}{\subds{CAT08}} & \subds{0.72} & \subds{\underline{1.14}} & \subds{\underline{1.22}} & \subds{\underline{1.64}} & \subds{\underline{1.95}} & \subds{0.87} & \subds{\underline{1.00}} & \subds{\underline{1.01}} & \subds{1.14} & \subds{\textbf{1.21}} &  \\[\bucketgapDataSize]
 & \multicolumn{5}{c}{\subds{\bucketrow{50.00}{25.00}{25.00}{0.00}}} & \multicolumn{5}{c}{\subds{\bucketrow{100.00}{0.00}{0.00}{0.00}}} & \\
\multicolumn{1}{r}{\subds{ImageCAS}} & \subds{0.59} & \subds{\underline{0.77}} & \subds{\textbf{0.81}} & \subds{\underline{1.04}} & \subds{\textbf{1.77}} & \subds{\textbf{0.59}} & \subds{\underline{0.90}} & \subds{3.76} & \subds{\underline{1.28}} & \subds{141.24} &  \\[\bucketgapDataSize]
 & \multicolumn{5}{c}{\subds{\bucketrow{78.00}{22.00}{0.00}{0.00}}} & \multicolumn{5}{c}{\subds{\bucketrow{82.00}{16.00}{0.00}{2.00}}} & \\
\addlinespace[\blockgapDataSize]
1000 & \textbf{0.53} & \textbf{0.74} & \textbf{0.83} & \textbf{1.02} & \underline{2.07} & \textbf{0.61} & \textbf{0.85} & \underline{0.98} & 1.32 & 2.51 & 55.6 \\[\bucketgapDataSize]
 & \multicolumn{5}{c}{\bucketrow{77.59}{20.69}{1.72}{0.00}} & \multicolumn{5}{c}{\bucketrow{84.48}{15.52}{0.00}{0.00}} & \\
\multicolumn{1}{r}{\subds{ASOCA}} & \subds{\underline{0.53}} & \subds{0.95} & \subds{0.95} & \subds{1.37} & \subds{1.45} & \subds{\underline{0.54}} & \subds{\underline{0.60}} & \subds{\underline{0.58}} & \subds{\underline{0.64}} & \subds{\underline{0.71}} & \\[\bucketgapDataSize]
 & \multicolumn{5}{c}{\subds{\bucketrow{50.00}{50.00}{0.00}{0.00}}} & \multicolumn{5}{c}{\subds{\bucketrow{100.00}{0.00}{0.00}{0.00}}} & \\
\multicolumn{1}{r}{\subds{CAT08}} & \subds{\textbf{0.60}} & \subds{\textbf{1.02}} & \subds{\textbf{1.04}} & \subds{\textbf{1.46}} & \subds{\textbf{1.75}} & \subds{\textbf{0.59}} & \subds{1.04} & \subds{1.20} & \subds{1.64} & \subds{2.25} & \\[\bucketgapDataSize]
 & \multicolumn{5}{c}{\subds{\bucketrow{50.00}{50.00}{0.00}{0.00}}} & \multicolumn{5}{c}{\subds{\bucketrow{75.00}{25.00}{0.00}{0.00}}} & \\
\multicolumn{1}{r}{\subds{ImageCAS}} & \subds{\underline{0.54}} & \subds{\textbf{0.74}} & \subds{\textbf{0.81}} & \subds{\textbf{0.98}} & \subds{\underline{2.07}} & \subds{\underline{0.62}} & \subds{\textbf{0.89}} & \subds{\underline{1.00}} & \subds{1.32} & \subds{2.51} & \\[\bucketgapDataSize]
 & \multicolumn{5}{c}{\subds{\bucketrow{82.00}{16.00}{2.00}{0.00}}} & \multicolumn{5}{c}{\subds{\bucketrow{84.00}{16.00}{0.00}{0.00}}} & \\
\bottomrule
\end{tabular}
\end{table}

\subsection{Ablation: optimal respacing}
\label{subsec:app-respacing-optimal}
\begin{figure}[H]
\centering
\definecolor{cRight}{HTML}{B22222}
\definecolor{cLeft}{HTML}{000080}
\definecolor{cTime}{HTML}{555555}
\begin{tikzpicture}
\begin{axis}[
    name=top,
    width=\linewidth, height=5.4cm,
    ylabel={Euclidean distance $d_c$ (mm)},
    label style={font=\scriptsize}, tick label style={font=\scriptsize},
    xmin=0.5, xmax=4.5, xtick={1,2,3,4}, xticklabels={,,,},
    ymin=0.2, ymax=1.8, ytick={0.4,0.8,1.2,1.6},
    ymajorgrids=true, grid style={dashed, gray!30},
    legend style={at={(0.5,1.02)}, anchor=south, legend columns=2, draw=none, font=\scriptsize, /tikz/every even column/.append style={column sep=2pt}},
    legend cell align=left,
]
\addplot[draw=none, fill=cRight, fill opacity=0.18, forget plot] coordinates {(1,1.0158) (2,1.2439) (3,1.0240) (4,1.2759) (4,0.6901) (3,0.4493) (2,0.6036) (1,0.5272)} \closedcycle;
\addplot[draw=none, fill=cLeft,  fill opacity=0.18, forget plot] coordinates {(1,1.3243) (2,1.4534) (3,1.2943) (4,1.6175) (4,0.8311) (3,0.6944) (2,0.7553) (1,0.6087)} \closedcycle;
\addplot[color=cRight, mark=square*, mark size=1.3pt, mark options={solid}, thick] coordinates {(1,0.7395) (2,0.8437) (3,0.7431) (4,0.9029)}; \addlegendentry{Right}
\addplot[color=cLeft,  mark=triangle*, mark size=1.6pt, mark options={solid}, thick] coordinates {(1,0.8520) (2,1.0756) (3,1.0091) (4,1.2954)}; \addlegendentry{Left}
\end{axis}
\begin{axis}[
    at={(top.south)}, anchor=north, yshift=-0.35cm,
    width=\linewidth, height=2.6cm,
    xlabel={Training and inference spacing (mm)}, ylabel={Time (s)},
    label style={font=\scriptsize}, tick label style={font=\scriptsize},
    xmin=0.5, xmax=4.5, xtick={1,2,3,4}, xticklabels={0.5,0.75,1.0,1.5},
    ymin=0, ymax=70, ytick={0,25,50},
    ymajorgrids=true, grid style={dashed, gray!30},
]
\addplot[color=cTime, mark=*, mark size=1.3pt, mark options={solid}, thick, dashed] coordinates {(1,55.6) (2,26.3) (3,21.9) (4,23.1)};
\end{axis}
\end{tikzpicture}
\caption{Euclidean distance $d_c$ (mm) of \swinv2 trained and tested at the same isotropic spacing (equidistant axis). Top: medians (lines) and p25--p75 (bands) of the right and left ostium over the whole test set. Bottom: processing time per patient. See Section~\ref{subsec:evaluation-metrics}.}
\label{fig:app-respacing-optimal-plot}
\end{figure}
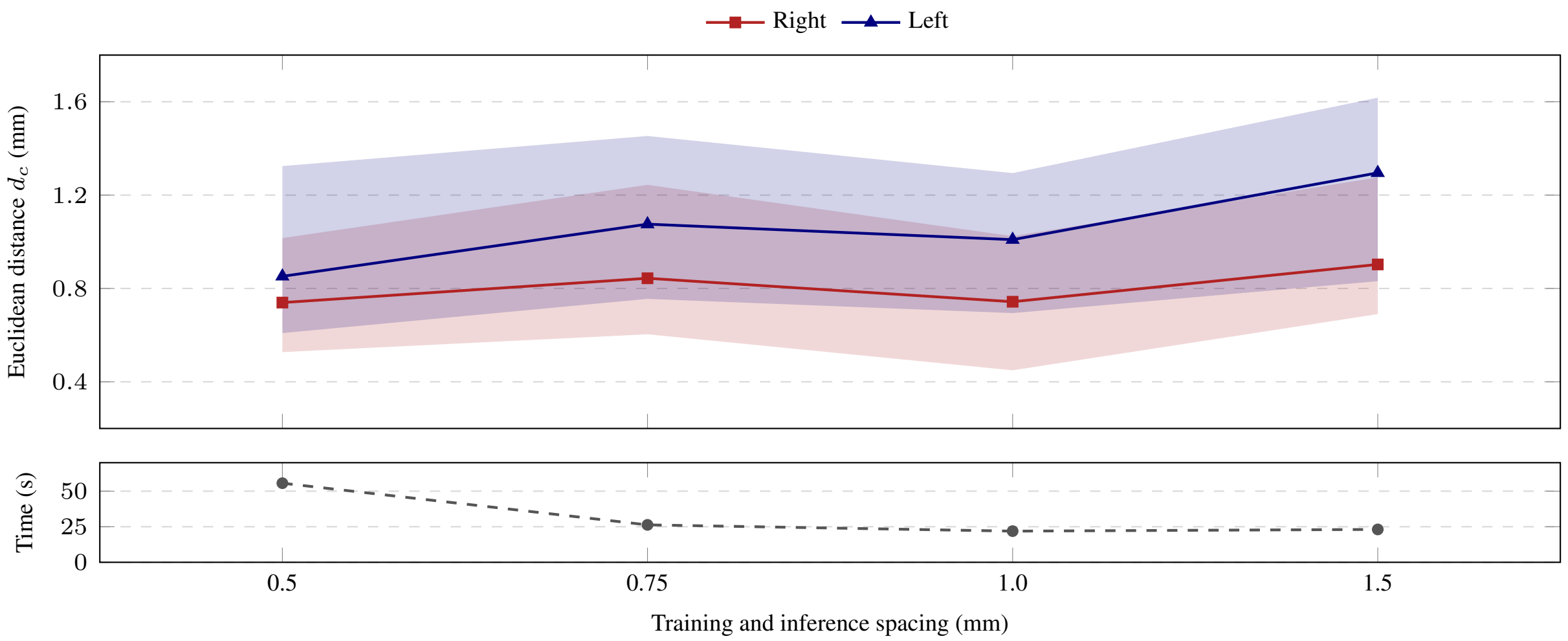

We trained \swinv2 at different isotropic spacings, namely 0.75, 1.0 and 1.5~mm in addition to the 0.5~mm of the final model, and tested each model on volumes resampled to the spacing it was trained at. All the settings expressed in voxels (the $128^3$ patch, the width of the Gaussian target and the 11-voxel peak window) were left unchanged, and therefore grow in millimeters with the spacing. Table~\ref{tab:app-respacing-optimal} and Figure~\ref{fig:app-respacing-optimal-plot} report the errors and the processing time per patient.
No localization failed at 0.5, 0.75, 1.0 and 1.5~mm. The median error of the right ostium was 0.74, 0.84, 0.74 and 0.90~mm, respectively, and that of the left ostium 0.85, 1.08, 1.01 and 1.30~mm; the mean error of the two ostia was 0.91, 1.02, 0.95 and 1.14~mm. The share of Optimal localizations was 77.6, 69.0, 77.6 and 60.3\% for the right ostium and 84.5, 74.1, 86.2 and 65.5\% for the left one, while that of Acceptable localizations was 1.7, 5.2, 8.6 and 6.9\% for the right ostium and 0, 0, 1.7 and 1.7\% for the left one; at patient level, 69.0, 55.2, 69.0 and 44.8\% of the cases were Optimal. The processing time per patient decreased from 55.6~s at 0.5~mm to 26.3~s at 0.75~mm and 21.9~s at 1.0~mm, and was 23.1~s at 1.5~mm.
\begin{table}[H]
\centering
\scriptsize
\setlength{\tabcolsep}{1.5pt}
\newcommand{\bucketgapRespOptimal}{-2pt}
\newcommand{\blockgapRespOptimal}{5pt}
\caption{Euclidean distance $d_c$ (mm) of \swinv2 trained and tested at the isotropic spacing (mm) reported in the first column. The line below each row gives the percentage of cases in each accuracy bucket for the right and left ostium: O = Optimal, G = Good, A = Acceptable, F = Failed. See Section~\ref{subsec:evaluation-metrics}. In each distance column, the best value is in bold and the second best is underlined, comparing the rows of the same kind (overall and, separately, each dataset).}
\label{tab:app-respacing-optimal}
\begin{tabular}{l *{10}{r} r}
\toprule
 & \multicolumn{5}{c}{Right ostium} & \multicolumn{5}{c}{Left ostium} & \\
\cmidrule(lr){2-6} \cmidrule(lr){7-11}
\makecell[l]{Spacing\\(mm)} & p25 & Med. & Mean & p75 & Max & p25 & Med. & Mean & p75 & Max & \makecell{Time\\(s)} \\
\midrule
0.5 & \underline{0.53} & \textbf{0.74} & \textbf{0.83} & \textbf{1.02} & \textbf{2.07} & \textbf{0.61} & \textbf{0.85} & \textbf{0.98} & \underline{1.32} & \underline{2.51} & 55.6 \\[\bucketgapRespOptimal]
 & \multicolumn{5}{c}{\bucketrow{77.59}{20.69}{1.72}{0.00}} & \multicolumn{5}{c}{\bucketrow{84.48}{15.52}{0.00}{0.00}} & \\
\multicolumn{1}{r}{\subds{ASOCA}} & \subds{\textbf{0.53}} & \subds{\underline{0.95}} & \subds{\underline{0.95}} & \subds{1.37} & \subds{\underline{1.45}} & \subds{\textbf{0.54}} & \subds{\textbf{0.60}} & \subds{\textbf{0.58}} & \subds{\textbf{0.64}} & \subds{\textbf{0.71}} & \\[\bucketgapRespOptimal]
 & \multicolumn{5}{c}{\subds{\bucketrow{50.00}{50.00}{0.00}{0.00}}} & \multicolumn{5}{c}{\subds{\bucketrow{100.00}{0.00}{0.00}{0.00}}} & \\
\multicolumn{1}{r}{\subds{CAT08}} & \subds{\textbf{0.60}} & \subds{\textbf{1.02}} & \subds{\textbf{1.04}} & \subds{\underline{1.46}} & \subds{\textbf{1.75}} & \subds{\textbf{0.59}} & \subds{\underline{1.04}} & \subds{1.20} & \subds{1.64} & \subds{2.25} & \\[\bucketgapRespOptimal]
 & \multicolumn{5}{c}{\subds{\bucketrow{50.00}{50.00}{0.00}{0.00}}} & \multicolumn{5}{c}{\subds{\bucketrow{75.00}{25.00}{0.00}{0.00}}} & \\
\multicolumn{1}{r}{\subds{ImageCAS}} & \subds{\underline{0.54}} & \subds{\underline{0.74}} & \subds{\textbf{0.81}} & \subds{\textbf{0.98}} & \subds{\textbf{2.07}} & \subds{\textbf{0.62}} & \subds{\textbf{0.89}} & \subds{\textbf{1.00}} & \subds{\underline{1.32}} & \subds{\underline{2.51}} & \\[\bucketgapRespOptimal]
 & \multicolumn{5}{c}{\subds{\bucketrow{82.00}{16.00}{2.00}{0.00}}} & \multicolumn{5}{c}{\subds{\bucketrow{84.00}{16.00}{0.00}{0.00}}} & \\
\addlinespace[\blockgapRespOptimal]
0.75 & 0.60 & \underline{0.84} & 0.94 & \underline{1.24} & \underline{2.12} & 0.76 & 1.08 & 1.09 & 1.45 & \textbf{2.39} & 26.3 \\[\bucketgapRespOptimal]
 & \multicolumn{5}{c}{\bucketrow{68.97}{25.86}{5.17}{0.00}} & \multicolumn{5}{c}{\bucketrow{74.14}{25.86}{0.00}{0.00}} & \\
\multicolumn{1}{r}{\subds{ASOCA}} & \subds{0.82} & \subds{1.17} & \subds{1.13} & \subds{1.48} & \subds{1.56} & \subds{\underline{0.65}} & \subds{\underline{0.81}} & \subds{0.80} & \subds{0.96} & \subds{1.00} & \\[\bucketgapRespOptimal]
 & \multicolumn{5}{c}{\subds{\bucketrow{50.00}{50.00}{0.00}{0.00}}} & \multicolumn{5}{c}{\subds{\bucketrow{100.00}{0.00}{0.00}{0.00}}} & \\
\multicolumn{1}{r}{\subds{CAT08}} & \subds{\underline{0.70}} & \subds{1.19} & \subds{1.14} & \subds{1.63} & \subds{2.07} & \subds{\underline{0.78}} & \subds{\textbf{0.83}} & \subds{\textbf{1.01}} & \subds{\textbf{1.06}} & \subds{\textbf{1.63}} & \\[\bucketgapRespOptimal]
 & \multicolumn{5}{c}{\subds{\bucketrow{50.00}{25.00}{25.00}{0.00}}} & \multicolumn{5}{c}{\subds{\bucketrow{75.00}{25.00}{0.00}{0.00}}} & \\
\multicolumn{1}{r}{\subds{ImageCAS}} & \subds{0.58} & \subds{0.79} & \subds{0.91} & \subds{\underline{1.16}} & \subds{\underline{2.12}} & \subds{0.77} & \subds{1.15} & \subds{1.12} & \subds{1.48} & \subds{\textbf{2.39}} & \\[\bucketgapRespOptimal]
 & \multicolumn{5}{c}{\subds{\bucketrow{72.00}{24.00}{4.00}{0.00}}} & \multicolumn{5}{c}{\subds{\bucketrow{72.00}{28.00}{0.00}{0.00}}} & \\
\addlinespace[\blockgapRespOptimal]
1.0 & \textbf{0.45} & \textbf{0.74} & \underline{0.86} & \textbf{1.02} & 2.20 & \underline{0.69} & \underline{1.01} & \underline{1.05} & \textbf{1.29} & 2.72 & 21.9 \\[\bucketgapRespOptimal]
 & \multicolumn{5}{c}{\bucketrow{77.59}{13.79}{8.62}{0.00}} & \multicolumn{5}{c}{\bucketrow{86.21}{12.07}{1.72}{0.00}} & \\
\multicolumn{1}{r}{\subds{ASOCA}} & \subds{\underline{0.74}} & \subds{\textbf{0.76}} & \subds{\textbf{0.82}} & \subds{\textbf{0.85}} & \subds{\textbf{1.07}} & \subds{0.68} & \subds{0.84} & \subds{\underline{0.76}} & \subds{\underline{0.91}} & \subds{\underline{0.95}} & \\[\bucketgapRespOptimal]
 & \multicolumn{5}{c}{\subds{\bucketrow{100.00}{0.00}{0.00}{0.00}}} & \multicolumn{5}{c}{\subds{\bucketrow{100.00}{0.00}{0.00}{0.00}}} & \\
\multicolumn{1}{r}{\subds{CAT08}} & \subds{0.81} & \subds{1.25} & \subds{1.14} & \subds{1.58} & \subds{\underline{1.76}} & \subds{0.80} & \subds{1.13} & \subds{\underline{1.18}} & \subds{1.51} & \subds{\underline{1.82}} & \\[\bucketgapRespOptimal]
 & \multicolumn{5}{c}{\subds{\bucketrow{50.00}{50.00}{0.00}{0.00}}} & \multicolumn{5}{c}{\subds{\bucketrow{75.00}{25.00}{0.00}{0.00}}} & \\
\multicolumn{1}{r}{\subds{ImageCAS}} & \subds{\textbf{0.44}} & \subds{\textbf{0.66}} & \subds{\underline{0.84}} & \subds{\textbf{0.98}} & \subds{2.20} & \subds{\underline{0.69}} & \subds{\underline{1.06}} & \subds{\underline{1.06}} & \subds{\textbf{1.29}} & \subds{2.72} & \\[\bucketgapRespOptimal]
 & \multicolumn{5}{c}{\subds{\bucketrow{78.00}{12.00}{10.00}{0.00}}} & \multicolumn{5}{c}{\subds{\bucketrow{86.00}{12.00}{2.00}{0.00}}} & \\
\addlinespace[\blockgapRespOptimal]
1.5 & 0.69 & 0.90 & 1.01 & 1.28 & 2.27 & 0.83 & 1.30 & 1.27 & 1.62 & 2.61 & 23.1 \\[\bucketgapRespOptimal]
 & \multicolumn{5}{c}{\bucketrow{60.34}{32.76}{6.90}{0.00}} & \multicolumn{5}{c}{\bucketrow{65.52}{32.76}{1.72}{0.00}} & \\
\multicolumn{1}{r}{\subds{ASOCA}} & \subds{0.88} & \subds{1.09} & \subds{1.12} & \subds{\underline{1.33}} & \subds{1.50} & \subds{0.71} & \subds{0.93} & \subds{0.93} & \subds{1.15} & \subds{1.29} & \\[\bucketgapRespOptimal]
 & \multicolumn{5}{c}{\subds{\bucketrow{50.00}{50.00}{0.00}{0.00}}} & \multicolumn{5}{c}{\subds{\bucketrow{100.00}{0.00}{0.00}{0.00}}} & \\
\multicolumn{1}{r}{\subds{CAT08}} & \subds{0.76} & \subds{\underline{1.09}} & \subds{\underline{1.08}} & \subds{\textbf{1.41}} & \subds{1.80} & \subds{1.19} & \subds{1.29} & \subds{1.40} & \subds{\underline{1.49}} & \subds{2.07} & \\[\bucketgapRespOptimal]
 & \multicolumn{5}{c}{\subds{\bucketrow{50.00}{50.00}{0.00}{0.00}}} & \multicolumn{5}{c}{\subds{\bucketrow{75.00}{25.00}{0.00}{0.00}}} & \\
\multicolumn{1}{r}{\subds{ImageCAS}} & \subds{0.68} & \subds{0.89} & \subds{0.99} & \subds{1.22} & \subds{2.27} & \subds{0.83} & \subds{1.36} & \subds{1.29} & \subds{1.65} & \subds{2.61} & \\[\bucketgapRespOptimal]
 & \multicolumn{5}{c}{\subds{\bucketrow{62.00}{30.00}{8.00}{0.00}}} & \multicolumn{5}{c}{\subds{\bucketrow{62.00}{36.00}{2.00}{0.00}}} & \\
\bottomrule
\end{tabular}
\end{table}

\begin{figure}[H]
\centering
\definecolor{cRight}{HTML}{B22222}
\definecolor{cLeft}{HTML}{000080}
\definecolor{cTime}{HTML}{555555}
\begin{tikzpicture}
\begin{axis}[
    name=top,
    width=\linewidth, height=5.4cm,
    ylabel={Euclidean distance $d_c$ (mm)},
    label style={font=\scriptsize}, tick label style={font=\scriptsize},
    xmin=0.5, xmax=4.5, xtick={1,2,3,4}, xticklabels={,,,},
    ymin=0.2, ymax=1.8, ytick={0.4,0.8,1.2,1.6},
    ymajorgrids=true, grid style={dashed, gray!30},
    legend style={at={(0.5,1.02)}, anchor=south, legend columns=2, draw=none, font=\scriptsize, /tikz/every even column/.append style={column sep=2pt}},
    legend cell align=left,
]
\addplot[draw=none, fill=cRight, fill opacity=0.18, forget plot] coordinates {(1,1.1337) (2,0.9831) (3,1.0970) (4,1.5509) (4,0.7667) (3,0.4728) (2,0.5427) (1,0.5167)} \closedcycle;
\addplot[draw=none, fill=cLeft,  fill opacity=0.18, forget plot] coordinates {(1,1.2077) (2,1.2361) (3,1.1863) (4,1.4694) (4,0.8861) (3,0.6097) (2,0.6018) (1,0.6758)} \closedcycle;
\addplot[color=cRight, mark=square*, mark size=1.3pt, mark options={solid}, thick] coordinates {(1,0.7022) (2,0.7474) (3,0.6541) (4,1.0778)}; \addlegendentry{Right}
\addplot[color=cLeft,  mark=triangle*, mark size=1.6pt, mark options={solid}, thick] coordinates {(1,0.8692) (2,0.9002) (3,0.9356) (4,1.1923)}; \addlegendentry{Left}
\end{axis}
\begin{axis}[
    at={(top.south)}, anchor=north, yshift=-0.35cm,
    width=\linewidth, height=2.6cm,
    xlabel={Training and inference spacing (mm)}, ylabel={Time (s)},
    label style={font=\scriptsize}, tick label style={font=\scriptsize},
    xmin=0.5, xmax=4.5, xtick={1,2,3,4}, xticklabels={0.5,0.75,1.0,1.5},
    ymin=0, ymax=70, ytick={0,25,50},
    ymajorgrids=true, grid style={dashed, gray!30},
]
\addplot[color=cTime, mark=*, mark size=1.3pt, mark options={solid}, thick, dashed] coordinates {(1,41.1) (2,26.2) (3,21.3) (4,18.3)};
\end{axis}
\end{tikzpicture}
\caption{Euclidean distance $d_c$ (mm) of \nnresunet trained and tested at the same isotropic spacing (equidistant axis). Top: medians (lines) and p25--p75 (bands) of the right and left ostium over the whole test set. Bottom: processing time per patient. See Section~\ref{subsec:evaluation-metrics}.}
\label{fig:app-respacing-optimal-nn-plot}
\end{figure}
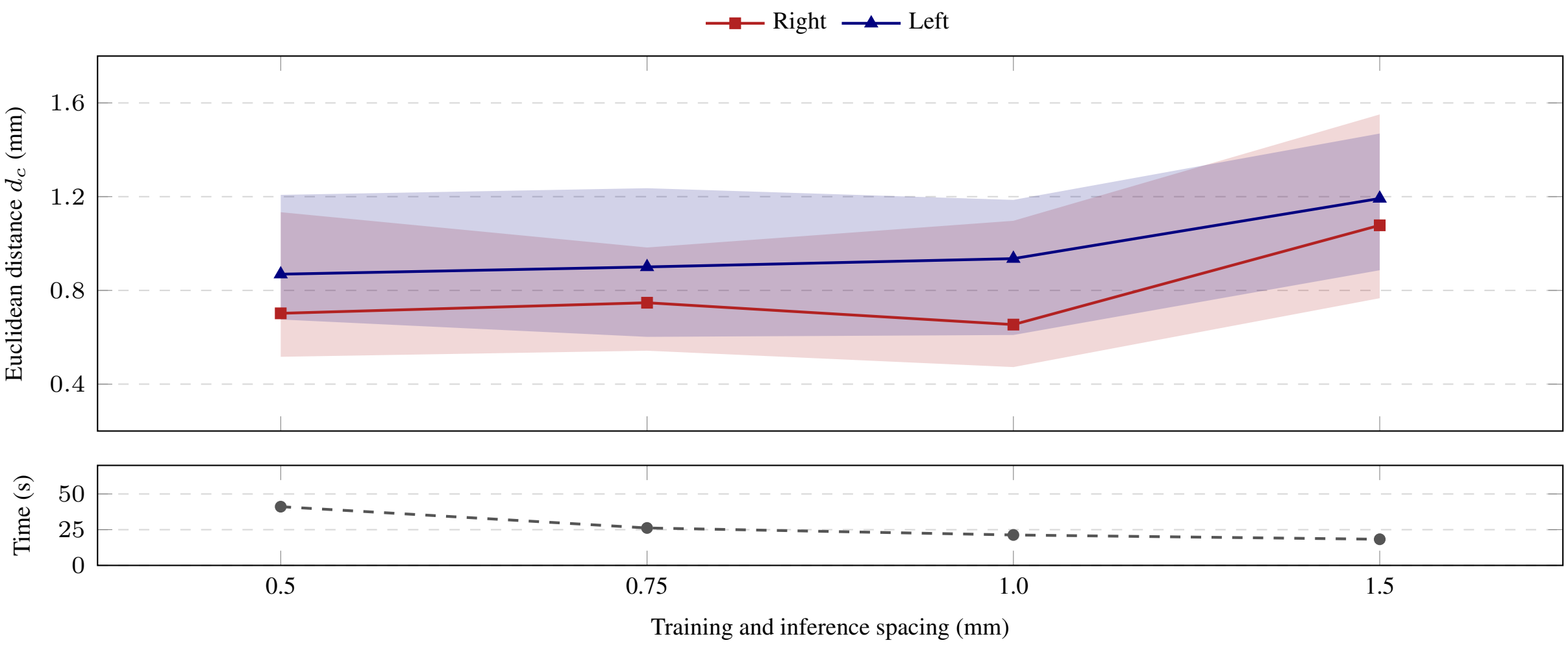

The same experiment was repeated for \nnresunet, with the same spacings and the same unchanged settings in voxels (Table~\ref{tab:app-respacing-optimal-nn} and Figure~\ref{fig:app-respacing-optimal-nn-plot}). Again, no localization failed at any spacing. The median error of the right ostium was 0.70, 0.75, 0.65 and 1.08~mm at 0.5, 0.75, 1.0 and 1.5~mm, respectively, and that of the left ostium 0.87, 0.90, 0.94 and 1.19~mm; the mean error of the two ostia was 0.89, 0.91, 0.89 and 1.20~mm. The share of Optimal localizations was 72.4, 81.0, 74.1 and 50.0\% for the right ostium and 89.7, 81.0, 89.7 and 74.1\% for the left one, while that of Acceptable localizations was 3.4, 5.2, 3.4 and 12.1\% for the right ostium and 0, 1.7, 1.7 and 0\% for the left one; at patient level, 67.2, 69.0, 69.0 and 37.9\% of the cases were Optimal. The processing time per patient decreased steadily, from 41.1~s at 0.5~mm to 26.2, 21.3 and 18.3~s at 0.75, 1.0 and 1.5~mm.
\begin{table}[H]
\centering
\scriptsize
\setlength{\tabcolsep}{1.5pt}
\newcommand{\bucketgapRespOptimalNN}{-2pt}
\newcommand{\blockgapRespOptimalNN}{5pt}
\caption{Euclidean distance $d_c$ (mm) of \nnresunet trained and tested at the isotropic spacing (mm) reported in the first column. The line below each row gives the percentage of cases in each accuracy bucket for the right and left ostium: O = Optimal, G = Good, A = Acceptable, F = Failed. See Section~\ref{subsec:evaluation-metrics}. In each distance column, the best value is in bold and the second best is underlined, comparing the rows of the same kind (overall and, separately, each dataset).}
\label{tab:app-respacing-optimal-nn}
\begin{tabular}{l *{10}{r} r}
\toprule
 & \multicolumn{5}{c}{Right ostium} & \multicolumn{5}{c}{Left ostium} & \\
\cmidrule(lr){2-6} \cmidrule(lr){7-11}
\makecell[l]{Spacing\\(mm)} & p25 & Med. & Mean & p75 & Max & p25 & Med. & Mean & p75 & Max & \makecell{Time\\(s)} \\
\midrule
0.5 & \underline{0.52} & \underline{0.70} & \textbf{0.83} & 1.13 & \textbf{2.03} & 0.68 & \textbf{0.87} & \textbf{0.96} & \underline{1.21} & \underline{2.32} & 41.1 \\[\bucketgapRespOptimalNN]
 & \multicolumn{5}{c}{\bucketrow{72.41}{24.14}{3.45}{0.00}} & \multicolumn{5}{c}{\bucketrow{89.66}{10.34}{0.00}{0.00}} & \\
\multicolumn{1}{r}{\subds{ASOCA}} & \subds{\underline{0.61}} & \subds{\underline{0.67}} & \subds{\underline{0.77}} & \subds{\underline{0.83}} & \subds{1.20} & \subds{0.58} & \subds{0.74} & \subds{\underline{0.67}} & \subds{\underline{0.83}} & \subds{\underline{1.06}} & \\[\bucketgapRespOptimalNN]
 & \multicolumn{5}{c}{\subds{\bucketrow{75.00}{25.00}{0.00}{0.00}}} & \multicolumn{5}{c}{\subds{\bucketrow{100.00}{0.00}{0.00}{0.00}}} & \\
\multicolumn{1}{r}{\subds{CAT08}} & \subds{\textbf{0.68}} & \subds{\underline{1.22}} & \subds{\underline{1.17}} & \subds{1.71} & \subds{\underline{1.74}} & \subds{1.03} & \subds{1.17} & \subds{\underline{1.05}} & \subds{1.20} & \subds{\textbf{1.23}} & \\[\bucketgapRespOptimalNN]
 & \multicolumn{5}{c}{\subds{\bucketrow{50.00}{50.00}{0.00}{0.00}}} & \multicolumn{5}{c}{\subds{\bucketrow{100.00}{0.00}{0.00}{0.00}}} & \\
\multicolumn{1}{r}{\subds{ImageCAS}} & \subds{\underline{0.48}} & \subds{\underline{0.70}} & \subds{\textbf{0.80}} & \subds{1.07} & \subds{\textbf{2.03}} & \subds{0.68} & \subds{\textbf{0.87}} & \subds{\textbf{0.97}} & \subds{\underline{1.23}} & \subds{\underline{2.32}} & \\[\bucketgapRespOptimalNN]
 & \multicolumn{5}{c}{\subds{\bucketrow{74.00}{22.00}{4.00}{0.00}}} & \multicolumn{5}{c}{\subds{\bucketrow{88.00}{12.00}{0.00}{0.00}}} & \\
\addlinespace[\blockgapRespOptimalNN]
0.75 & 0.54 & 0.75 & \textbf{0.83} & \textbf{0.98} & \underline{2.05} & \textbf{0.60} & \underline{0.90} & \underline{0.98} & 1.24 & 2.66 & 26.2 \\[\bucketgapRespOptimalNN]
 & \multicolumn{5}{c}{\bucketrow{81.03}{13.79}{5.17}{0.00}} & \multicolumn{5}{c}{\bucketrow{81.03}{17.24}{1.72}{0.00}} & \\
\multicolumn{1}{r}{\subds{ASOCA}} & \subds{\textbf{0.51}} & \subds{\textbf{0.66}} & \subds{\textbf{0.67}} & \subds{\textbf{0.82}} & \subds{\textbf{0.95}} & \subds{\underline{0.51}} & \subds{\underline{0.68}} & \subds{0.72} & \subds{0.88} & \subds{1.16} & \\[\bucketgapRespOptimalNN]
 & \multicolumn{5}{c}{\subds{\bucketrow{100.00}{0.00}{0.00}{0.00}}} & \multicolumn{5}{c}{\subds{\bucketrow{100.00}{0.00}{0.00}{0.00}}} & \\
\multicolumn{1}{r}{\subds{CAT08}} & \subds{\underline{0.77}} & \subds{\textbf{1.12}} & \subds{\textbf{1.14}} & \subds{\underline{1.49}} & \subds{2.05} & \subds{\underline{0.90}} & \subds{\underline{0.95}} & \subds{1.07} & \subds{\textbf{1.12}} & \subds{\underline{1.58}} & \\[\bucketgapRespOptimalNN]
 & \multicolumn{5}{c}{\subds{\bucketrow{50.00}{25.00}{25.00}{0.00}}} & \multicolumn{5}{c}{\subds{\bucketrow{75.00}{25.00}{0.00}{0.00}}} & \\
\multicolumn{1}{r}{\subds{ImageCAS}} & \subds{0.56} & \subds{0.74} & \subds{0.82} & \subds{\textbf{0.98}} & \subds{\underline{2.05}} & \subds{\textbf{0.60}} & \subds{\underline{0.90}} & \subds{1.00} & \subds{1.26} & \subds{2.66} & \\[\bucketgapRespOptimalNN]
 & \multicolumn{5}{c}{\subds{\bucketrow{82.00}{14.00}{4.00}{0.00}}} & \multicolumn{5}{c}{\subds{\bucketrow{80.00}{18.00}{2.00}{0.00}}} & \\
\addlinespace[\blockgapRespOptimalNN]
1.0 & \textbf{0.47} & \textbf{0.65} & \textbf{0.83} & \underline{1.10} & 2.17 & \underline{0.61} & 0.94 & \textbf{0.96} & \textbf{1.19} & 3.44 & 21.3 \\[\bucketgapRespOptimalNN]
 & \multicolumn{5}{c}{\bucketrow{74.14}{22.41}{3.45}{0.00}} & \multicolumn{5}{c}{\bucketrow{89.66}{8.62}{1.72}{0.00}} & \\
\multicolumn{1}{r}{\subds{ASOCA}} & \subds{0.63} & \subds{0.70} & \subds{\underline{0.77}} & \subds{0.84} & \subds{\underline{1.08}} & \subds{\textbf{0.41}} & \subds{\textbf{0.58}} & \subds{\textbf{0.64}} & \subds{\textbf{0.81}} & \subds{\textbf{0.98}} & \\[\bucketgapRespOptimalNN]
 & \multicolumn{5}{c}{\subds{\bucketrow{100.00}{0.00}{0.00}{0.00}}} & \multicolumn{5}{c}{\subds{\bucketrow{100.00}{0.00}{0.00}{0.00}}} & \\
\multicolumn{1}{r}{\subds{CAT08}} & \subds{1.04} & \subds{1.33} & \subds{\underline{1.17}} & \subds{\textbf{1.47}} & \subds{\textbf{1.70}} & \subds{\textbf{0.63}} & \subds{\textbf{0.78}} & \subds{\textbf{0.99}} & \subds{\underline{1.14}} & \subds{1.82} & \\[\bucketgapRespOptimalNN]
 & \multicolumn{5}{c}{\subds{\bucketrow{25.00}{75.00}{0.00}{0.00}}} & \multicolumn{5}{c}{\subds{\bucketrow{75.00}{25.00}{0.00}{0.00}}} & \\
\multicolumn{1}{r}{\subds{ImageCAS}} & \subds{\textbf{0.45}} & \subds{\textbf{0.64}} & \subds{\underline{0.81}} & \subds{\underline{1.04}} & \subds{2.17} & \subds{\underline{0.61}} & \subds{0.97} & \subds{\underline{0.98}} & \subds{\textbf{1.22}} & \subds{3.44} & \\[\bucketgapRespOptimalNN]
 & \multicolumn{5}{c}{\subds{\bucketrow{76.00}{20.00}{4.00}{0.00}}} & \multicolumn{5}{c}{\subds{\bucketrow{90.00}{8.00}{2.00}{0.00}}} & \\
\addlinespace[\blockgapRespOptimalNN]
1.5 & 0.77 & 1.08 & \underline{1.20} & 1.55 & 3.12 & 0.89 & 1.19 & 1.20 & 1.47 & \textbf{2.16} & 18.3 \\[\bucketgapRespOptimalNN]
 & \multicolumn{5}{c}{\bucketrow{50.00}{37.93}{12.07}{0.00}} & \multicolumn{5}{c}{\bucketrow{74.14}{25.86}{0.00}{0.00}} & \\
\multicolumn{1}{r}{\subds{ASOCA}} & \subds{0.74} & \subds{0.86} & \subds{0.96} & \subds{1.08} & \subds{1.44} & \subds{0.54} & \subds{0.82} & \subds{0.79} & \subds{1.08} & \subds{1.17} & \\[\bucketgapRespOptimalNN]
 & \multicolumn{5}{c}{\subds{\bucketrow{75.00}{25.00}{0.00}{0.00}}} & \multicolumn{5}{c}{\subds{\bucketrow{100.00}{0.00}{0.00}{0.00}}} & \\
\multicolumn{1}{r}{\subds{CAT08}} & \subds{1.33} & \subds{1.93} & \subds{1.99} & \subds{2.60} & \subds{3.12} & \subds{1.11} & \subds{1.34} & \subds{1.19} & \subds{1.42} & \subds{1.59} & \\[\bucketgapRespOptimalNN]
 & \multicolumn{5}{c}{\subds{\bucketrow{25.00}{25.00}{50.00}{0.00}}} & \multicolumn{5}{c}{\subds{\bucketrow{75.00}{25.00}{0.00}{0.00}}} & \\
\multicolumn{1}{r}{\subds{ImageCAS}} & \subds{0.77} & \subds{1.08} & \subds{1.16} & \subds{1.55} & \subds{2.75} & \subds{0.91} & \subds{1.23} & \subds{1.23} & \subds{1.50} & \subds{\textbf{2.16}} & \\[\bucketgapRespOptimalNN]
 & \multicolumn{5}{c}{\subds{\bucketrow{50.00}{40.00}{10.00}{0.00}}} & \multicolumn{5}{c}{\subds{\bucketrow{72.00}{28.00}{0.00}{0.00}}} & \\
\bottomrule
\end{tabular}
\end{table}

\paragraph{Interpretation} When the model is trained at the voxel size at which it is used, most of the accuracy is retained. For \swinv2, up to 1.0~mm the median error of the right ostium is identical to that at 0.5~mm, the share of Optimal cases at patient level is the same, and the processing time per patient falls from 55.6 to 21.9~s; the cost is concentrated in the left ostium (median of 1.01~mm against 0.85~mm) and in the tail of the distribution, with the share of Acceptable right ostia rising from 1.7 to 8.6\% and the maximum errors from 2.07 and 2.51~mm to 2.20 and 2.72~mm. The model trained at 0.75~mm was less accurate than the one trained at 1.0~mm, so the accuracy is not monotonic in the spacing; with a single training run per spacing and 58 test volumes we cannot tell whether this reflects the spacing or the variability between training runs, as for the models trained with different numbers of images, but \nnresunet shows no such dip, which favors the latter reading. \nnresunet confirms the trend: no localization fails at any spacing, up to 1.0~mm the accuracy matches that at 0.5~mm while the processing time per patient halves, and only at 1.5~mm, where it is 2.2 times faster than at 0.5~mm, does it degrade more than \swinv2. At 1.5~mm the errors of both models are clearly larger, and the processing time of \swinv2 no longer decreases with respect to 1.0~mm, while that of \nnresunet still falls by 3~s; the part of the processing that does not depend on the size of the volume, such as reading and resampling it, probably becomes dominant, but we did not measure this. We did not train at coarser spacings, as we expect no benefit from them: the time gained beyond 1.0~mm is already small, while the errors keep growing and the voxel, already comparable in size to the Anatomical thresholds (1.08 and 1.45~mm), would become larger than them. Finally, the spacing of 0.5~mm adopted for the benchmark was not necessary to obtain an accuracy of the same order, but the experiment changes, together with the spacing, the extent of the patch (64~mm at 0.5~mm, 128~mm at 1.0~mm) and the width of the target and of the peak window in millimeters, because these are expressed in voxels, so that their effects cannot be separated. The implications of this ablation are summarized in Section~\ref{sec:discussion-respacing-optimal}.

\subsection{Ablation: respacing mismatch}
\label{subsec:app-respacing-mismatch}
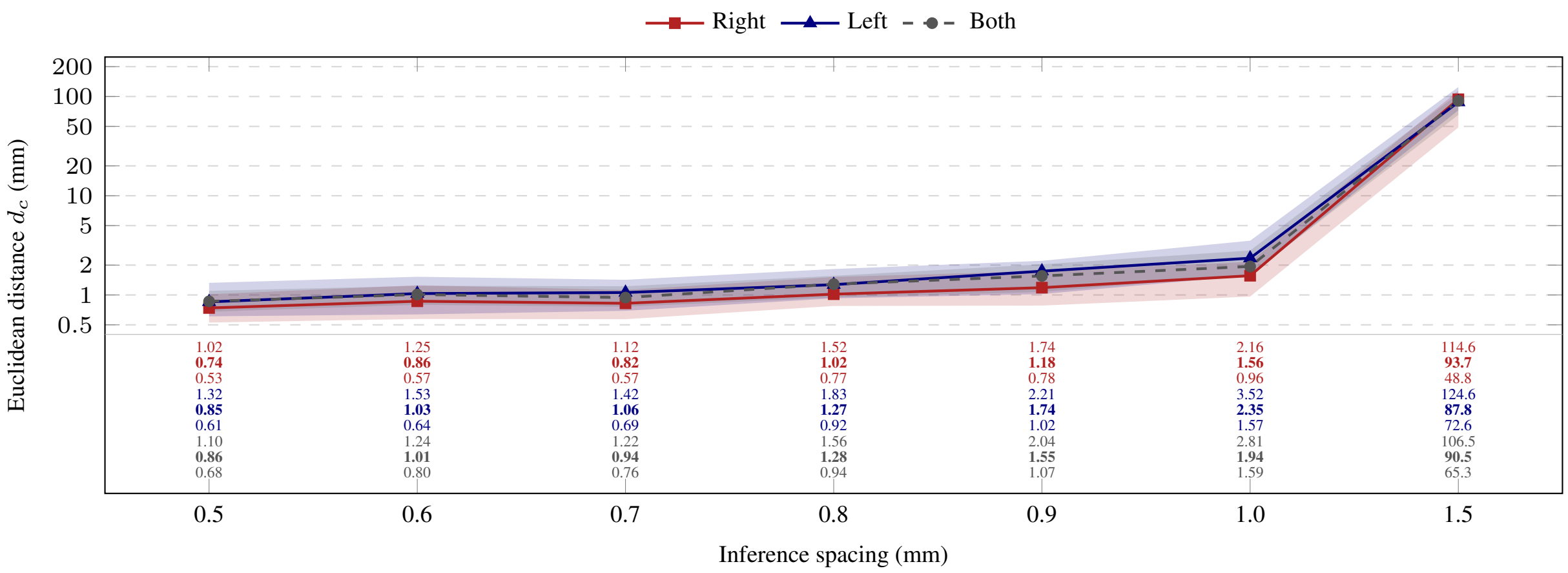
\begin{figure}[H]
\centering
\definecolor{cRight}{HTML}{B22222}
\definecolor{cLeft}{HTML}{000080}
\definecolor{cBoth}{HTML}{555555}
\begin{tikzpicture}
\begin{axis}[
    width=\linewidth, height=6.05cm,
    xlabel={Inference spacing (mm)}, ylabel={Euclidean distance $d_c$ (mm)},
    label style={font=\scriptsize}, tick label style={font=\scriptsize},
    xmin=0.5, xmax=7.5, xtick={1,2,3,4,5,6,7}, xticklabels={0.5,0.6,0.7,0.8,0.9,1.0,1.5},
    ymode=log, ymin=0.010, ymax=250, ytick={0.5,1,2,5,10,20,50,100,200}, log ticks with fixed point,
    ymajorgrids=true, grid style={dashed, gray!30},
    legend style={at={(0.5,1.02)}, anchor=south, legend columns=3, draw=none, font=\scriptsize, /tikz/every even column/.append style={column sep=2pt}},
    legend cell align=left,
]
\addplot[draw=none, fill=cRight, fill opacity=0.18, forget plot] coordinates {(1,1.015844) (2,1.251772) (3,1.123711) (4,1.517658) (5,1.740001) (6,2.156159) (7,114.59209) (7,48.80847) (6,0.963513) (5,0.781972) (4,0.774282) (3,0.569926) (2,0.572156) (1,0.527204) } \closedcycle;
\addplot[draw=none, fill=cLeft,  fill opacity=0.18, forget plot] coordinates {(1,1.324319) (2,1.525439) (3,1.42237) (4,1.825583) (5,2.208616) (6,3.520708) (7,124.58712) (7,72.554981) (6,1.571378) (5,1.020506) (4,0.924762) (3,0.692268) (2,0.636585) (1,0.608716) } \closedcycle;
\addplot[draw=none, fill=cBoth,  fill opacity=0.18, forget plot] coordinates {(1,1.09951) (2,1.235025) (3,1.223934) (4,1.558188) (5,2.035602) (6,2.806845) (7,106.489649) (7,65.261259) (6,1.594973) (5,1.072261) (4,0.942872) (3,0.761243) (2,0.796377) (1,0.677562) } \closedcycle;
\addplot[color=cRight, mark=square*, mark size=1.3pt, mark options={solid}, thick] coordinates {(1,0.739533) (2,0.864434) (3,0.823831) (4,1.018633) (5,1.184793) (6,1.563368) (7,93.747971) }; \addlegendentry{Right}
\addplot[color=cLeft,  mark=triangle*, mark size=1.6pt, mark options={solid}, thick] coordinates {(1,0.852039) (2,1.032951) (3,1.058042) (4,1.269914) (5,1.737773) (6,2.348891) (7,87.826485) }; \addlegendentry{Left}
\addplot[color=cBoth,  mark=*, mark size=1.3pt, mark options={solid}, thick, dashed] coordinates {(1,0.862462) (2,1.01026) (3,0.940862) (4,1.28457) (5,1.554769) (6,1.937831) (7,90.484669) }; \addlegendentry{Both}
\draw[gray!60, very thin] (axis cs:0.5,0.4) -- (axis cs:7.5,0.4);
\tikzset{nn/.style={font=\fontsize{4.5}{5}\selectfont, inner sep=0pt}}
\node[nn, text=cRight] at (axis cs:1,0.30000) {1.02};
\node[nn, text=cRight] at (axis cs:1,0.20851) {\textbf{0.74}};
\node[nn, text=cRight] at (axis cs:1,0.14492) {0.53};
\node[nn, text=cLeft] at (axis cs:1,0.10072) {1.32};
\node[nn, text=cLeft] at (axis cs:1,0.07000) {\textbf{0.85}};
\node[nn, text=cLeft] at (axis cs:1,0.04865) {0.61};
\node[nn, text=cBoth] at (axis cs:1,0.03382) {1.10};
\node[nn, text=cBoth] at (axis cs:1,0.02350) {\textbf{0.86}};
\node[nn, text=cBoth] at (axis cs:1,0.01634) {0.68};
\node[nn, text=cRight] at (axis cs:2,0.30000) {1.25};
\node[nn, text=cRight] at (axis cs:2,0.20851) {\textbf{0.86}};
\node[nn, text=cRight] at (axis cs:2,0.14492) {0.57};
\node[nn, text=cLeft] at (axis cs:2,0.10072) {1.53};
\node[nn, text=cLeft] at (axis cs:2,0.07000) {\textbf{1.03}};
\node[nn, text=cLeft] at (axis cs:2,0.04865) {0.64};
\node[nn, text=cBoth] at (axis cs:2,0.03382) {1.24};
\node[nn, text=cBoth] at (axis cs:2,0.02350) {\textbf{1.01}};
\node[nn, text=cBoth] at (axis cs:2,0.01634) {0.80};
\node[nn, text=cRight] at (axis cs:3,0.30000) {1.12};
\node[nn, text=cRight] at (axis cs:3,0.20851) {\textbf{0.82}};
\node[nn, text=cRight] at (axis cs:3,0.14492) {0.57};
\node[nn, text=cLeft] at (axis cs:3,0.10072) {1.42};
\node[nn, text=cLeft] at (axis cs:3,0.07000) {\textbf{1.06}};
\node[nn, text=cLeft] at (axis cs:3,0.04865) {0.69};
\node[nn, text=cBoth] at (axis cs:3,0.03382) {1.22};
\node[nn, text=cBoth] at (axis cs:3,0.02350) {\textbf{0.94}};
\node[nn, text=cBoth] at (axis cs:3,0.01634) {0.76};
\node[nn, text=cRight] at (axis cs:4,0.30000) {1.52};
\node[nn, text=cRight] at (axis cs:4,0.20851) {\textbf{1.02}};
\node[nn, text=cRight] at (axis cs:4,0.14492) {0.77};
\node[nn, text=cLeft] at (axis cs:4,0.10072) {1.83};
\node[nn, text=cLeft] at (axis cs:4,0.07000) {\textbf{1.27}};
\node[nn, text=cLeft] at (axis cs:4,0.04865) {0.92};
\node[nn, text=cBoth] at (axis cs:4,0.03382) {1.56};
\node[nn, text=cBoth] at (axis cs:4,0.02350) {\textbf{1.28}};
\node[nn, text=cBoth] at (axis cs:4,0.01634) {0.94};
\node[nn, text=cRight] at (axis cs:5,0.30000) {1.74};
\node[nn, text=cRight] at (axis cs:5,0.20851) {\textbf{1.18}};
\node[nn, text=cRight] at (axis cs:5,0.14492) {0.78};
\node[nn, text=cLeft] at (axis cs:5,0.10072) {2.21};
\node[nn, text=cLeft] at (axis cs:5,0.07000) {\textbf{1.74}};
\node[nn, text=cLeft] at (axis cs:5,0.04865) {1.02};
\node[nn, text=cBoth] at (axis cs:5,0.03382) {2.04};
\node[nn, text=cBoth] at (axis cs:5,0.02350) {\textbf{1.55}};
\node[nn, text=cBoth] at (axis cs:5,0.01634) {1.07};
\node[nn, text=cRight] at (axis cs:6,0.30000) {2.16};
\node[nn, text=cRight] at (axis cs:6,0.20851) {\textbf{1.56}};
\node[nn, text=cRight] at (axis cs:6,0.14492) {0.96};
\node[nn, text=cLeft] at (axis cs:6,0.10072) {3.52};
\node[nn, text=cLeft] at (axis cs:6,0.07000) {\textbf{2.35}};
\node[nn, text=cLeft] at (axis cs:6,0.04865) {1.57};
\node[nn, text=cBoth] at (axis cs:6,0.03382) {2.81};
\node[nn, text=cBoth] at (axis cs:6,0.02350) {\textbf{1.94}};
\node[nn, text=cBoth] at (axis cs:6,0.01634) {1.59};
\node[nn, text=cRight] at (axis cs:7,0.30000) {114.6};
\node[nn, text=cRight] at (axis cs:7,0.20851) {\textbf{93.7}};
\node[nn, text=cRight] at (axis cs:7,0.14492) {48.8};
\node[nn, text=cLeft] at (axis cs:7,0.10072) {124.6};
\node[nn, text=cLeft] at (axis cs:7,0.07000) {\textbf{87.8}};
\node[nn, text=cLeft] at (axis cs:7,0.04865) {72.6};
\node[nn, text=cBoth] at (axis cs:7,0.03382) {106.5};
\node[nn, text=cBoth] at (axis cs:7,0.02350) {\textbf{90.5}};
\node[nn, text=cBoth] at (axis cs:7,0.01634) {65.3};
\end{axis}
\end{tikzpicture}
\caption{Euclidean distance $d_c$ (mm) of \swinv2 trained at 0.5~mm vs. inference spacing (equidistant axis, logarithmic distance). Lines: medians; bands: p25--p75 over the whole test set. Numbers at the bottom: p75, median (bold) and p25 of the right (red), left (blue) and both (gray) ostia. See Section~\ref{subsec:evaluation-metrics}.}
\label{fig:app-respacing-plot}
\end{figure}

We evaluated the impact of a mismatch between the training and the inference spacing. \swinv2, trained at 0.5~mm isotropic spacing, was tested on volumes resampled at isotropic spacings from 0.6 to 1.5~mm; the results are reported in Table~\ref{tab:app-respacing-mismatch} and Figure~\ref{fig:app-respacing-plot}.

Up to 0.6~mm, the median errors increase only slightly (0.86~mm for the right and 1.03~mm for the left ostium at 0.6~mm, against 0.74~mm and 0.85~mm at 0.5~mm) and no localization fails. From 0.7~mm onwards, failed localizations appear, first on the left ostium (1.7\% at 0.7~mm, 3.4\% at 0.8~mm, 6.9\% at 0.9~mm and 22.4\% at 1.0~mm) and, from 0.9~mm, also on the right ostium (1.7\% at 0.9~mm and 3.4\% at 1.0~mm). As a consequence, the mean errors become progressively larger than the medians; at 1.0~mm, for instance, the left ostium has a mean error of 19.5~mm and a median error of 2.35~mm. At 1.5~mm, all the ostia fail. The accuracy also degrades before failures appear: the share of cases that are Optimal at patient level falls from 69.0\% at 0.5~mm to 50.0, 51.7, 31.0, 25.9, 8.6 and 0\% at 0.6, 0.7, 0.8, 0.9, 1.0 and 1.5~mm, respectively. In the meantime, the processing time per patient decreases from 55.6~s at 0.5~mm to 44.5, 39.5, 35.3, 33.4, 32.1 and 28.2~s at 0.6, 0.7, 0.8, 0.9, 1.0 and 1.5~mm, respectively.

\paragraph{Interpretation}
The mismatch has a much larger effect than the spacing itself. At 1.0~mm, the model trained at 0.5~mm fails on 22.4\% of the left and 3.4\% of the right ostia, whereas the model trained at 1.0~mm does not fail at all (\ref{subsec:app-respacing-optimal}): the information needed to locate the ostia is still present in the coarser volume, and the failures are caused by the mismatch rather than by the loss of resolution. Resampling to a coarser grid without retraining changes the apparent size of the anatomy in voxels: at 1.0~mm every structure is half as large as during training and the $128^3$ patch covers twice the field of view, and at 1.5~mm a third as large. The features learned by the network at a fixed scale are not invariant to such a change, and the degradation grows with it: at 0.6~mm, where the structures appear about 17\% smaller, no localization fails yet, but the share of Optimal cases already drops. The left ostium fails first. It already has the larger errors at the training spacing and the less defined interface with the aorta, reflected in the larger variability between annotators (Table~\ref{tab:annotator-stats}), so it possibly has less margin before the heatmap peak moves to a wrong structure. The gain in time does not compensate for the loss, since the model trained at 1.0~mm is faster (21.9 against 32.1~s per patient) and never fails. A correct voxel spacing in the image header is therefore a requirement of the pipeline. Only coarser inference spacings were tested, on a single model; a mismatch toward finer spacings, which would enlarge the anatomy instead, was not evaluated.
\begin{table}[H]
\centering
\fontsize{6.3}{8}\selectfont
\setlength{\tabcolsep}{1.5pt}
\newcommand{\bucketgapRespMismatch}{-2pt}
\newcommand{\blockgapRespMismatch}{5pt}
\caption{Euclidean distance $d_c$ (mm) of \swinv2 trained at 0.5~mm and tested at the isotropic spacing (mm) reported in the first column. The line below each row gives the percentage of cases in each accuracy bucket for the right and left ostium: O = Optimal, G = Good, A = Acceptable, F = Failed. See Section~\ref{subsec:evaluation-metrics}. In each distance column, the best value is in bold and the second best is underlined, comparing the rows of the same kind (overall and, separately, each dataset).}
\label{tab:app-respacing-mismatch}
\begin{tabular}{l *{10}{r} r}
\toprule
 & \multicolumn{5}{c}{Right ostium} & \multicolumn{5}{c}{Left ostium} & \\
\cmidrule(lr){2-6} \cmidrule(lr){7-11}
\makecell[l]{Inference\\spacing (mm)} & p25 & Med. & Mean & p75 & Max & p25 & Med. & Mean & p75 & Max & \makecell{Time\\(s)} \\
\midrule
0.5 & \textbf{0.53} & \textbf{0.74} & \textbf{0.83} & \textbf{1.02} & \underline{2.07} & \textbf{0.61} & \textbf{0.85} & \textbf{0.98} & \textbf{1.32} & \textbf{2.51} & 55.6 \\[\bucketgapRespMismatch]
 & \multicolumn{5}{c}{\bucketrow{77.59}{20.69}{1.72}{0.00}} & \multicolumn{5}{c}{\bucketrow{84.48}{15.52}{0.00}{0.00}} & \\
\multicolumn{1}{r}{\subds{ASOCA}} & \subds{\textbf{0.53}} & \subds{0.95} & \subds{0.95} & \subds{1.37} & \subds{\underline{1.45}} & \subds{\textbf{0.54}} & \subds{\textbf{0.60}} & \subds{\textbf{0.58}} & \subds{\textbf{0.64}} & \subds{\underline{0.71}} & \\[\bucketgapRespMismatch]
 & \multicolumn{5}{c}{\subds{\bucketrow{50.00}{50.00}{0.00}{0.00}}} & \multicolumn{5}{c}{\subds{\bucketrow{100.00}{0.00}{0.00}{0.00}}} & \\
\multicolumn{1}{r}{\subds{CAT08}} & \subds{0.60} & \subds{\underline{1.02}} & \subds{\underline{1.04}} & \subds{1.46} & \subds{\underline{1.75}} & \subds{\textbf{0.59}} & \subds{\underline{1.04}} & \subds{1.20} & \subds{1.64} & \subds{2.25} & \\[\bucketgapRespMismatch]
 & \multicolumn{5}{c}{\subds{\bucketrow{50.00}{50.00}{0.00}{0.00}}} & \multicolumn{5}{c}{\subds{\bucketrow{75.00}{25.00}{0.00}{0.00}}} & \\
\multicolumn{1}{r}{\subds{ImageCAS}} & \subds{\textbf{0.54}} & \subds{\textbf{0.74}} & \subds{\textbf{0.81}} & \subds{\textbf{0.98}} & \subds{\underline{2.07}} & \subds{\textbf{0.62}} & \subds{\textbf{0.89}} & \subds{\textbf{1.00}} & \subds{\textbf{1.32}} & \subds{\textbf{2.51}} & \\[\bucketgapRespMismatch]
 & \multicolumn{5}{c}{\subds{\bucketrow{82.00}{16.00}{2.00}{0.00}}} & \multicolumn{5}{c}{\subds{\bucketrow{84.00}{16.00}{0.00}{0.00}}} & \\
\addlinespace[\blockgapRespMismatch]
0.6 & \underline{0.57} & 0.86 & 0.95 & 1.25 & 2.09 & \underline{0.64} & \underline{1.03} & \underline{1.13} & 1.53 & \underline{3.18} & 44.5 \\[\bucketgapRespMismatch]
 & \multicolumn{5}{c}{\bucketrow{67.24}{24.14}{8.62}{0.00}} & \multicolumn{5}{c}{\bucketrow{72.41}{25.86}{1.72}{0.00}} & \\
\multicolumn{1}{r}{\subds{ASOCA}} & \subds{0.67} & \subds{\underline{0.90}} & \subds{\underline{0.90}} & \subds{\textbf{1.12}} & \subds{1.49} & \subds{\underline{0.57}} & \subds{\underline{0.62}} & \subds{\underline{0.59}} & \subds{\textbf{0.64}} & \subds{\textbf{0.64}} & \\[\bucketgapRespMismatch]
 & \multicolumn{5}{c}{\subds{\bucketrow{75.00}{25.00}{0.00}{0.00}}} & \multicolumn{5}{c}{\subds{\bucketrow{100.00}{0.00}{0.00}{0.00}}} & \\
\multicolumn{1}{r}{\subds{CAT08}} & \subds{0.81} & \subds{1.11} & \subds{1.13} & \subds{\underline{1.43}} & \subds{1.99} & \subds{\underline{0.84}} & \subds{\underline{1.04}} & \subds{\textbf{1.00}} & \subds{\textbf{1.20}} & \subds{\textbf{1.28}} & \\[\bucketgapRespMismatch]
 & \multicolumn{5}{c}{\subds{\bucketrow{50.00}{25.00}{25.00}{0.00}}} & \multicolumn{5}{c}{\subds{\bucketrow{100.00}{0.00}{0.00}{0.00}}} & \\
\multicolumn{1}{r}{\subds{ImageCAS}} & \subds{0.57} & \subds{0.86} & \subds{0.93} & \subds{1.23} & \subds{2.09} & \subds{0.70} & \subds{1.08} & \subds{\underline{1.19}} & \subds{1.62} & \subds{\underline{3.18}} & \\[\bucketgapRespMismatch]
 & \multicolumn{5}{c}{\subds{\bucketrow{68.00}{24.00}{8.00}{0.00}}} & \multicolumn{5}{c}{\subds{\bucketrow{68.00}{30.00}{2.00}{0.00}}} & \\
\addlinespace[\blockgapRespMismatch]
0.7 & \underline{0.57} & \underline{0.82} & \underline{0.85} & \underline{1.12} & \textbf{2.05} & 0.69 & 1.06 & 2.04 & \underline{1.42} & 54.2 & 39.5 \\[\bucketgapRespMismatch]
 & \multicolumn{5}{c}{\bucketrow{72.41}{25.86}{1.72}{0.00}} & \multicolumn{5}{c}{\bucketrow{75.86}{20.69}{1.72}{1.72}} & \\
\multicolumn{1}{r}{\subds{ASOCA}} & \subds{\underline{0.61}} & \subds{\textbf{0.88}} & \subds{\textbf{0.88}} & \subds{\underline{1.14}} & \subds{\textbf{1.18}} & \subds{0.76} & \subds{0.97} & \subds{1.28} & \subds{1.49} & \subds{2.53} & \\[\bucketgapRespMismatch]
 & \multicolumn{5}{c}{\subds{\bucketrow{50.00}{50.00}{0.00}{0.00}}} & \multicolumn{5}{c}{\subds{\bucketrow{75.00}{25.00}{0.00}{0.00}}} & \\
\multicolumn{1}{r}{\subds{CAT08}} & \subds{\underline{0.53}} & \subds{\textbf{0.99}} & \subds{\textbf{1.01}} & \subds{1.47} & \subds{\textbf{1.67}} & \subds{1.11} & \subds{1.19} & \subds{\underline{1.15}} & \subds{1.23} & \subds{\textbf{1.28}} & \\[\bucketgapRespMismatch]
 & \multicolumn{5}{c}{\subds{\bucketrow{50.00}{50.00}{0.00}{0.00}}} & \multicolumn{5}{c}{\subds{\bucketrow{100.00}{0.00}{0.00}{0.00}}} & \\
\multicolumn{1}{r}{\subds{ImageCAS}} & \subds{\underline{0.56}} & \subds{\underline{0.82}} & \subds{\underline{0.83}} & \subds{\underline{1.05}} & \subds{\textbf{2.05}} & \subds{\underline{0.66}} & \subds{\underline{0.98}} & \subds{2.17} & \subds{\underline{1.45}} & \subds{54.2} & \\[\bucketgapRespMismatch]
 & \multicolumn{5}{c}{\subds{\bucketrow{76.00}{22.00}{2.00}{0.00}}} & \multicolumn{5}{c}{\subds{\bucketrow{74.00}{22.00}{2.00}{2.00}}} & \\
\addlinespace[\blockgapRespMismatch]
0.8 & 0.77 & 1.02 & 1.21 & 1.52 & 3.34 & 0.92 & 1.27 & 3.01 & 1.83 & 55.0 & 35.3 \\[\bucketgapRespMismatch]
 & \multicolumn{5}{c}{\bucketrow{51.72}{34.48}{13.79}{0.00}} & \multicolumn{5}{c}{\bucketrow{60.34}{32.76}{3.45}{3.45}} & \\
\multicolumn{1}{r}{\subds{ASOCA}} & \subds{1.33} & \subds{1.72} & \subds{1.76} & \subds{2.14} & \subds{2.68} & \subds{1.15} & \subds{1.23} & \subds{1.24} & \subds{\underline{1.32}} & \subds{1.49} & \\[\bucketgapRespMismatch]
 & \multicolumn{5}{c}{\subds{\bucketrow{25.00}{25.00}{50.00}{0.00}}} & \multicolumn{5}{c}{\subds{\bucketrow{75.00}{25.00}{0.00}{0.00}}} & \\
\multicolumn{1}{r}{\subds{CAT08}} & \subds{0.76} & \subds{1.04} & \subds{1.15} & \subds{\textbf{1.42}} & \subds{2.03} & \subds{0.87} & \subds{\textbf{0.94}} & \subds{1.16} & \subds{\underline{1.22}} & \subds{\underline{1.96}} & \\[\bucketgapRespMismatch]
 & \multicolumn{5}{c}{\subds{\bucketrow{50.00}{25.00}{25.00}{0.00}}} & \multicolumn{5}{c}{\subds{\bucketrow{75.00}{25.00}{0.00}{0.00}}} & \\
\multicolumn{1}{r}{\subds{ImageCAS}} & \subds{0.77} & \subds{1.01} & \subds{1.17} & \subds{1.50} & \subds{3.34} & \subds{0.93} & \subds{1.33} & \subds{3.30} & \subds{1.89} & \subds{55.0} & \\[\bucketgapRespMismatch]
 & \multicolumn{5}{c}{\subds{\bucketrow{54.00}{36.00}{10.00}{0.00}}} & \multicolumn{5}{c}{\subds{\bucketrow{58.00}{34.00}{4.00}{4.00}}} & \\
\addlinespace[\blockgapRespMismatch]
0.9 & 0.78 & 1.18 & 3.56 & 1.74 & 132.1 & 1.02 & 1.74 & 4.21 & 2.21 & 55.2 & 33.4 \\[\bucketgapRespMismatch]
 & \multicolumn{5}{c}{\bucketrow{46.55}{31.03}{20.69}{1.72}} & \multicolumn{5}{c}{\bucketrow{43.10}{39.66}{10.34}{6.90}} & \\
\multicolumn{1}{r}{\subds{ASOCA}} & \subds{0.90} & \subds{1.46} & \subds{1.51} & \subds{2.07} & \subds{2.29} & \subds{1.34} & \subds{1.47} & \subds{1.49} & \subds{1.62} & \subds{1.83} & \\[\bucketgapRespMismatch]
 & \multicolumn{5}{c}{\subds{\bucketrow{50.00}{0.00}{50.00}{0.00}}} & \multicolumn{5}{c}{\subds{\bucketrow{50.00}{50.00}{0.00}{0.00}}} & \\
\multicolumn{1}{r}{\subds{CAT08}} & \subds{\textbf{0.41}} & \subds{1.04} & \subds{1.19} & \subds{1.82} & \subds{2.45} & \subds{1.06} & \subds{1.56} & \subds{1.55} & \subds{2.05} & \subds{2.43} & \\[\bucketgapRespMismatch]
 & \multicolumn{5}{c}{\subds{\bucketrow{50.00}{25.00}{25.00}{0.00}}} & \multicolumn{5}{c}{\subds{\bucketrow{50.00}{50.00}{0.00}{0.00}}} & \\
\multicolumn{1}{r}{\subds{ImageCAS}} & \subds{0.78} & \subds{1.18} & \subds{3.92} & \subds{1.73} & \subds{132.1} & \subds{0.99} & \subds{1.80} & \subds{4.64} & \subds{2.42} & \subds{55.2} & \\[\bucketgapRespMismatch]
 & \multicolumn{5}{c}{\subds{\bucketrow{46.00}{34.00}{18.00}{2.00}}} & \multicolumn{5}{c}{\subds{\bucketrow{42.00}{38.00}{12.00}{8.00}}} & \\
\addlinespace[\blockgapRespMismatch]
1.0 & 0.96 & 1.56 & 4.52 & 2.16 & 131.6 & 1.57 & 2.35 & 19.5 & 3.52 & 160.6 & 32.1 \\[\bucketgapRespMismatch]
 & \multicolumn{5}{c}{\bucketrow{29.31}{29.31}{37.93}{3.45}} & \multicolumn{5}{c}{\bucketrow{22.41}{36.21}{18.97}{22.41}} & \\
\multicolumn{1}{r}{\subds{ASOCA}} & \subds{0.67} & \subds{1.06} & \subds{1.15} & \subds{1.53} & \subds{2.14} & \subds{1.72} & \subds{1.87} & \subds{2.19} & \subds{2.34} & \subds{3.72} & \\[\bucketgapRespMismatch]
 & \multicolumn{5}{c}{\subds{\bucketrow{50.00}{25.00}{25.00}{0.00}}} & \multicolumn{5}{c}{\subds{\bucketrow{25.00}{50.00}{25.00}{0.00}}} & \\
\multicolumn{1}{r}{\subds{CAT08}} & \subds{0.97} & \subds{1.20} & \subds{1.25} & \subds{1.48} & \subds{2.27} & \subds{0.86} & \subds{1.50} & \subds{32.5} & \subds{33.1} & \subds{126.2} & \\[\bucketgapRespMismatch]
 & \multicolumn{5}{c}{\subds{\bucketrow{25.00}{50.00}{25.00}{0.00}}} & \multicolumn{5}{c}{\subds{\bucketrow{50.00}{25.00}{0.00}{25.00}}} & \\
\multicolumn{1}{r}{\subds{ImageCAS}} & \subds{1.01} & \subds{1.58} & \subds{5.05} & \subds{2.18} & \subds{131.6} & \subds{1.60} & \subds{2.41} & \subds{19.9} & \subds{3.52} & \subds{160.6} & \\[\bucketgapRespMismatch]
 & \multicolumn{5}{c}{\subds{\bucketrow{28.00}{28.00}{40.00}{4.00}}} & \multicolumn{5}{c}{\subds{\bucketrow{20.00}{36.00}{20.00}{24.00}}} & \\
\addlinespace[\blockgapRespMismatch]
1.5 & 48.8 & 93.7 & 84.6 & 114.6 & 153.9 & 72.6 & 87.8 & 94.2 & 124.6 & 155.4 & 28.2 \\[\bucketgapRespMismatch]
 & \multicolumn{5}{c}{\bucketrow{0.00}{0.00}{0.00}{100.00}} & \multicolumn{5}{c}{\bucketrow{0.00}{0.00}{0.00}{100.00}} & \\
\multicolumn{1}{r}{\subds{ASOCA}} & \subds{41.3} & \subds{69.4} & \subds{72.3} & \subds{100.3} & \subds{112.9} & \subds{73.9} & \subds{78.5} & \subds{79.6} & \subds{84.3} & \subds{88.0} & \\[\bucketgapRespMismatch]
 & \multicolumn{5}{c}{\subds{\bucketrow{0.00}{0.00}{0.00}{100.00}}} & \multicolumn{5}{c}{\subds{\bucketrow{0.00}{0.00}{0.00}{100.00}}} & \\
\multicolumn{1}{r}{\subds{CAT08}} & \subds{99.9} & \subds{102.9} & \subds{104.8} & \subds{107.8} & \subds{114.3} & \subds{63.4} & \subds{74.3} & \subds{67.7} & \subds{78.6} & \subds{89.3} & \\[\bucketgapRespMismatch]
 & \multicolumn{5}{c}{\subds{\bucketrow{0.00}{0.00}{0.00}{100.00}}} & \multicolumn{5}{c}{\subds{\bucketrow{0.00}{0.00}{0.00}{100.00}}} & \\
\multicolumn{1}{r}{\subds{ImageCAS}} & \subds{48.8} & \subds{87.8} & \subds{84.0} & \subds{117.7} & \subds{153.9} & \subds{71.7} & \subds{94.8} & \subds{97.5} & \subds{128.7} & \subds{155.4} & \\[\bucketgapRespMismatch]
 & \multicolumn{5}{c}{\subds{\bucketrow{0.00}{0.00}{0.00}{100.00}}} & \multicolumn{5}{c}{\subds{\bucketrow{0.00}{0.00}{0.00}{100.00}}} & \\
\bottomrule
\end{tabular}
\end{table}

\subsection{Ablation: peak window size}
\label{subsec:app-window}
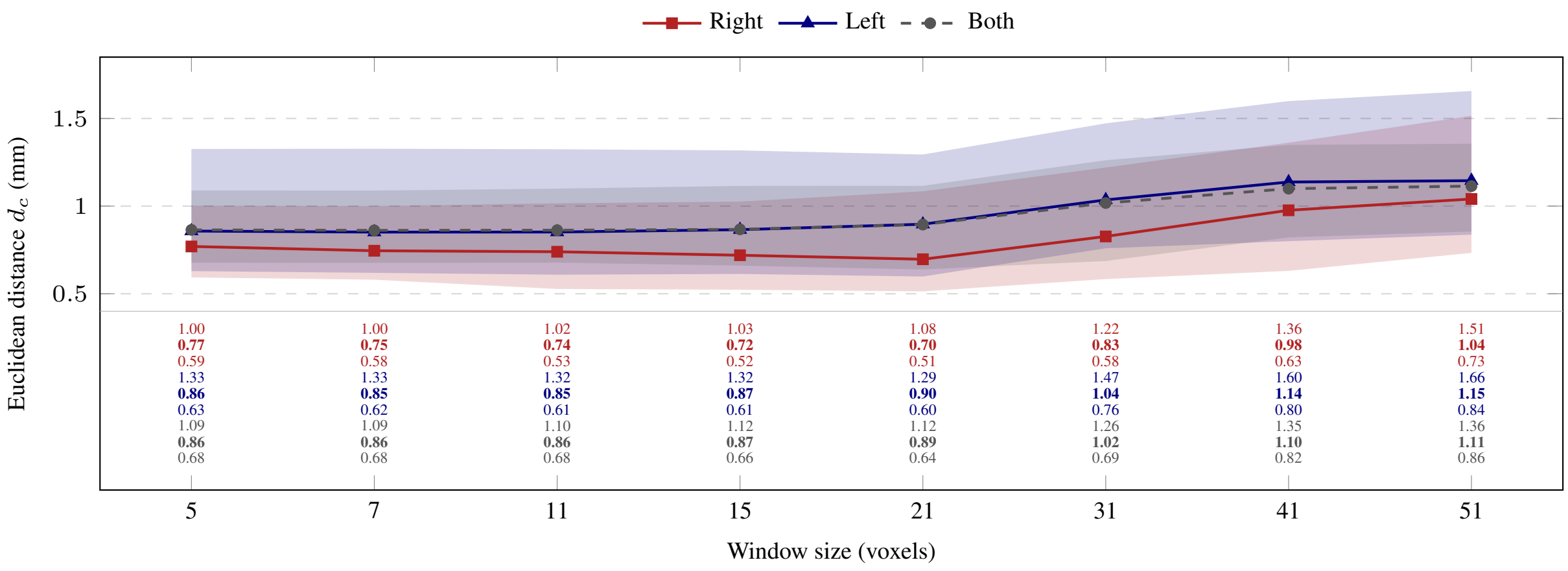
\begin{figure}[H]
\centering
\definecolor{cRight}{HTML}{B22222}
\definecolor{cLeft}{HTML}{000080}
\definecolor{cBoth}{HTML}{555555}
\begin{tikzpicture}
\begin{axis}[
    width=\linewidth, height=6.0cm,
    xlabel={Window size (voxels)}, ylabel={Euclidean distance $d_c$ (mm)},
    label style={font=\scriptsize}, tick label style={font=\scriptsize},
    xmin=0.5, xmax=8.5, xtick={1,2,3,4,5,6,7,8}, xticklabels={5,7,11,15,21,31,41,51},
    ymin=-0.62, ymax=1.85, ytick={0.5,1.0,1.5},
    ymajorgrids=true, grid style={dashed, gray!30},
    legend style={at={(0.5,1.02)}, anchor=south, legend columns=3, draw=none, font=\scriptsize, /tikz/every even column/.append style={column sep=2pt}},
    legend cell align=left,
]
\addplot[draw=none, fill=cRight, fill opacity=0.18, forget plot] coordinates {(1,1.002795) (2,1.00048) (3,1.015844) (4,1.025632) (5,1.084256) (6,1.22) (7,1.361565) (8,1.514907) (8,0.732903) (7,0.630477) (6,0.583564) (5,0.513609) (4,0.522763) (3,0.527204) (2,0.580174) (1,0.593087) } \closedcycle;
\addplot[draw=none, fill=cLeft,  fill opacity=0.18, forget plot] coordinates {(1,1.325792) (2,1.327176) (3,1.324319) (4,1.317685) (5,1.294161) (6,1.472009) (7,1.599015) (8,1.656509) (8,0.836758) (7,0.799781) (6,0.759296) (5,0.598274) (4,0.612925) (3,0.608716) (2,0.619276) (1,0.628975) } \closedcycle;
\addplot[draw=none, fill=cBoth,  fill opacity=0.18, forget plot] coordinates {(1,1.088942) (2,1.088585) (3,1.09951) (4,1.115111) (5,1.115531) (6,1.261592) (7,1.348887) (8,1.355842) (8,0.855309) (7,0.821701) (6,0.686529) (5,0.638355) (4,0.660538) (3,0.677562) (2,0.67628) (1,0.677229) } \closedcycle;
\addplot[color=cRight, mark=square*, mark size=1.3pt, mark options={solid}, thick] coordinates {(1,0.769764) (2,0.745283) (3,0.739533) (4,0.719616) (5,0.696726) (6,0.826865) (7,0.976073) (8,1.040532) }; \addlegendentry{Right}
\addplot[color=cLeft,  mark=triangle*, mark size=1.6pt, mark options={solid}, thick] coordinates {(1,0.857619) (2,0.85166) (3,0.852039) (4,0.865499) (5,0.896589) (6,1.035324) (7,1.137367) (8,1.145047) }; \addlegendentry{Left}
\addplot[color=cBoth,  mark=*, mark size=1.3pt, mark options={solid}, thick, dashed] coordinates {(1,0.864356) (2,0.861571) (3,0.862462) (4,0.867452) (5,0.89457) (6,1.017607) (7,1.099374) (8,1.114047) }; \addlegendentry{Both}
\draw[gray!60, very thin] (axis cs:0.5,0.4) -- (axis cs:8.5,0.4);
\tikzset{nn/.style={font=\fontsize{4.5}{5}\selectfont, inner sep=0pt}}
\node[nn, text=cRight] at (axis cs:1,0.3000) {1.00};
\node[nn, text=cRight] at (axis cs:1,0.2080) {\textbf{0.77}};
\node[nn, text=cRight] at (axis cs:1,0.1160) {0.59};
\node[nn, text=cLeft] at (axis cs:1,0.0240) {1.33};
\node[nn, text=cLeft] at (axis cs:1,-0.0680) {\textbf{0.86}};
\node[nn, text=cLeft] at (axis cs:1,-0.1600) {0.63};
\node[nn, text=cBoth] at (axis cs:1,-0.2520) {1.09};
\node[nn, text=cBoth] at (axis cs:1,-0.3440) {\textbf{0.86}};
\node[nn, text=cBoth] at (axis cs:1,-0.4360) {0.68};
\node[nn, text=cRight] at (axis cs:2,0.3000) {1.00};
\node[nn, text=cRight] at (axis cs:2,0.2080) {\textbf{0.75}};
\node[nn, text=cRight] at (axis cs:2,0.1160) {0.58};
\node[nn, text=cLeft] at (axis cs:2,0.0240) {1.33};
\node[nn, text=cLeft] at (axis cs:2,-0.0680) {\textbf{0.85}};
\node[nn, text=cLeft] at (axis cs:2,-0.1600) {0.62};
\node[nn, text=cBoth] at (axis cs:2,-0.2520) {1.09};
\node[nn, text=cBoth] at (axis cs:2,-0.3440) {\textbf{0.86}};
\node[nn, text=cBoth] at (axis cs:2,-0.4360) {0.68};
\node[nn, text=cRight] at (axis cs:3,0.3000) {1.02};
\node[nn, text=cRight] at (axis cs:3,0.2080) {\textbf{0.74}};
\node[nn, text=cRight] at (axis cs:3,0.1160) {0.53};
\node[nn, text=cLeft] at (axis cs:3,0.0240) {1.32};
\node[nn, text=cLeft] at (axis cs:3,-0.0680) {\textbf{0.85}};
\node[nn, text=cLeft] at (axis cs:3,-0.1600) {0.61};
\node[nn, text=cBoth] at (axis cs:3,-0.2520) {1.10};
\node[nn, text=cBoth] at (axis cs:3,-0.3440) {\textbf{0.86}};
\node[nn, text=cBoth] at (axis cs:3,-0.4360) {0.68};
\node[nn, text=cRight] at (axis cs:4,0.3000) {1.03};
\node[nn, text=cRight] at (axis cs:4,0.2080) {\textbf{0.72}};
\node[nn, text=cRight] at (axis cs:4,0.1160) {0.52};
\node[nn, text=cLeft] at (axis cs:4,0.0240) {1.32};
\node[nn, text=cLeft] at (axis cs:4,-0.0680) {\textbf{0.87}};
\node[nn, text=cLeft] at (axis cs:4,-0.1600) {0.61};
\node[nn, text=cBoth] at (axis cs:4,-0.2520) {1.12};
\node[nn, text=cBoth] at (axis cs:4,-0.3440) {\textbf{0.87}};
\node[nn, text=cBoth] at (axis cs:4,-0.4360) {0.66};
\node[nn, text=cRight] at (axis cs:5,0.3000) {1.08};
\node[nn, text=cRight] at (axis cs:5,0.2080) {\textbf{0.70}};
\node[nn, text=cRight] at (axis cs:5,0.1160) {0.51};
\node[nn, text=cLeft] at (axis cs:5,0.0240) {1.29};
\node[nn, text=cLeft] at (axis cs:5,-0.0680) {\textbf{0.90}};
\node[nn, text=cLeft] at (axis cs:5,-0.1600) {0.60};
\node[nn, text=cBoth] at (axis cs:5,-0.2520) {1.12};
\node[nn, text=cBoth] at (axis cs:5,-0.3440) {\textbf{0.89}};
\node[nn, text=cBoth] at (axis cs:5,-0.4360) {0.64};
\node[nn, text=cRight] at (axis cs:6,0.3000) {1.22};
\node[nn, text=cRight] at (axis cs:6,0.2080) {\textbf{0.83}};
\node[nn, text=cRight] at (axis cs:6,0.1160) {0.58};
\node[nn, text=cLeft] at (axis cs:6,0.0240) {1.47};
\node[nn, text=cLeft] at (axis cs:6,-0.0680) {\textbf{1.04}};
\node[nn, text=cLeft] at (axis cs:6,-0.1600) {0.76};
\node[nn, text=cBoth] at (axis cs:6,-0.2520) {1.26};
\node[nn, text=cBoth] at (axis cs:6,-0.3440) {\textbf{1.02}};
\node[nn, text=cBoth] at (axis cs:6,-0.4360) {0.69};
\node[nn, text=cRight] at (axis cs:7,0.3000) {1.36};
\node[nn, text=cRight] at (axis cs:7,0.2080) {\textbf{0.98}};
\node[nn, text=cRight] at (axis cs:7,0.1160) {0.63};
\node[nn, text=cLeft] at (axis cs:7,0.0240) {1.60};
\node[nn, text=cLeft] at (axis cs:7,-0.0680) {\textbf{1.14}};
\node[nn, text=cLeft] at (axis cs:7,-0.1600) {0.80};
\node[nn, text=cBoth] at (axis cs:7,-0.2520) {1.35};
\node[nn, text=cBoth] at (axis cs:7,-0.3440) {\textbf{1.10}};
\node[nn, text=cBoth] at (axis cs:7,-0.4360) {0.82};
\node[nn, text=cRight] at (axis cs:8,0.3000) {1.51};
\node[nn, text=cRight] at (axis cs:8,0.2080) {\textbf{1.04}};
\node[nn, text=cRight] at (axis cs:8,0.1160) {0.73};
\node[nn, text=cLeft] at (axis cs:8,0.0240) {1.66};
\node[nn, text=cLeft] at (axis cs:8,-0.0680) {\textbf{1.15}};
\node[nn, text=cLeft] at (axis cs:8,-0.1600) {0.84};
\node[nn, text=cBoth] at (axis cs:8,-0.2520) {1.36};
\node[nn, text=cBoth] at (axis cs:8,-0.3440) {\textbf{1.11}};
\node[nn, text=cBoth] at (axis cs:8,-0.4360) {0.86};
\end{axis}
\end{tikzpicture}
\caption{Euclidean distance $d_c$ (mm) of \swinv2 vs. centroid window size (equidistant axis; default 11). Lines: medians; bands: p25--p75 over the whole test set. Numbers at the bottom: p75, median (bold) and p25 of the right (red), left (blue) and both (gray) ostia. See Section~\ref{subsec:evaluation-metrics}.}
\label{fig:app-window-plot}
\end{figure}

The final ostium coordinates are computed as a probability-weighted centroid within a cubic window centered on the heatmap peak (see Section~\ref{subsec:postprocessing}). We varied the side of this window from 5 to 51 voxels (2.5 to 25.5~mm at 0.5~mm isotropic spacing), keeping the trained model fixed; the default value, used in all the other experiments, is 11 voxels. The results are reported in Table~\ref{tab:app-window} and Figure~\ref{fig:app-window-plot}.

No localization failed with any of the tested window sizes, and the processing time was unaffected (between 55.6 and 56.2~s per patient). For windows of up to 21 voxels, the errors were essentially stable: the mean error ranged between 0.82 and 0.84~mm for the right ostium and between 0.98 and 1.01~mm for the left ostium, and the median error between 0.70 and 0.77~mm and between 0.85 and 0.90~mm, respectively. Accordingly, the share of Optimal cases at patient level was 62.1--70.7\% for these windows, against 69.0\% for the default one. Larger windows resulted in progressively larger errors: with 31, 41 and 51 voxels, the mean error was 0.88, 1.01 and 1.10~mm for the right ostium, and 1.16, 1.27 and 1.30~mm for the left ostium, and the share of Optimal cases at patient level dropped to 51.7, 39.7 and 36.2\%, mostly moving to the Good bucket (46.6, 55.2 and 56.9\%).

\paragraph{Interpretation}
The differences among windows of up to 21 voxels (10.5~mm) are negligible for a test set of this size, so the default of 11 voxels, fixed before this experiment, lies in a flat region and is not a finely tuned parameter. Since the window is centered on the maximum of the heatmap, its size decides only how the coordinates are refined around a peak that has already been found, not which peak is found: it moves the estimate by at most half the side of the window, and no tested size caused a failure. Larger windows probably degrade the estimate because they include probability mass that does not belong to the peak, which pulls the weighted centroid away from it. The time per patient is unaffected, since the centroid is computed on a small region.
\begin{table}[H]
\centering
\scriptsize
\setlength{\tabcolsep}{1.5pt}
\newcommand{\bucketgapWindow}{-2pt}
\newcommand{\blockgapWindow}{5pt}
\caption{Euclidean distance $d_c$ (mm) of \swinv2 for different sizes of the centroid window (voxels; default 11). The line below each row gives the percentage of cases in each accuracy bucket for the right and left ostium: O = Optimal, G = Good, A = Acceptable, F = Failed. See Section~\ref{subsec:evaluation-metrics}. In each distance column, the best value is in bold and the second best is underlined, comparing the rows of the same kind (overall and, separately, each dataset).}
\label{tab:app-window}
\begin{tabular}{l *{10}{r} r}
\toprule
 & \multicolumn{5}{c}{Right ostium} & \multicolumn{5}{c}{Left ostium} & \\
\cmidrule(lr){2-6} \cmidrule(lr){7-11}
\makecell[l]{Window\\(vox)} & p25 & Med. & Mean & p75 & Max & p25 & Med. & Mean & p75 & Max & \makecell{Time\\(s)} \\
\midrule
5 & 0.59 & 0.77 & 0.84 & \textbf{1.00} & 2.04 & 0.63 & \underline{0.86} & \textbf{0.98} & 1.33 & 2.54 & 55.8 \\[\bucketgapWindow]
 & \multicolumn{5}{c}{\bucketrow{77.59}{20.69}{1.72}{0.00}} & \multicolumn{5}{c}{\bucketrow{84.48}{15.52}{0.00}{0.00}} & \\
\multicolumn{1}{r}{\subds{ASOCA}} & \subds{0.56} & \subds{0.94} & \subds{0.95} & \subds{1.33} & \subds{1.43} & \subds{0.53} & \subds{\underline{0.60}} & \subds{\textbf{0.58}} & \subds{\underline{0.65}} & \subds{\textbf{0.70}} & \\[\bucketgapWindow]
 & \multicolumn{5}{c}{\subds{\bucketrow{50.00}{50.00}{0.00}{0.00}}} & \multicolumn{5}{c}{\subds{\bucketrow{100.00}{0.00}{0.00}{0.00}}} & \\
\multicolumn{1}{r}{\subds{CAT08}} & \subds{0.70} & \subds{1.08} & \subds{1.07} & \subds{\underline{1.45}} & \subds{1.74} & \subds{\textbf{0.59}} & \subds{\textbf{1.00}} & \subds{\underline{1.19}} & \subds{\textbf{1.60}} & \subds{2.29} & \\[\bucketgapWindow]
 & \multicolumn{5}{c}{\subds{\bucketrow{50.00}{50.00}{0.00}{0.00}}} & \multicolumn{5}{c}{\subds{\bucketrow{75.00}{25.00}{0.00}{0.00}}} & \\
\multicolumn{1}{r}{\subds{ImageCAS}} & \subds{0.62} & \subds{0.75} & \subds{0.81} & \subds{\underline{0.96}} & \subds{2.04} & \subds{0.64} & \subds{\underline{0.89}} & \subds{\textbf{1.00}} & \subds{1.33} & \subds{2.54} & \\[\bucketgapWindow]
 & \multicolumn{5}{c}{\subds{\bucketrow{82.00}{16.00}{2.00}{0.00}}} & \multicolumn{5}{c}{\subds{\bucketrow{84.00}{16.00}{0.00}{0.00}}} & \\
\addlinespace[\blockgapWindow]
7 & 0.58 & 0.75 & 0.84 & \textbf{1.00} & 2.06 & 0.62 & \textbf{0.85} & \textbf{0.98} & 1.33 & 2.53 & 56.1 \\[\bucketgapWindow]
 & \multicolumn{5}{c}{\bucketrow{79.31}{18.97}{1.72}{0.00}} & \multicolumn{5}{c}{\bucketrow{84.48}{15.52}{0.00}{0.00}} & \\
\multicolumn{1}{r}{\subds{ASOCA}} & \subds{0.54} & \subds{0.93} & \subds{0.94} & \subds{1.33} & \subds{1.44} & \subds{0.55} & \subds{\underline{0.60}} & \subds{\underline{0.59}} & \subds{\textbf{0.64}} & \subds{\underline{0.71}} & \\[\bucketgapWindow]
 & \multicolumn{5}{c}{\subds{\bucketrow{50.00}{50.00}{0.00}{0.00}}} & \multicolumn{5}{c}{\subds{\bucketrow{100.00}{0.00}{0.00}{0.00}}} & \\
\multicolumn{1}{r}{\subds{CAT08}} & \subds{0.64} & \subds{1.04} & \subds{1.06} & \subds{\underline{1.45}} & \subds{1.75} & \subds{\textbf{0.59}} & \subds{\underline{1.01}} & \subds{\underline{1.19}} & \subds{\underline{1.61}} & \subds{2.27} & \\[\bucketgapWindow]
 & \multicolumn{5}{c}{\subds{\bucketrow{50.00}{50.00}{0.00}{0.00}}} & \multicolumn{5}{c}{\subds{\bucketrow{75.00}{25.00}{0.00}{0.00}}} & \\
\multicolumn{1}{r}{\subds{ImageCAS}} & \subds{0.61} & \subds{0.75} & \subds{0.81} & \subds{\textbf{0.95}} & \subds{2.06} & \subds{0.63} & \subds{\textbf{0.88}} & \subds{\textbf{1.00}} & \subds{1.33} & \subds{2.53} & \\[\bucketgapWindow]
 & \multicolumn{5}{c}{\subds{\bucketrow{84.00}{14.00}{2.00}{0.00}}} & \multicolumn{5}{c}{\subds{\bucketrow{84.00}{16.00}{0.00}{0.00}}} & \\
\addlinespace[\blockgapWindow]
11 & 0.53 & 0.74 & \underline{0.83} & \underline{1.02} & 2.07 & \underline{0.61} & \textbf{0.85} & \textbf{0.98} & \underline{1.32} & \textbf{2.51} & 55.6 \\[\bucketgapWindow]
 & \multicolumn{5}{c}{\bucketrow{77.59}{20.69}{1.72}{0.00}} & \multicolumn{5}{c}{\bucketrow{84.48}{15.52}{0.00}{0.00}} & \\
\multicolumn{1}{r}{\subds{ASOCA}} & \subds{0.53} & \subds{0.95} & \subds{0.95} & \subds{1.37} & \subds{1.45} & \subds{0.54} & \subds{\underline{0.60}} & \subds{\textbf{0.58}} & \subds{\textbf{0.64}} & \subds{\underline{0.71}} & \\[\bucketgapWindow]
 & \multicolumn{5}{c}{\subds{\bucketrow{50.00}{50.00}{0.00}{0.00}}} & \multicolumn{5}{c}{\subds{\bucketrow{100.00}{0.00}{0.00}{0.00}}} & \\
\multicolumn{1}{r}{\subds{CAT08}} & \subds{\underline{0.60}} & \subds{\underline{1.02}} & \subds{\underline{1.04}} & \subds{1.46} & \subds{1.75} & \subds{\textbf{0.59}} & \subds{1.04} & \subds{1.20} & \subds{1.64} & \subds{2.25} & \\[\bucketgapWindow]
 & \multicolumn{5}{c}{\subds{\bucketrow{50.00}{50.00}{0.00}{0.00}}} & \multicolumn{5}{c}{\subds{\bucketrow{75.00}{25.00}{0.00}{0.00}}} & \\
\multicolumn{1}{r}{\subds{ImageCAS}} & \subds{0.54} & \subds{0.74} & \subds{0.81} & \subds{0.98} & \subds{2.07} & \subds{\underline{0.62}} & \subds{\underline{0.89}} & \subds{\textbf{1.00}} & \subds{\underline{1.32}} & \subds{\textbf{2.51}} & \\[\bucketgapWindow]
 & \multicolumn{5}{c}{\subds{\bucketrow{82.00}{16.00}{2.00}{0.00}}} & \multicolumn{5}{c}{\subds{\bucketrow{84.00}{16.00}{0.00}{0.00}}} & \\
\addlinespace[\blockgapWindow]
15 & \underline{0.52} & \underline{0.72} & \underline{0.83} & 1.03 & 2.08 & \underline{0.61} & 0.87 & \underline{0.99} & \underline{1.32} & \underline{2.52} & 56.1 \\[\bucketgapWindow]
 & \multicolumn{5}{c}{\bucketrow{77.59}{20.69}{1.72}{0.00}} & \multicolumn{5}{c}{\bucketrow{82.76}{17.24}{0.00}{0.00}} & \\
\multicolumn{1}{r}{\subds{ASOCA}} & \subds{\underline{0.52}} & \subds{0.93} & \subds{0.93} & \subds{1.35} & \subds{1.42} & \subds{\textbf{0.49}} & \subds{\textbf{0.58}} & \subds{\underline{0.59}} & \subds{0.68} & \subds{0.73} & \\[\bucketgapWindow]
 & \multicolumn{5}{c}{\subds{\bucketrow{50.00}{50.00}{0.00}{0.00}}} & \multicolumn{5}{c}{\subds{\bucketrow{100.00}{0.00}{0.00}{0.00}}} & \\
\multicolumn{1}{r}{\subds{CAT08}} & \subds{\underline{0.60}} & \subds{\underline{1.02}} & \subds{\underline{1.04}} & \subds{1.46} & \subds{1.75} & \subds{\textbf{0.59}} & \subds{1.04} & \subds{1.20} & \subds{1.66} & \subds{2.24} & \\[\bucketgapWindow]
 & \multicolumn{5}{c}{\subds{\bucketrow{50.00}{50.00}{0.00}{0.00}}} & \multicolumn{5}{c}{\subds{\bucketrow{50.00}{50.00}{0.00}{0.00}}} & \\
\multicolumn{1}{r}{\subds{ImageCAS}} & \subds{\underline{0.53}} & \subds{\underline{0.72}} & \subds{\underline{0.80}} & \subds{1.00} & \subds{2.08} & \subds{\underline{0.62}} & \subds{\underline{0.89}} & \subds{\textbf{1.00}} & \subds{\underline{1.32}} & \subds{\underline{2.52}} & \\[\bucketgapWindow]
 & \multicolumn{5}{c}{\subds{\bucketrow{82.00}{16.00}{2.00}{0.00}}} & \multicolumn{5}{c}{\subds{\bucketrow{84.00}{16.00}{0.00}{0.00}}} & \\
\addlinespace[\blockgapWindow]
21 & \textbf{0.51} & \textbf{0.70} & \textbf{0.82} & 1.08 & 2.06 & \textbf{0.60} & 0.90 & 1.01 & \textbf{1.29} & \underline{2.52} & 56.2 \\[\bucketgapWindow]
 & \multicolumn{5}{c}{\bucketrow{74.14}{24.14}{1.72}{0.00}} & \multicolumn{5}{c}{\bucketrow{81.03}{18.97}{0.00}{0.00}} & \\
\multicolumn{1}{r}{\subds{ASOCA}} & \subds{\textbf{0.51}} & \subds{\textbf{0.88}} & \subds{\textbf{0.87}} & \subds{\textbf{1.25}} & \subds{\textbf{1.28}} & \subds{\underline{0.52}} & \subds{0.62} & \subds{0.60} & \subds{0.69} & \subds{0.72} & \\[\bucketgapWindow]
 & \multicolumn{5}{c}{\subds{\bucketrow{50.00}{50.00}{0.00}{0.00}}} & \multicolumn{5}{c}{\subds{\bucketrow{100.00}{0.00}{0.00}{0.00}}} & \\
\multicolumn{1}{r}{\subds{CAT08}} & \subds{\textbf{0.55}} & \subds{\textbf{0.98}} & \subds{\textbf{1.02}} & \subds{\underline{1.45}} & \subds{1.73} & \subds{\underline{0.61}} & \subds{1.08} & \subds{\underline{1.19}} & \subds{1.66} & \subds{\underline{2.10}} & \\[\bucketgapWindow]
 & \multicolumn{5}{c}{\subds{\bucketrow{50.00}{50.00}{0.00}{0.00}}} & \multicolumn{5}{c}{\subds{\bucketrow{50.00}{50.00}{0.00}{0.00}}} & \\
\multicolumn{1}{r}{\subds{ImageCAS}} & \subds{\textbf{0.51}} & \subds{\textbf{0.70}} & \subds{\textbf{0.79}} & \subds{1.03} & \subds{2.06} & \subds{\textbf{0.61}} & \subds{1.01} & \subds{\underline{1.03}} & \subds{\textbf{1.29}} & \subds{\underline{2.52}} & \\[\bucketgapWindow]
 & \multicolumn{5}{c}{\subds{\bucketrow{78.00}{20.00}{2.00}{0.00}}} & \multicolumn{5}{c}{\subds{\bucketrow{82.00}{18.00}{0.00}{0.00}}} & \\
\addlinespace[\blockgapWindow]
31 & 0.58 & 0.83 & 0.88 & 1.22 & \textbf{1.81} & 0.76 & 1.04 & 1.16 & 1.47 & 2.61 & 56.2 \\[\bucketgapWindow]
 & \multicolumn{5}{c}{\bucketrow{67.24}{32.76}{0.00}{0.00}} & \multicolumn{5}{c}{\bucketrow{74.14}{24.14}{1.72}{0.00}} & \\
\multicolumn{1}{r}{\subds{ASOCA}} & \subds{0.55} & \subds{\underline{0.92}} & \subds{\underline{0.90}} & \subds{\underline{1.27}} & \subds{\underline{1.33}} & \subds{0.77} & \subds{0.83} & \subds{0.82} & \subds{0.88} & \subds{1.01} & \\[\bucketgapWindow]
 & \multicolumn{5}{c}{\subds{\bucketrow{50.00}{50.00}{0.00}{0.00}}} & \multicolumn{5}{c}{\subds{\bucketrow{100.00}{0.00}{0.00}{0.00}}} & \\
\multicolumn{1}{r}{\subds{CAT08}} & \subds{0.76} & \subds{1.05} & \subds{1.11} & \subds{\textbf{1.39}} & \subds{\underline{1.68}} & \subds{0.73} & \subds{1.19} & \subds{\textbf{1.18}} & \subds{1.64} & \subds{\textbf{1.87}} & \\[\bucketgapWindow]
 & \multicolumn{5}{c}{\subds{\bucketrow{50.00}{50.00}{0.00}{0.00}}} & \multicolumn{5}{c}{\subds{\bucketrow{50.00}{50.00}{0.00}{0.00}}} & \\
\multicolumn{1}{r}{\subds{ImageCAS}} & \subds{0.58} & \subds{0.83} & \subds{0.86} & \subds{1.16} & \subds{\textbf{1.81}} & \subds{0.76} & \subds{1.06} & \subds{1.18} & \subds{1.47} & \subds{2.61} & \\[\bucketgapWindow]
 & \multicolumn{5}{c}{\subds{\bucketrow{70.00}{30.00}{0.00}{0.00}}} & \multicolumn{5}{c}{\subds{\bucketrow{74.00}{24.00}{2.00}{0.00}}} & \\
\addlinespace[\blockgapWindow]
41 & 0.63 & 0.98 & 1.01 & 1.36 & \underline{1.97} & 0.80 & 1.14 & 1.27 & 1.60 & 2.80 & 56.2 \\[\bucketgapWindow]
 & \multicolumn{5}{c}{\bucketrow{55.17}{43.10}{1.72}{0.00}} & \multicolumn{5}{c}{\bucketrow{68.97}{27.59}{3.45}{0.00}} & \\
\multicolumn{1}{r}{\subds{ASOCA}} & \subds{0.68} & \subds{1.04} & \subds{1.03} & \subds{1.39} & \subds{1.55} & \subds{0.90} & \subds{1.08} & \subds{0.98} & \subds{1.16} & \subds{1.19} & \\[\bucketgapWindow]
 & \multicolumn{5}{c}{\subds{\bucketrow{50.00}{50.00}{0.00}{0.00}}} & \multicolumn{5}{c}{\subds{\bucketrow{100.00}{0.00}{0.00}{0.00}}} & \\
\multicolumn{1}{r}{\subds{CAT08}} & \subds{1.14} & \subds{1.34} & \subds{1.31} & \subds{1.51} & \subds{\textbf{1.64}} & \subds{0.85} & \subds{1.31} & \subds{1.29} & \subds{1.75} & \subds{2.13} & \\[\bucketgapWindow]
 & \multicolumn{5}{c}{\subds{\bucketrow{25.00}{75.00}{0.00}{0.00}}} & \multicolumn{5}{c}{\subds{\bucketrow{50.00}{50.00}{0.00}{0.00}}} & \\
\multicolumn{1}{r}{\subds{ImageCAS}} & \subds{0.61} & \subds{0.93} & \subds{0.98} & \subds{1.35} & \subds{\underline{1.97}} & \subds{0.80} & \subds{1.14} & \subds{1.29} & \subds{1.60} & \subds{2.80} & \\[\bucketgapWindow]
 & \multicolumn{5}{c}{\subds{\bucketrow{58.00}{40.00}{2.00}{0.00}}} & \multicolumn{5}{c}{\subds{\bucketrow{68.00}{28.00}{4.00}{0.00}}} & \\
\addlinespace[\blockgapWindow]
51 & 0.73 & 1.04 & 1.10 & 1.51 & 2.26 & 0.84 & 1.15 & 1.30 & 1.66 & 2.84 & 55.7 \\[\bucketgapWindow]
 & \multicolumn{5}{c}{\bucketrow{51.72}{41.38}{6.90}{0.00}} & \multicolumn{5}{c}{\bucketrow{68.97}{29.31}{1.72}{0.00}} & \\
\multicolumn{1}{r}{\subds{ASOCA}} & \subds{0.87} & \subds{1.11} & \subds{1.11} & \subds{1.35} & \subds{1.82} & \subds{0.69} & \subds{0.87} & \subds{0.84} & \subds{1.02} & \subds{1.08} & \\[\bucketgapWindow]
 & \multicolumn{5}{c}{\subds{\bucketrow{50.00}{50.00}{0.00}{0.00}}} & \multicolumn{5}{c}{\subds{\bucketrow{100.00}{0.00}{0.00}{0.00}}} & \\
\multicolumn{1}{r}{\subds{CAT08}} & \subds{1.11} & \subds{1.39} & \subds{1.35} & \subds{1.62} & \subds{\underline{1.68}} & \subds{1.00} & \subds{1.39} & \subds{1.46} & \subds{1.84} & \subds{2.52} & \\[\bucketgapWindow]
 & \multicolumn{5}{c}{\subds{\bucketrow{25.00}{75.00}{0.00}{0.00}}} & \multicolumn{5}{c}{\subds{\bucketrow{50.00}{50.00}{0.00}{0.00}}} & \\
\multicolumn{1}{r}{\subds{ImageCAS}} & \subds{0.72} & \subds{1.01} & \subds{1.08} & \subds{1.48} & \subds{2.26} & \subds{0.88} & \subds{1.18} & \subds{1.32} & \subds{1.71} & \subds{2.84} & \\[\bucketgapWindow]
 & \multicolumn{5}{c}{\subds{\bucketrow{54.00}{38.00}{8.00}{0.00}}} & \multicolumn{5}{c}{\subds{\bucketrow{68.00}{30.00}{2.00}{0.00}}} & \\
\bottomrule
\end{tabular}
\end{table}

\subsection{Ablation: sliding-window overlap}
\label{subsec:app-overlap}
\begin{figure}[H]
\centering
\definecolor{cRight}{HTML}{B22222}
\definecolor{cLeft}{HTML}{000080}
\definecolor{cBoth}{HTML}{555555}
\begin{tikzpicture}
\begin{axis}[
    width=\linewidth, height=6.0cm,
    xlabel={Overlap (voxels)}, ylabel={Euclidean distance $d_c$ (mm)},
    label style={font=\scriptsize}, tick label style={font=\scriptsize},
    xmin=0.5, xmax=9.5, xtick={1,2,3,4,6,7,8,9}, xticklabels={0,32,64,96,0,32,64,96},
    ymin=-0.62, ymax=1.85, ytick={0.5,1.0,1.5},
    ymajorgrids=true, grid style={dashed, gray!30},
    legend style={at={(0.5,1.02)}, anchor=south, legend columns=3, draw=none, font=\scriptsize, /tikz/every even column/.append style={column sep=2pt}},
    legend cell align=left,
]
\addplot[draw=none, fill=cRight, fill opacity=0.18, forget plot] coordinates {(1,1.09477) (2,1.06637) (3,1.01584) (4,0.973186) (4,0.508197) (3,0.527204) (2,0.516919) (1,0.519586)} \closedcycle;
\addplot[draw=none, fill=cRight, fill opacity=0.18, forget plot] coordinates {(6,1.09054) (7,1.06182) (8,1.13372) (9,0.968661) (9,0.503187) (8,0.516688) (7,0.569257) (6,0.517183)} \closedcycle;
\addplot[draw=none, fill=cLeft, fill opacity=0.18, forget plot] coordinates {(1,1.46026) (2,1.34988) (3,1.32432) (4,1.356) (4,0.61349) (3,0.608716) (2,0.538732) (1,0.668868)} \closedcycle;
\addplot[draw=none, fill=cLeft, fill opacity=0.18, forget plot] coordinates {(6,1.43179) (7,1.18414) (8,1.20769) (9,1.21301) (9,0.664359) (8,0.67582) (7,0.646091) (6,0.701761)} \closedcycle;
\addplot[draw=none, fill=cBoth, fill opacity=0.18, forget plot] coordinates {(1,1.23032) (2,1.12091) (3,1.09951) (4,1.08333) (4,0.643949) (3,0.677562) (2,0.672452) (1,0.728623)} \closedcycle;
\addplot[draw=none, fill=cBoth, fill opacity=0.18, forget plot] coordinates {(6,1.14053) (7,1.01036) (8,1.01369) (9,1.02613) (9,0.61579) (8,0.684554) (7,0.685243) (6,0.664589)} \closedcycle;
\addplot[color=cRight, mark=square*, mark size=1.3pt, mark options={solid}, thick] coordinates {(1,0.766014) (2,0.673507) (3,0.739533) (4,0.656554)}; \addlegendentry{Right}
\addplot[color=cRight, mark=square*, mark size=1.3pt, mark options={solid}, thick, forget plot] coordinates {(6,0.757378) (7,0.756918) (8,0.702182) (9,0.705202)};
\addplot[color=cLeft, mark=triangle*, mark size=1.6pt, mark options={solid}, thick] coordinates {(1,0.966845) (2,0.860934) (3,0.852039) (4,0.872673)}; \addlegendentry{Left}
\addplot[color=cLeft, mark=triangle*, mark size=1.6pt, mark options={solid}, thick, forget plot] coordinates {(6,0.931156) (7,0.88166) (8,0.869216) (9,0.875225)};
\addplot[color=cBoth, mark=*, mark size=1.3pt, mark options={solid}, thick, dashed] coordinates {(1,0.895353) (2,0.806059) (3,0.862462) (4,0.815122)}; \addlegendentry{Both}
\addplot[color=cBoth, mark=*, mark size=1.3pt, mark options={solid}, thick, dashed, forget plot] coordinates {(6,0.911887) (7,0.828272) (8,0.80668) (9,0.834964)};
\draw[gray!60, very thin] (axis cs:0.5,0.4) -- (axis cs:9.5,0.4);
\draw[gray!60, very thin, dashed] (axis cs:5,-0.62) -- (axis cs:5,1.85);
\tikzset{nn/.style={font=\fontsize{4.5}{5}\selectfont, inner sep=0pt}}
\node[font=\scriptsize\itshape, anchor=north] at (axis cs:2.5,1.83) {\swinv2};
\node[font=\scriptsize\itshape, anchor=north] at (axis cs:7.5,1.83) {\nnresunet};
\node[nn, text=cRight] at (axis cs:1,0.3000) {1.09};
\node[nn, text=cRight] at (axis cs:1,0.2080) {\textbf{0.77}};
\node[nn, text=cRight] at (axis cs:1,0.1160) {0.52};
\node[nn, text=cLeft] at (axis cs:1,0.0240) {1.46};
\node[nn, text=cLeft] at (axis cs:1,-0.0680) {\textbf{0.97}};
\node[nn, text=cLeft] at (axis cs:1,-0.1600) {0.67};
\node[nn, text=cBoth] at (axis cs:1,-0.2520) {1.23};
\node[nn, text=cBoth] at (axis cs:1,-0.3440) {\textbf{0.90}};
\node[nn, text=cBoth] at (axis cs:1,-0.4360) {0.73};
\node[nn, text=cRight] at (axis cs:2,0.3000) {1.07};
\node[nn, text=cRight] at (axis cs:2,0.2080) {\textbf{0.67}};
\node[nn, text=cRight] at (axis cs:2,0.1160) {0.52};
\node[nn, text=cLeft] at (axis cs:2,0.0240) {1.35};
\node[nn, text=cLeft] at (axis cs:2,-0.0680) {\textbf{0.86}};
\node[nn, text=cLeft] at (axis cs:2,-0.1600) {0.54};
\node[nn, text=cBoth] at (axis cs:2,-0.2520) {1.12};
\node[nn, text=cBoth] at (axis cs:2,-0.3440) {\textbf{0.81}};
\node[nn, text=cBoth] at (axis cs:2,-0.4360) {0.67};
\node[nn, text=cRight] at (axis cs:3,0.3000) {1.02};
\node[nn, text=cRight] at (axis cs:3,0.2080) {\textbf{0.74}};
\node[nn, text=cRight] at (axis cs:3,0.1160) {0.53};
\node[nn, text=cLeft] at (axis cs:3,0.0240) {1.32};
\node[nn, text=cLeft] at (axis cs:3,-0.0680) {\textbf{0.85}};
\node[nn, text=cLeft] at (axis cs:3,-0.1600) {0.61};
\node[nn, text=cBoth] at (axis cs:3,-0.2520) {1.10};
\node[nn, text=cBoth] at (axis cs:3,-0.3440) {\textbf{0.86}};
\node[nn, text=cBoth] at (axis cs:3,-0.4360) {0.68};
\node[nn, text=cRight] at (axis cs:4,0.3000) {0.97};
\node[nn, text=cRight] at (axis cs:4,0.2080) {\textbf{0.66}};
\node[nn, text=cRight] at (axis cs:4,0.1160) {0.51};
\node[nn, text=cLeft] at (axis cs:4,0.0240) {1.36};
\node[nn, text=cLeft] at (axis cs:4,-0.0680) {\textbf{0.87}};
\node[nn, text=cLeft] at (axis cs:4,-0.1600) {0.61};
\node[nn, text=cBoth] at (axis cs:4,-0.2520) {1.08};
\node[nn, text=cBoth] at (axis cs:4,-0.3440) {\textbf{0.82}};
\node[nn, text=cBoth] at (axis cs:4,-0.4360) {0.64};
\node[nn, text=cRight] at (axis cs:6,0.3000) {1.09};
\node[nn, text=cRight] at (axis cs:6,0.2080) {\textbf{0.76}};
\node[nn, text=cRight] at (axis cs:6,0.1160) {0.52};
\node[nn, text=cLeft] at (axis cs:6,0.0240) {1.43};
\node[nn, text=cLeft] at (axis cs:6,-0.0680) {\textbf{0.93}};
\node[nn, text=cLeft] at (axis cs:6,-0.1600) {0.70};
\node[nn, text=cBoth] at (axis cs:6,-0.2520) {1.14};
\node[nn, text=cBoth] at (axis cs:6,-0.3440) {\textbf{0.91}};
\node[nn, text=cBoth] at (axis cs:6,-0.4360) {0.66};
\node[nn, text=cRight] at (axis cs:7,0.3000) {1.06};
\node[nn, text=cRight] at (axis cs:7,0.2080) {\textbf{0.76}};
\node[nn, text=cRight] at (axis cs:7,0.1160) {0.57};
\node[nn, text=cLeft] at (axis cs:7,0.0240) {1.18};
\node[nn, text=cLeft] at (axis cs:7,-0.0680) {\textbf{0.88}};
\node[nn, text=cLeft] at (axis cs:7,-0.1600) {0.65};
\node[nn, text=cBoth] at (axis cs:7,-0.2520) {1.01};
\node[nn, text=cBoth] at (axis cs:7,-0.3440) {\textbf{0.83}};
\node[nn, text=cBoth] at (axis cs:7,-0.4360) {0.69};
\node[nn, text=cRight] at (axis cs:8,0.3000) {1.13};
\node[nn, text=cRight] at (axis cs:8,0.2080) {\textbf{0.70}};
\node[nn, text=cRight] at (axis cs:8,0.1160) {0.52};
\node[nn, text=cLeft] at (axis cs:8,0.0240) {1.21};
\node[nn, text=cLeft] at (axis cs:8,-0.0680) {\textbf{0.87}};
\node[nn, text=cLeft] at (axis cs:8,-0.1600) {0.68};
\node[nn, text=cBoth] at (axis cs:8,-0.2520) {1.01};
\node[nn, text=cBoth] at (axis cs:8,-0.3440) {\textbf{0.81}};
\node[nn, text=cBoth] at (axis cs:8,-0.4360) {0.68};
\node[nn, text=cRight] at (axis cs:9,0.3000) {0.97};
\node[nn, text=cRight] at (axis cs:9,0.2080) {\textbf{0.71}};
\node[nn, text=cRight] at (axis cs:9,0.1160) {0.50};
\node[nn, text=cLeft] at (axis cs:9,0.0240) {1.21};
\node[nn, text=cLeft] at (axis cs:9,-0.0680) {\textbf{0.88}};
\node[nn, text=cLeft] at (axis cs:9,-0.1600) {0.66};
\node[nn, text=cBoth] at (axis cs:9,-0.2520) {1.03};
\node[nn, text=cBoth] at (axis cs:9,-0.3440) {\textbf{0.83}};
\node[nn, text=cBoth] at (axis cs:9,-0.4360) {0.62};
\end{axis}
\end{tikzpicture}
\caption{Euclidean distance $d_c$ (mm) of \swinv2 and \nnresunet vs. sliding-window overlap at inference (equidistant axis; default 64 voxels, $128^3$ patches). Lines: medians; bands: p25--p75 over the whole test set. Numbers at the bottom: p75, median (bold) and p25 of the right (red), left (blue) and both (gray) ostia. See Section~\ref{subsec:evaluation-metrics}.}
\label{fig:app-overlap-plot}
\end{figure}
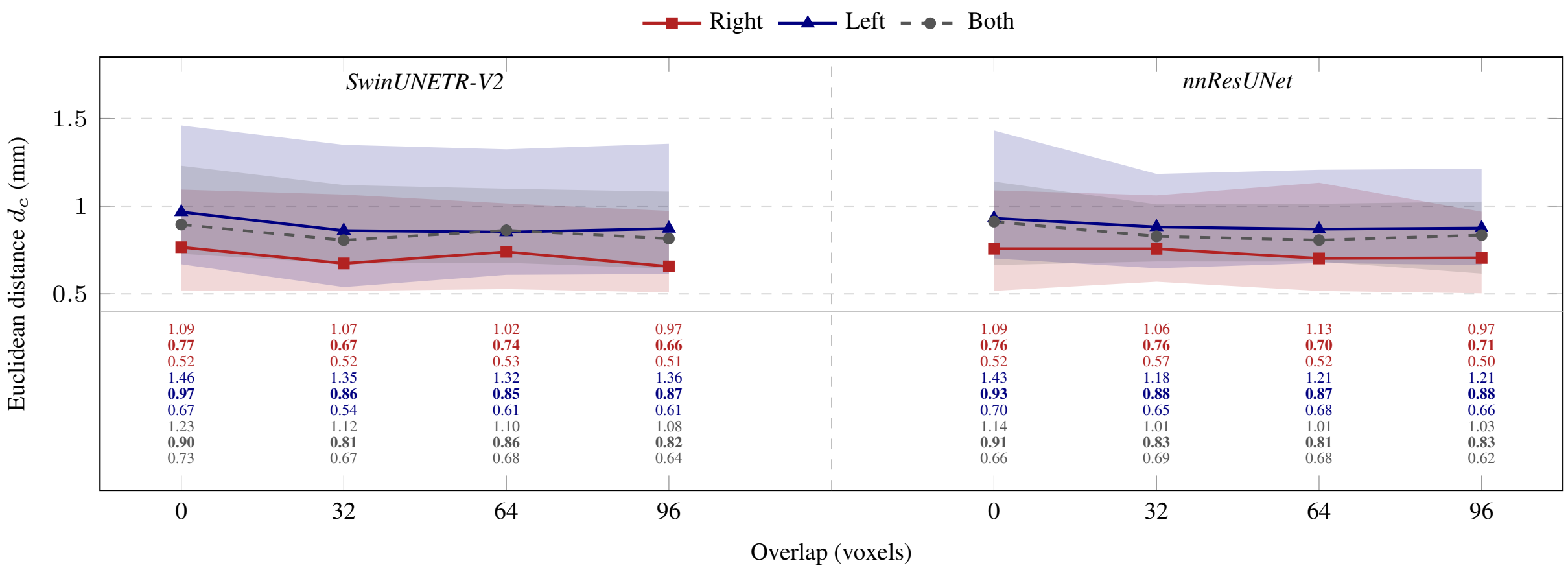

At inference, the volumes are covered by overlapping $128^3$ patches whose logits are averaged where the patches overlap (see Section~\ref{subsec:postprocessing}). We varied the overlap between adjacent patches from 0 to 96 voxels, i.e., from no overlap to 75\%, keeping the trained models fixed; the default, used in all the other experiments, is 64 voxels (50\%), and an overlap of 128 voxels is not possible, as it equals the patch size. The experiment was run on \swinv2 and \nnresunet, and the results are reported in Table~\ref{tab:app-overlap} and Figure~\ref{fig:app-overlap-plot}.

With an overlap of 32 voxels or more the errors were comparable. For \swinv2, the median error ranged between 0.66 and 0.74~mm for the right ostium and between 0.85 and 0.87~mm for the left one, and for \nnresunet between 0.70 and 0.76~mm and between 0.87 and 0.88~mm; no localization failed in these settings. Without overlap the accuracy was lower. For \swinv2, the left ostium had a median error of 0.97~mm (mean 1.12~mm), the share of Optimal cases at patient level dropped from 69.0\% to 50.0\% and that of Acceptable cases rose from 1.7\% to 10.3\%. \nnresunet had one failed right ostium (1.7\% of the cases, with a mean error of 1.11~mm against 0.83~mm) and 62.1\% of Optimal cases at patient level, against 67.2\%. Increasing the overlap to 96 voxels brought no consistent improvement: the share of Optimal cases at patient level was 63.8\% for \swinv2 (against 69.0\% with 64 voxels) and 70.7\% for \nnresunet (against 67.2\%), while the time per patient increased from 55.6 to 119.3~s and from 41.1 to 93.4~s, respectively. Reducing the overlap to 32 voxels shortened the time per patient to 33.3~s for \swinv2 and 31.3~s for \nnresunet, with 62.1\% and 69.0\% of Optimal cases at patient level.

\paragraph{Interpretation}
Without overlap, each voxel is predicted by a single window and nothing is blended: an ostium that falls close to a patch border is located with less context and without any averaging. This is probably what lies behind the larger errors of both models and the single failure of \nnresunet, which point to a loss of robustness rather than to a uniform loss of precision. From 32 voxels upwards the differences show no consistent trend: the best share of Optimal cases at patient level is obtained with 64 voxels by \swinv2 and with 96 voxels by \nnresunet, and the differences, up to 7 percentage points (four volumes), are within the variability expected on this test set. The processing time, instead, grows steeply: with 96 voxels (stride of 32) it is more than twice that with 64 voxels (stride of 64) for both models. Each setting was run once, on two models, and the time includes the loading of the volume and the export of the results, so differences of a few seconds are not meaningful.
\begin{table}[H]
\centering
\scriptsize
\setlength{\tabcolsep}{1.5pt}
\newcommand{\bucketgapOverlap}{-2pt}
\newcommand{\blockgapOverlap}{5pt}
\caption{Euclidean distance $d_c$ (mm) of \swinv2 and \nnresunet for different overlaps between adjacent sliding-window patches at inference (voxels, with $128^3$ patches; default 64). The line below each row gives the percentage of cases in each accuracy bucket for the right and left ostium: O = Optimal, G = Good, A = Acceptable, F = Failed. See Section~\ref{subsec:evaluation-metrics}. In each distance column, the best value is in bold and the second best is underlined, comparing the rows of the same kind (overall and, separately, each dataset).}
\label{tab:app-overlap}
\begin{tabular}{l *{10}{r} r}
\toprule
 & \multicolumn{5}{c}{Right ostium} & \multicolumn{5}{c}{Left ostium} & \\
\cmidrule(lr){2-6} \cmidrule(lr){7-11}
\makecell[l]{Overlap\\(vox)} & p25 & Med. & Mean & p75 & Max & p25 & Med. & Mean & p75 & Max & \makecell{Time\\(s)} \\
\midrule
\multicolumn{12}{l}{\textit{\swinv2}} \\
\midrule
0 & \underline{0.52} & 0.77 & 0.86 & 1.09 & 2.09 & 0.67 & 0.97 & 1.12 & 1.46 & 3.02 & 30.1 \\[\bucketgapOverlap]
 & \multicolumn{5}{c}{\bucketrow{72.41}{20.69}{6.90}{0.00}} & \multicolumn{5}{c}{\bucketrow{74.14}{22.41}{3.45}{0.00}} & \\
\multicolumn{1}{r}{\subds{ASOCA}} & \subds{0.66} & \subds{0.85} & \subds{\textbf{0.81}} & \subds{\underline{1.00}} & \subds{\textbf{1.10}} & \subds{0.57} & \subds{0.64} & \subds{0.61} & \subds{0.68} & \subds{0.75} &  \\[\bucketgapOverlap]
 & \multicolumn{5}{c}{\subds{\bucketrow{75.00}{25.00}{0.00}{0.00}}} & \multicolumn{5}{c}{\subds{\bucketrow{100.00}{0.00}{0.00}{0.00}}} & \\
\multicolumn{1}{r}{\subds{CAT08}} & \subds{\textbf{0.60}} & \subds{\underline{1.03}} & \subds{\underline{1.07}} & \subds{\underline{1.51}} & \subds{\underline{1.86}} & \subds{1.05} & \subds{1.32} & \subds{1.38} & \subds{1.64} & \subds{2.24} &  \\[\bucketgapOverlap]
 & \multicolumn{5}{c}{\subds{\bucketrow{50.00}{25.00}{25.00}{0.00}}} & \multicolumn{5}{c}{\subds{\bucketrow{75.00}{25.00}{0.00}{0.00}}} & \\
\multicolumn{1}{r}{\subds{ImageCAS}} & \subds{\underline{0.52}} & \subds{0.77} & \subds{0.84} & \subds{1.07} & \subds{2.09} & \subds{0.69} & \subds{0.99} & \subds{1.14} & \subds{1.56} & \subds{3.02} &  \\[\bucketgapOverlap]
 & \multicolumn{5}{c}{\subds{\bucketrow{74.00}{20.00}{6.00}{0.00}}} & \multicolumn{5}{c}{\subds{\bucketrow{72.00}{24.00}{4.00}{0.00}}} & \\
\addlinespace[\blockgapOverlap]
32 & \underline{0.52} & \underline{0.67} & 0.84 & 1.07 & 2.58 & \textbf{0.54} & \underline{0.86} & \textbf{0.98} & \underline{1.35} & \underline{2.48} & 33.3 \\[\bucketgapOverlap]
 & \multicolumn{5}{c}{\bucketrow{74.14}{20.69}{5.17}{0.00}} & \multicolumn{5}{c}{\bucketrow{81.03}{18.97}{0.00}{0.00}} & \\
\multicolumn{1}{r}{\subds{ASOCA}} & \subds{\underline{0.55}} & \subds{\textbf{0.75}} & \subds{\underline{0.82}} & \subds{1.03} & \subds{1.47} & \subds{0.57} & \subds{0.64} & \subds{0.61} & \subds{0.67} & \subds{\textbf{0.71}} &  \\[\bucketgapOverlap]
 & \multicolumn{5}{c}{\subds{\bucketrow{75.00}{25.00}{0.00}{0.00}}} & \multicolumn{5}{c}{\subds{\bucketrow{100.00}{0.00}{0.00}{0.00}}} & \\
\multicolumn{1}{r}{\subds{CAT08}} & \subds{0.67} & \subds{1.08} & \subds{1.22} & \subds{1.62} & \subds{2.08} & \subds{1.05} & \subds{1.34} & \subds{1.26} & \subds{\textbf{1.55}} & \subds{\textbf{1.84}} &  \\[\bucketgapOverlap]
 & \multicolumn{5}{c}{\subds{\bucketrow{50.00}{25.00}{25.00}{0.00}}} & \multicolumn{5}{c}{\subds{\bucketrow{50.00}{50.00}{0.00}{0.00}}} & \\
\multicolumn{1}{r}{\subds{ImageCAS}} & \subds{\textbf{0.50}} & \subds{\underline{0.66}} & \subds{\underline{0.81}} & \subds{\underline{0.98}} & \subds{2.58} & \subds{\textbf{0.54}} & \subds{\textbf{0.87}} & \subds{\textbf{0.99}} & \subds{\underline{1.35}} & \subds{\underline{2.48}} &  \\[\bucketgapOverlap]
 & \multicolumn{5}{c}{\subds{\bucketrow{76.00}{20.00}{4.00}{0.00}}} & \multicolumn{5}{c}{\subds{\bucketrow{82.00}{18.00}{0.00}{0.00}}} & \\
\addlinespace[\blockgapOverlap]
64 & 0.53 & 0.74 & \underline{0.83} & \underline{1.02} & \underline{2.07} & \underline{0.61} & \textbf{0.85} & \textbf{0.98} & \textbf{1.32} & 2.51 & 55.6 \\[\bucketgapOverlap]
 & \multicolumn{5}{c}{\bucketrow{77.59}{20.69}{1.72}{0.00}} & \multicolumn{5}{c}{\bucketrow{84.48}{15.52}{0.00}{0.00}} & \\
\multicolumn{1}{r}{\subds{ASOCA}} & \subds{\textbf{0.53}} & \subds{0.95} & \subds{0.95} & \subds{1.37} & \subds{1.45} & \subds{\textbf{0.54}} & \subds{\textbf{0.60}} & \subds{\textbf{0.58}} & \subds{\textbf{0.64}} & \subds{\textbf{0.71}} &  \\[\bucketgapOverlap]
 & \multicolumn{5}{c}{\subds{\bucketrow{50.00}{50.00}{0.00}{0.00}}} & \multicolumn{5}{c}{\subds{\bucketrow{100.00}{0.00}{0.00}{0.00}}} & \\
\multicolumn{1}{r}{\subds{CAT08}} & \subds{\textbf{0.60}} & \subds{\textbf{1.02}} & \subds{\textbf{1.04}} & \subds{\textbf{1.46}} & \subds{\textbf{1.75}} & \subds{\textbf{0.59}} & \subds{\textbf{1.04}} & \subds{\textbf{1.20}} & \subds{1.64} & \subds{2.25} &  \\[\bucketgapOverlap]
 & \multicolumn{5}{c}{\subds{\bucketrow{50.00}{50.00}{0.00}{0.00}}} & \multicolumn{5}{c}{\subds{\bucketrow{75.00}{25.00}{0.00}{0.00}}} & \\
\multicolumn{1}{r}{\subds{ImageCAS}} & \subds{0.54} & \subds{0.74} & \subds{\underline{0.81}} & \subds{\underline{0.98}} & \subds{\underline{2.07}} & \subds{0.62} & \subds{\underline{0.89}} & \subds{\underline{1.00}} & \subds{\textbf{1.32}} & \subds{2.51} &  \\[\bucketgapOverlap]
 & \multicolumn{5}{c}{\subds{\bucketrow{82.00}{16.00}{2.00}{0.00}}} & \multicolumn{5}{c}{\subds{\bucketrow{84.00}{16.00}{0.00}{0.00}}} & \\
\addlinespace[\blockgapOverlap]
96 & \textbf{0.51} & \textbf{0.66} & \textbf{0.80} & \textbf{0.97} & \textbf{2.06} & \underline{0.61} & 0.87 & \underline{1.00} & 1.36 & \textbf{2.46} & 119.3 \\[\bucketgapOverlap]
 & \multicolumn{5}{c}{\bucketrow{79.31}{17.24}{3.45}{0.00}} & \multicolumn{5}{c}{\bucketrow{77.59}{22.41}{0.00}{0.00}} & \\
\multicolumn{1}{r}{\subds{ASOCA}} & \subds{0.60} & \subds{\underline{0.77}} & \subds{\textbf{0.81}} & \subds{\textbf{0.99}} & \subds{\underline{1.12}} & \subds{\underline{0.56}} & \subds{\underline{0.62}} & \subds{\underline{0.59}} & \subds{\underline{0.65}} & \subds{\underline{0.72}} &  \\[\bucketgapOverlap]
 & \multicolumn{5}{c}{\subds{\bucketrow{75.00}{25.00}{0.00}{0.00}}} & \multicolumn{5}{c}{\subds{\bucketrow{100.00}{0.00}{0.00}{0.00}}} & \\
\multicolumn{1}{r}{\subds{CAT08}} & \subds{\underline{0.66}} & \subds{1.06} & \subds{1.16} & \subds{1.56} & \subds{1.88} & \subds{\underline{0.84}} & \subds{\underline{1.18}} & \subds{\underline{1.22}} & \subds{\underline{1.57}} & \subds{\underline{1.88}} &  \\[\bucketgapOverlap]
 & \multicolumn{5}{c}{\subds{\bucketrow{50.00}{25.00}{25.00}{0.00}}} & \multicolumn{5}{c}{\subds{\bucketrow{50.00}{50.00}{0.00}{0.00}}} & \\
\multicolumn{1}{r}{\subds{ImageCAS}} & \subds{\textbf{0.50}} & \subds{\textbf{0.64}} & \subds{\textbf{0.77}} & \subds{\textbf{0.95}} & \subds{\textbf{2.06}} & \subds{\underline{0.58}} & \subds{\underline{0.89}} & \subds{1.01} & \subds{1.36} & \subds{\textbf{2.46}} &  \\[\bucketgapOverlap]
 & \multicolumn{5}{c}{\subds{\bucketrow{82.00}{16.00}{2.00}{0.00}}} & \multicolumn{5}{c}{\subds{\bucketrow{78.00}{22.00}{0.00}{0.00}}} & \\
\midrule
\multicolumn{12}{l}{\textit{\nnresunet}} \\
\midrule
0 & \underline{0.52} & 0.76 & 1.11 & 1.09 & 16.93 & 0.70 & 0.93 & 1.11 & 1.43 & 2.98 & 29.3 \\[\bucketgapOverlap]
 & \multicolumn{5}{c}{\bucketrow{74.14}{20.69}{3.45}{1.72}} & \multicolumn{5}{c}{\bucketrow{75.86}{20.69}{3.45}{0.00}} & \\
\multicolumn{1}{r}{\subds{ASOCA}} & \subds{\textbf{0.54}} & \subds{\underline{0.74}} & \subds{\textbf{0.69}} & \subds{0.89} & \subds{\underline{0.97}} & \subds{0.59} & \subds{0.74} & \subds{\textbf{0.62}} & \subds{\textbf{0.77}} & \subds{\textbf{0.85}} &  \\[\bucketgapOverlap]
 & \multicolumn{5}{c}{\subds{\bucketrow{100.00}{0.00}{0.00}{0.00}}} & \multicolumn{5}{c}{\subds{\bucketrow{100.00}{0.00}{0.00}{0.00}}} & \\
\multicolumn{1}{r}{\subds{CAT08}} & \subds{0.94} & \subds{1.33} & \subds{1.29} & \subds{\underline{1.68}} & \subds{\textbf{1.74}} & \subds{1.06} & \subds{1.22} & \subds{1.22} & \subds{1.38} & \subds{1.63} &  \\[\bucketgapOverlap]
 & \multicolumn{5}{c}{\subds{\bucketrow{50.00}{50.00}{0.00}{0.00}}} & \multicolumn{5}{c}{\subds{\bucketrow{75.00}{25.00}{0.00}{0.00}}} & \\
\multicolumn{1}{r}{\subds{ImageCAS}} & \subds{0.50} & \subds{\underline{0.71}} & \subds{1.13} & \subds{1.08} & \subds{16.93} & \subds{0.67} & \subds{0.93} & \subds{\underline{1.14}} & \subds{1.49} & \subds{2.98} &  \\[\bucketgapOverlap]
 & \multicolumn{5}{c}{\subds{\bucketrow{74.00}{20.00}{4.00}{2.00}}} & \multicolumn{5}{c}{\subds{\bucketrow{74.00}{22.00}{4.00}{0.00}}} & \\
\addlinespace[\blockgapOverlap]
32 & 0.57 & 0.76 & 0.85 & \underline{1.06} & 2.07 & \textbf{0.65} & \underline{0.88} & \textbf{0.95} & \textbf{1.18} & \underline{2.27} & 31.3 \\[\bucketgapOverlap]
 & \multicolumn{5}{c}{\bucketrow{75.86}{20.69}{3.45}{0.00}} & \multicolumn{5}{c}{\bucketrow{87.93}{12.07}{0.00}{0.00}} & \\
\multicolumn{1}{r}{\subds{ASOCA}} & \subds{\underline{0.61}} & \subds{0.75} & \subds{0.88} & \subds{1.02} & \subds{1.44} & \subds{\textbf{0.55}} & \subds{\underline{0.72}} & \subds{0.68} & \subds{0.84} & \subds{\underline{1.04}} &  \\[\bucketgapOverlap]
 & \multicolumn{5}{c}{\subds{\bucketrow{75.00}{25.00}{0.00}{0.00}}} & \multicolumn{5}{c}{\subds{\bucketrow{100.00}{0.00}{0.00}{0.00}}} & \\
\multicolumn{1}{r}{\subds{CAT08}} & \subds{\underline{0.74}} & \subds{\underline{1.20}} & \subds{1.22} & \subds{\underline{1.68}} & \subds{\textbf{1.74}} & \subds{\textbf{0.90}} & \subds{\textbf{1.07}} & \subds{\textbf{0.99}} & \subds{\textbf{1.17}} & \subds{1.31} &  \\[\bucketgapOverlap]
 & \multicolumn{5}{c}{\subds{\bucketrow{50.00}{50.00}{0.00}{0.00}}} & \multicolumn{5}{c}{\subds{\bucketrow{100.00}{0.00}{0.00}{0.00}}} & \\
\multicolumn{1}{r}{\subds{ImageCAS}} & \subds{0.51} & \subds{0.76} & \subds{0.82} & \subds{\underline{1.03}} & \subds{2.07} & \subds{\textbf{0.65}} & \subds{\underline{0.88}} & \subds{\textbf{0.97}} & \subds{\textbf{1.21}} & \subds{\underline{2.27}} &  \\[\bucketgapOverlap]
 & \multicolumn{5}{c}{\subds{\bucketrow{78.00}{18.00}{4.00}{0.00}}} & \multicolumn{5}{c}{\subds{\bucketrow{86.00}{14.00}{0.00}{0.00}}} & \\
\addlinespace[\blockgapOverlap]
64 & \underline{0.52} & \textbf{0.70} & \underline{0.83} & 1.13 & \underline{2.03} & 0.68 & \textbf{0.87} & \underline{0.96} & \underline{1.21} & 2.32 & 41.1 \\[\bucketgapOverlap]
 & \multicolumn{5}{c}{\bucketrow{72.41}{24.14}{3.45}{0.00}} & \multicolumn{5}{c}{\bucketrow{89.66}{10.34}{0.00}{0.00}} & \\
\multicolumn{1}{r}{\subds{ASOCA}} & \subds{\underline{0.61}} & \subds{\textbf{0.67}} & \subds{0.77} & \subds{\textbf{0.83}} & \subds{1.20} & \subds{\underline{0.58}} & \subds{0.74} & \subds{0.67} & \subds{0.83} & \subds{1.06} &  \\[\bucketgapOverlap]
 & \multicolumn{5}{c}{\subds{\bucketrow{75.00}{25.00}{0.00}{0.00}}} & \multicolumn{5}{c}{\subds{\bucketrow{100.00}{0.00}{0.00}{0.00}}} & \\
\multicolumn{1}{r}{\subds{CAT08}} & \subds{\textbf{0.68}} & \subds{1.22} & \subds{\textbf{1.17}} & \subds{1.71} & \subds{\textbf{1.74}} & \subds{1.03} & \subds{1.17} & \subds{\underline{1.05}} & \subds{\underline{1.20}} & \subds{\underline{1.23}} &  \\[\bucketgapOverlap]
 & \multicolumn{5}{c}{\subds{\bucketrow{50.00}{50.00}{0.00}{0.00}}} & \multicolumn{5}{c}{\subds{\bucketrow{100.00}{0.00}{0.00}{0.00}}} & \\
\multicolumn{1}{r}{\subds{ImageCAS}} & \subds{\textbf{0.48}} & \subds{\textbf{0.70}} & \subds{\underline{0.80}} & \subds{1.07} & \subds{\underline{2.03}} & \subds{0.68} & \subds{\textbf{0.87}} & \subds{\textbf{0.97}} & \subds{\underline{1.23}} & \subds{2.32} &  \\[\bucketgapOverlap]
 & \multicolumn{5}{c}{\subds{\bucketrow{74.00}{22.00}{4.00}{0.00}}} & \multicolumn{5}{c}{\subds{\bucketrow{88.00}{12.00}{0.00}{0.00}}} & \\
\addlinespace[\blockgapOverlap]
96 & \textbf{0.50} & \underline{0.71} & \textbf{0.81} & \textbf{0.97} & \textbf{2.01} & \underline{0.66} & \underline{0.88} & \underline{0.96} & \underline{1.21} & \textbf{2.11} & 93.4 \\[\bucketgapOverlap]
 & \multicolumn{5}{c}{\bucketrow{77.59}{18.97}{3.45}{0.00}} & \multicolumn{5}{c}{\bucketrow{87.93}{12.07}{0.00}{0.00}} & \\
\multicolumn{1}{r}{\subds{ASOCA}} & \subds{\underline{0.61}} & \subds{0.75} & \subds{\underline{0.74}} & \subds{\underline{0.88}} & \subds{\textbf{0.94}} & \subds{\textbf{0.55}} & \subds{\textbf{0.69}} & \subds{\underline{0.65}} & \subds{\underline{0.79}} & \subds{\underline{1.04}} &  \\[\bucketgapOverlap]
 & \multicolumn{5}{c}{\subds{\bucketrow{100.00}{0.00}{0.00}{0.00}}} & \multicolumn{5}{c}{\subds{\bucketrow{100.00}{0.00}{0.00}{0.00}}} & \\
\multicolumn{1}{r}{\subds{CAT08}} & \subds{0.75} & \subds{\textbf{1.15}} & \subds{\underline{1.21}} & \subds{\textbf{1.61}} & \subds{\underline{1.83}} & \subds{\underline{0.99}} & \subds{\underline{1.15}} & \subds{\underline{1.05}} & \subds{1.21} & \subds{\textbf{1.21}} &  \\[\bucketgapOverlap]
 & \multicolumn{5}{c}{\subds{\bucketrow{50.00}{50.00}{0.00}{0.00}}} & \multicolumn{5}{c}{\subds{\bucketrow{100.00}{0.00}{0.00}{0.00}}} & \\
\multicolumn{1}{r}{\subds{ImageCAS}} & \subds{\underline{0.49}} & \subds{\textbf{0.70}} & \subds{\textbf{0.79}} & \subds{\textbf{0.97}} & \subds{\textbf{2.01}} & \subds{\underline{0.66}} & \subds{\underline{0.88}} & \subds{\textbf{0.97}} & \subds{1.26} & \subds{\textbf{2.11}} &  \\[\bucketgapOverlap]
 & \multicolumn{5}{c}{\subds{\bucketrow{78.00}{18.00}{4.00}{0.00}}} & \multicolumn{5}{c}{\subds{\bucketrow{86.00}{14.00}{0.00}{0.00}}} & \\
\bottomrule
\end{tabular}
\end{table}

\subsection{Ablation: training on ImageCAS only (external validation)}
\label{subsec:app-imagecas-only}
The models of the benchmark were trained on ImageCAS, CAT08 and ASOCA, and 64 of the 72 volumes of CAT08 and ASOCA were used for training or validation, so their results on these two datasets do not measure the generalization to an unseen domain. To measure it, \swinv2 and \nnresunet were trained again with the recipe of the benchmark (same architecture, loss, spacing of 0.5~mm and hyperparameters) on the 950 ImageCAS images of the training and validation sets only; no volume of ASOCA or CAT08 was used for training, validation or model selection. The models were tested on all the 40 ASOCA and 32 CAT08 volumes, and on the 50 ImageCAS test volumes as the in-domain reference. The 8 volumes of ASOCA and CAT08 that belong to the test set of the benchmark were not used for training by any model, so on these volumes the two kinds of training can be compared directly. The results are reported in Table~\ref{tab:app-imagecas-only}.

\swinv2 trained on ImageCAS only localized all the ostia of the two external datasets but one. On CAT08, the median error was 0.77~mm for the right and 0.89~mm for the left ostium, with no failure and 62.50\% and 81.25\% of Optimal localizations, respectively. On ASOCA the median errors were 1.18~mm (right) and 1.03~mm (left), 40.00\% and 80.00\% of the localizations were Optimal, and one left ostium failed (2.50\%, with an error of 118.08~mm, which raises the mean error of the left ostium to 4.05~mm); at patient level, 37.50\% of the ASOCA volumes were Optimal and 80.00\% were Optimal or Good. On the ImageCAS test volumes the median errors were 0.72~mm and 0.91~mm, with 86.00\% and 88.00\% of Optimal localizations and no failure, almost identical to those of the model trained on the three datasets (0.74~mm and 0.89~mm). On the 8 volumes unseen by every model, the median errors of the model trained on ImageCAS only were 1.08~mm (right) and 0.67~mm (left) for ASOCA and 1.42~mm and 1.00~mm for CAT08, against 0.95~mm and 0.60~mm, and 1.02~mm and 1.04~mm, for the model trained on the three datasets; no ostium failed in either case.

\nnresunet trained on ImageCAS only gave errors of the same order on the external datasets, with two failures. On CAT08, the median errors were 0.89~mm for the right and 0.90~mm for the left ostium, with 65.62\% and 71.88\% of Optimal localizations; one right ostium failed (3.12\%, with an error of 169.4~mm, which raises the mean error of the right ostium to 6.23~mm). On ASOCA the median errors were 1.14~mm (right) and 0.97~mm (left), 42.50\% and 72.50\% of the localizations were Optimal, and one left ostium failed (2.50\%, with an error of 141.1~mm, mean 4.64~mm). The two failures belong to different volumes, so 70 of the 72 external volumes were localized without any failure. On the ImageCAS test volumes the median errors were 0.66~mm and 0.84~mm, with 80.00\% and 82.00\% of Optimal localizations and no failure, against 0.70~mm and 0.87~mm for the model trained on the three datasets. On the 8 volumes unseen by every model, the median errors of the model trained on ImageCAS only were 0.84~mm (right) and 0.62~mm (left) for ASOCA and 1.57~mm and 0.79~mm for CAT08, against 0.67~mm and 0.74~mm, and 1.22~mm and 1.17~mm, for the model trained on the three datasets; no ostium failed in either case.

\paragraph{Interpretation} The errors on CAT08 and ASOCA were of the same order as on ImageCAS, although the model had never seen a volume of these datasets, which come from other scanners and centers and whose ostia were defined with other operational procedures (Section~\ref{subsec:ostia-dataset}): the annotations of ImageCAS transfer to them. The errors were slightly larger than on ImageCAS, and the right ostium of ASOCA and of CAT08 was less often Optimal, as expected for a change of domain. The failures were isolated: a single left ostium of ASOCA for \swinv2, the dataset that differs most from ImageCAS in acquisition and the one that failed most often in the training set size ablation (\ref{subsec:app-datasize}), and one left ostium of ASOCA and one right ostium of CAT08 for \nnresunet. In all three cases the error was of the order of a hundred millimeters or more, which suggests that a different structure was located, not that the ostium was imprecisely placed, while the remaining ostia of the same volumes and of the other volumes were localized correctly. The two models behaved similarly, without a consistent advantage of either one across datasets and sides. The comparison on the 8 unseen volumes is only descriptive, since each dataset contributes four volumes: the differences between the two kinds of training, of a few tenths of a millimeter, are within the variability expected with so few cases and with a single training per model. The experiment also suggests that, on ImageCAS, the volumes of CAT08 and ASOCA add little to the accuracy.
\begin{table}[H]
\centering
\scriptsize
\setlength{\tabcolsep}{1.5pt}
\newcommand{\bucketgapImageCASOnly}{-2pt}
\newcommand{\blockgapImageCASOnly}{5pt}
\caption{External validation: Euclidean distance $d_c$ (mm) of \swinv2 and \nnresunet trained on the ImageCAS images only (training and validation sets), with no volume of ASOCA or CAT08 used for training, validation or model selection. Upper part: all the volumes of ASOCA (40) and CAT08 (32), and the 50 ImageCAS test volumes as the in-domain reference. Lower part: the 8 volumes of ASOCA (4) and CAT08 (4) that belong to the test set of the benchmark, compared with the models of the benchmark trained on the three datasets (Table~\ref{tab:res-1000}); these 8 volumes were not used for training by either kind of model. The line below each row gives the share of cases in each accuracy bucket (O = Optimal, G = Good, A = Acceptable, F = Failed).}
\label{tab:app-imagecas-only}
\begin{tabular}{l *{10}{r}}
\toprule
 & \multicolumn{5}{c}{Right ostium} & \multicolumn{5}{c}{Left ostium} \\
\cmidrule(lr){2-6} \cmidrule(lr){7-11}
 & p25 & Med. & Mean & p75 & Max & p25 & Med. & Mean & p75 & Max \\
\midrule
\multicolumn{11}{l}{\textit{Trained on ImageCAS only, all external volumes}} \\
\addlinespace[2pt]
\multicolumn{11}{l}{\swinv2} \\
\multicolumn{1}{r}{ASOCA (40)} & 0.90 & 1.18 & 1.27 & 1.55 & 2.65 & 0.67 & 1.03 & 4.05 & 1.35 & 118.1 \\[\bucketgapImageCASOnly]
 & \multicolumn{5}{c}{\bucketrow{40.00}{47.50}{12.50}{0.00}} & \multicolumn{5}{c}{\bucketrow{80.00}{12.50}{5.00}{2.50}} \\
\multicolumn{1}{r}{CAT08 (32)} & 0.61 & 0.77 & 0.96 & 1.20 & 2.65 & 0.74 & 0.89 & 1.18 & 1.41 & 3.49 \\[\bucketgapImageCASOnly]
 & \multicolumn{5}{c}{\bucketrow{62.50}{31.25}{6.25}{0.00}} & \multicolumn{5}{c}{\bucketrow{81.25}{9.38}{9.38}{0.00}} \\
\multicolumn{1}{r}{ImageCAS test (50)} & 0.56 & 0.72 & 0.78 & 0.88 & 1.68 & 0.48 & 0.91 & 0.92 & 1.18 & 2.34 \\[\bucketgapImageCASOnly]
 & \multicolumn{5}{c}{\bucketrow{86.00}{14.00}{0.00}{0.00}} & \multicolumn{5}{c}{\bucketrow{88.00}{12.00}{0.00}{0.00}} \\
\addlinespace[\blockgapImageCASOnly]
\multicolumn{11}{l}{\nnresunet} \\
\multicolumn{1}{r}{ASOCA (40)} & 0.86 & 1.14 & 1.20 & 1.43 & 2.30 & 0.64 & 0.97 & 4.64 & 1.48 & 141.1 \\[\bucketgapImageCASOnly]
 & \multicolumn{5}{c}{\bucketrow{42.50}{47.50}{10.00}{0.00}} & \multicolumn{5}{c}{\bucketrow{72.50}{17.50}{7.50}{2.50}} \\
\multicolumn{1}{r}{CAT08 (32)} & 0.63 & 0.89 & 6.23 & 1.27 & 169.4 & 0.64 & 0.90 & 1.12 & 1.46 & 3.46 \\[\bucketgapImageCASOnly]
 & \multicolumn{5}{c}{\bucketrow{65.62}{28.12}{3.12}{3.12}} & \multicolumn{5}{c}{\bucketrow{71.88}{21.88}{6.25}{0.00}} \\
\multicolumn{1}{r}{ImageCAS test (50)} & 0.49 & 0.66 & 0.76 & 0.94 & 1.87 & 0.56 & 0.84 & 0.93 & 1.10 & 2.85 \\[\bucketgapImageCASOnly]
 & \multicolumn{5}{c}{\bucketrow{80.00}{18.00}{2.00}{0.00}} & \multicolumn{5}{c}{\bucketrow{82.00}{16.00}{2.00}{0.00}} \\
\midrule
\multicolumn{11}{l}{\textit{The 8 volumes unseen by every model (ASOCA 4, CAT08 4)}} \\
\addlinespace[2pt]
\multicolumn{11}{l}{\swinv2} \\
\multicolumn{1}{r}{ASOCA, ImageCAS only} & 0.93 & 1.08 & 1.15 & 1.30 & 1.62 & 0.59 & 0.67 & 0.67 & 0.75 & 0.92 \\[\bucketgapImageCASOnly]
 & \multicolumn{5}{c}{\bucketrow{50.00}{50.00}{0.00}{0.00}} & \multicolumn{5}{c}{\bucketrow{100.00}{0.00}{0.00}{0.00}} \\
\multicolumn{1}{r}{ASOCA, three datasets} & 0.53 & 0.95 & 0.95 & 1.37 & 1.45 & 0.54 & 0.60 & 0.58 & 0.64 & 0.71 \\[\bucketgapImageCASOnly]
 & \multicolumn{5}{c}{\bucketrow{50.00}{50.00}{0.00}{0.00}} & \multicolumn{5}{c}{\bucketrow{100.00}{0.00}{0.00}{0.00}} \\
\multicolumn{1}{r}{CAT08, ImageCAS only} & 1.05 & 1.42 & 1.36 & 1.73 & 1.95 & 0.66 & 1.00 & 0.96 & 1.30 & 1.48 \\[\bucketgapImageCASOnly]
 & \multicolumn{5}{c}{\bucketrow{25.00}{50.00}{25.00}{0.00}} & \multicolumn{5}{c}{\bucketrow{75.00}{25.00}{0.00}{0.00}} \\
\multicolumn{1}{r}{CAT08, three datasets} & 0.60 & 1.02 & 1.04 & 1.46 & 1.75 & 0.59 & 1.04 & 1.20 & 1.64 & 2.25 \\[\bucketgapImageCASOnly]
 & \multicolumn{5}{c}{\bucketrow{50.00}{50.00}{0.00}{0.00}} & \multicolumn{5}{c}{\bucketrow{75.00}{25.00}{0.00}{0.00}} \\
\addlinespace[\blockgapImageCASOnly]
\multicolumn{11}{l}{\nnresunet} \\
\multicolumn{1}{r}{ASOCA, ImageCAS only} & 0.70 & 0.84 & 0.96 & 1.09 & 1.55 & 0.57 & 0.62 & 0.62 & 0.67 & 0.74 \\[\bucketgapImageCASOnly]
 & \multicolumn{5}{c}{\bucketrow{75.00}{25.00}{0.00}{0.00}} & \multicolumn{5}{c}{\bucketrow{100.00}{0.00}{0.00}{0.00}} \\
\multicolumn{1}{r}{ASOCA, three datasets} & 0.61 & 0.67 & 0.77 & 0.83 & 1.20 & 0.58 & 0.74 & 0.67 & 0.83 & 1.06 \\[\bucketgapImageCASOnly]
 & \multicolumn{5}{c}{\bucketrow{75.00}{25.00}{0.00}{0.00}} & \multicolumn{5}{c}{\bucketrow{100.00}{0.00}{0.00}{0.00}} \\
\multicolumn{1}{r}{CAT08, ImageCAS only} & 1.30 & 1.57 & 1.38 & 1.64 & 1.68 & 0.62 & 0.79 & 0.89 & 1.06 & 1.50 \\[\bucketgapImageCASOnly]
 & \multicolumn{5}{c}{\bucketrow{25.00}{75.00}{0.00}{0.00}} & \multicolumn{5}{c}{\bucketrow{75.00}{25.00}{0.00}{0.00}} \\
\multicolumn{1}{r}{CAT08, three datasets} & 0.68 & 1.22 & 1.17 & 1.71 & 1.74 & 1.03 & 1.17 & 1.05 & 1.20 & 1.23 \\[\bucketgapImageCASOnly]
 & \multicolumn{5}{c}{\bucketrow{50.00}{50.00}{0.00}{0.00}} & \multicolumn{5}{c}{\bucketrow{100.00}{0.00}{0.00}{0.00}} \\
\bottomrule
\end{tabular}
\end{table}

\FloatBarrier
\section{Comparison with Published Methods}
\label{sec:app-literature}
Table~\ref{tab:app-comparison} lists the errors reported in the literature for coronary ostium localization. We did not reimplement the models or the pipelines of these works: the values are copied from their manuscripts, solely to give the reader a reference. The works were evaluated on different data, mostly private cohorts of different size and acquisition protocols, and with different operational definitions of the ostium, so that a direct comparison with the results of this work is of limited meaning. A public benchmark with a shared test set, such as the one released with this work (Table~\ref{tab:res-1000}), is therefore needed. The mean error of the two ostia combined (Both) of the published methods ranges between 1.2 and 2.1~mm on the data of the respective works, while the earlier LocNeXt model, which was trained and tested on the small CAT08 and ASOCA datasets, reached 7.93~mm (median 5.24~mm).

\begin{table}[H]
\centering
\fontsize{8}{10}\selectfont
\setlength{\tabcolsep}{3pt}
\caption{Euclidean distance $d_c$ (mm), as mean and median (Med.), reported in the manuscripts of previously published methods for coronary ostium localization. The data used by each work, which differ from ours and are mostly private, are indicated in gray below the reference (number of volumes in brackets). The methods were not reimplemented and the values are copied from the respective works; a dash indicates a statistic that the work does not report. Both: combined right and left values. The last row is our earlier published model, evaluated on its own test split.}
\label{tab:app-comparison}
\begin{tabular}{l *{6}{r}}
\toprule
 & \multicolumn{2}{c}{Right ostium} & \multicolumn{2}{c}{Left ostium} & \multicolumn{2}{c}{Both} \\
\cmidrule(lr){2-3} \cmidrule(lr){4-5} \cmidrule(lr){6-7}
\makecell[l]{Method} & Mean & Med. & Mean & Med. & Mean & Med. \\
\midrule
\makecell[tl]{\citet{zheng2011ostia} \\ \subds{\fontsize{6}{7}\selectfont Private (1360)}} & 1.73 & 1.31 & 1.66 & 1.45 & 1.70 & - \\
\makecell[tl]{\citet{jang2021ostia} \\ \subds{\fontsize{6}{7}\selectfont CAT08}} & 1.94 & 1.80 & 2.22 & 1.70 & 2.08 & 1.59 \\
\makecell[tl]{\citet{wolterink2020landmarks} \\ \subds{\fontsize{6}{7}\selectfont Private (672)}} & - & 1.45 & - & 1.55 & - & - \\
\makecell[tl]{\citet{gajowczyk2022ostia} \\ \subds{\fontsize{6}{7}\selectfont Private (201)}} & 1.29 & 0.98 & 1.18 & 1.14 & 1.24 & - \\
\makecell[tl]{\citet{locnext_embc} \\ \subds{\fontsize{6}{7}\selectfont ASOCA (40), CAT08 (32)}} & 5.28 & 4.16 & 10.57 & 7.79 & 7.93 & 5.24 \\
\bottomrule
\end{tabular}
\end{table}

\FloatBarrier
\section{Comparison with the ImageCAS-X Centerlines}
\label{sec:app-centerlines-imagecasx}
The extraction of Section~\ref{subsec:centerline-methods} was validated against ASOCA in Section~\ref{subsec:res-centerlines}. As a second check on a much larger sample, we ran the same extraction on the segmentation masks of the 800 ImageCAS scans that are part of ImageCAS-X~\citep{imagecasx2026}, and compared the result with the centerlines released with that dataset, using the same metrics as for ASOCA, with a fixed matching tolerance of 2~mm, since the ImageCAS-X centerlines carry no radius. The ImageCAS-X masks were manually corrected, but their centerlines were then derived from them automatically, by thinning followed by smoothing~\citep{imagecasx2026}; this comparison is therefore not independent of the masks, and it measures how closely two automatic procedures agree on the same curated masks, not the accuracy against an external ground truth.

The extraction completed in all 800 patients, and the agreement is reported in Table~\ref{tab:centerlines-imagecasx}. The extracted trees cover a median of 100\% of the reference length (recall; at least 98.4\% in every patient), and a median of 99.9\% of the extracted length is matched by the reference (precision; at least 98.6\% in every patient). The quality control of the comparison raised no discrepancy in any patient: no reference branch or distal portion of at least 10~mm was missed or truncated, no gap was found and no extra branch of at least 10~mm was found. The extracted centerlines lie at a median average distance of 0.23~mm from the reference, with a 95th-percentile Hausdorff distance of 0.50~mm, and are slightly shorter (median length ratio 0.97). The endpoints of the reference were paired with those of the extracted trees, and vice versa, in the large majority of the patients (median endpoint recall and precision of 100\%), at a median distance of 1.21~mm, and the median share of reference bifurcations found within 5~mm is 100\%. Results are similar for the left and the right tree.

Beyond the agreement of the paths, the two sets of centerlines differ in how they are rooted and connected, and in the masks they follow (Table~\ref{tab:imagecasx-differences}). The ImageCAS-X start points are the skeleton endpoints within 5~mm of an automatically segmented aorta~\citep{imagecasx2026}. They lie at a median of 2.45~mm (left) and 1.07~mm (right) from our annotated ostia; 55.5\% (left) and 83.5\% (right) of them lie within the interoperator thresholds of Table~\ref{tab:thresholds}, and 1.0\% and 0.4\% more than 5~mm away: the two origins agree closely on the right and less on the left, and only ours is an annotated landmark, the one on which the localization models are trained and evaluated.

Our centerlines are stored as graphs: the connectivity from the ostium to every endpoint is explicit, and every node carries its attributes, namely the radius, the topological type and the side. Polyline formats leave the connectivity implicit, to be reconstructed from the geometry: ASOCA stores one polyline from the inlet to each outlet, so that the shared proximal segments are repeated, while ImageCAS-X stores the segments between bifurcations separately, a segment being joined to its parent only where their ends coincide. In 15.6\% of the ImageCAS-X trees (28.6\% of the left and 2.5\% of the right ones), i.e.\ in 30.4\% of the scans, at least one segment starts more than 1~mm away from its parent, so that the part of the tree downstream of it, a median of 47.2\% of the tree length, cannot be reached from the start point. These gaps are short, with a median of 2.2~mm (5th--95th percentile 1.2--4.0~mm), and the ImageCAS-X mask is continuous across them for 97.7\% of the detached length; joining the ends closer than 3 or 5~mm reconnects all but 44 and 5 of the 249 split trees, respectively (Table~\ref{tab:imagecasx-gaps}). The reconstruction is therefore simple, but it depends on a tolerance chosen by the user, which a graph makes unnecessary.

The two sets of masks also differ. A median of 8.8\% of the ImageCAS-X centerline length lies outside the official masks, so that the ImageCAS-X centerlines cannot serve as the centerlines of those masks; conversely, a median of 18.7\% of the length extracted from the official masks lies outside the ImageCAS-X masks. On the official masks, the extracted trees are longer and branch more, markedly on the right (median length 251 against 198~mm, median 3 against 1 bifurcations; paired Wilcoxon signed-rank test, $p<0.001$), while they reach the same distal depth from the ostium. The extra length consists mostly of proximal side branches, with a median radius of 1.4~mm; a visual inspection of 14 of these branches, the three longest and four drawn at random on each side, judged 11 of them to be arteries, largely consistent with the side branches that the ImageCAS-X protocol leaves out by design, such as the septal perforators and the acute marginal and nodal arteries~\citep{imagecasx2026}. Users who train or evaluate on the official masks therefore find in our graphs the centerlines of those masks, side branches included.

Finally, the 200 scans excluded from ImageCAS-X as non-diagnostic are covered too. On them, the quality control flags 38\% of the patients, against 21\% of the other 800 (Fisher's exact test, $p<0.001$), and three of the four failed patients are among them (\ref{subsec:app-centerlines-imagecas}), so that the flags identify the cases to be used with care. The ostium localization, which does not depend on the masks, is not affected: on the 14 test scans among the excluded ones, none of the released models failed, and the median errors of \swinv2 were 0.65~mm (right) and 1.08~mm (left), against 0.77 and 0.85~mm on the other 36 test scans.

\begin{table}[tb]
\centering
\caption{Agreement between the centerlines extracted from the 800 ImageCAS-X masks and the ImageCAS-X centerlines, for both trees together and for each tree, as median [interquartile range] over the patients. The metrics are those of Table~\ref{tab:centerlines-asoca}; the overlap until first error and the bifurcation score are computed only over the patients in which they are defined (all of them for both trees together). The overlap of wide vessels and the radius errors are not reported, since the ImageCAS-X centerlines carry no radius.}
\label{tab:centerlines-imagecasx}
\fontsize{6.4}{7.5}\selectfont
\setlength{\tabcolsep}{3pt}
\begin{tabular}{lccc}
\toprule
 & Both trees & Left tree & Right tree \\
\midrule
Recall (\%)             & 100 [100--100] & 100 [100--100] & 100 [100--100] \\
Precision (\%)          & 99.9 [99.8--100] & 99.9 [99.7--100] & 100 [100--100] \\
OV (\%)                 & 99.9 [99.9--100] & 99.9 [99.8--100] & 100 [100--100] \\
OF (\%)                 & 100 [100--100] & 100 [100--100] & 100 [100--100] \\
AI (mm)                 & 0.23 [0.22--0.24] & 0.23 [0.22--0.25] & 0.22 [0.21--0.24] \\
HD95 (mm)               & 0.50 [0.47--0.53] & 0.50 [0.46--0.55] & 0.48 [0.44--0.52] \\
Length ratio            & 0.97 [0.97--0.98] & 0.97 [0.96--0.98] & 0.97 [0.96--0.98] \\
Endpoint recall (\%)    & 100 [100--100] & 100 [100--100] & 100 [100--100] \\
Endpoint precision (\%) & 100 [100--100] & 100 [100--100] & 100 [100--100] \\
Endpoint distance (mm)  & 1.21 [1.10--1.33] & 1.18 [1.05--1.33] & 1.30 [1.09--1.48] \\
Bifurcations $\leq$5~mm (\%) & 100 [100--100] & 100 [100--100] & 100 [100--100] \\
\bottomrule
\end{tabular}
\end{table}

\begin{table}[tb]
\centering
\caption{ImageCAS-X trees that remain split into pieces when the ends of their segments closer than the joining tolerance are joined, as a percentage (number) of the 800 scans and of the 800 left and 800 right trees.}
\label{tab:imagecasx-gaps}
\fontsize{6.4}{7.5}\selectfont
\setlength{\tabcolsep}{4pt}
\begin{tabular}{lccc}
\toprule
Joining tolerance (mm) & Scans & Left trees & Right trees \\
\midrule
1  & 30.4 (243) & 28.6 (229) & 2.5 (20) \\
2  & 18.1 (145) & 17.3 (138) & 1.3 (10) \\
3  & 5.4 (43)   & 5.1 (41)   & 0.4 (3) \\
5  & 0.6 (5)    & 0.6 (5)    & 0 (0) \\
10 & 0.1 (1)    & 0.1 (1)    & 0 (0) \\
\bottomrule
\end{tabular}
\end{table}

\FloatBarrier
\section{Data and Code Availability: Details}
\label{sec:app-data-availability}
This appendix gives the details of the statement in Section~\ref{sec:data-availability}.

\subsection{Third-party data}
\label{subsec:app-availability-thirdparty}
The CCTA volumes and coronary segmentation masks used in this work belong to the public ImageCAS~\citep{imagecas}, CAT08~\citep{cat08,cat08_2} and ASOCA~\citep{asoca1} datasets, and the alternative centerlines described below are extracted from the segmentation masks of ImageCAS-X~\citep{imagecasx2026}. They are not redistributed here. ImageCAS, ASOCA and ImageCAS-X must be obtained from their original sources, under their respective terms of use. As of 3~November 2026, ImageCAS is distributed on Kaggle under the CC BY-NC 4.0 license, ImageCAS-X (version~3) on Zenodo under the CC BY 4.0 license, and ASOCA through the UK Data Service, on request, under its End User Licence. The annotations, centerline graphs and model weights released here derive from these data, contain none of their images or masks, credit the original sources (the alternative centerlines require attribution to~\citet{imagecasx2026}), and are released under the CC BY-NC 4.0 license, in line with the non-commercial terms of ImageCAS. CAT08 is no longer distributed through its official website; however, having been openly available for almost two decades, it is still in use in the research community.

\subsection{Where the resources are}
\label{subsec:app-availability-where}
All the resources of this work, namely the code, the datasets and the pre-trained model weights, can be found at \url{https://hal9000-lab.github.io/}, where the page of this project (Awakening ImageCAS) gathers the links to each of them. The ostium and centerline annotations are deposited on Zenodo~\citep{awakening_dataset}, \url{https://doi.org/10.5281/zenodo.23243093}, the model weights in a separate Zenodo record~\citep{awakening_weights}, \url{https://doi.org/10.5281/zenodo.23399623}, and the code in a public GitHub repository~\citep{awakening_code}, \url{https://github.com/AAMIASoftwares-research/HCATOstiaLocalizationRelease}.

\subsection{Released material and licenses}
\label{subsec:app-availability-material}
The following material is released with this work: (i) the left and right coronary ostium annotations of all the 1000 ImageCAS volumes, stored as one file per volume that reports each landmark both in world and in voxel coordinates, along with the affine matrices relating the two; (ii) the coronary centerline graphs derived from the ImageCAS segmentation masks, one \texttt{networkx}-compatible GML file per patient, with the same trees in VTK format and as 3D Slicer markups, and a table reporting, for every patient, the outcome of the quality control and the recommended graph (\ref{subsec:app-centerline-release}); (iii) an alternative set of centerline graphs, in the same formats, extracted with the same procedure from the segmentation masks of ImageCAS-X~\citep{imagecasx2026} for its 800 patients, intended for users of the ImageCAS-X masks and recommended where the graph derived from the ImageCAS mask is defective; (iv) the weights of the \swinv2, ResNet50, ResNet101 and \nnresunet benchmark models (the two MedNeXt models are part of the benchmark, but, since they did not reach a usable accuracy, their weights are not released); (v) the code for training, inference and evaluation, and the lists of the training, validation and test volumes, so that every reported result can be reproduced and new methods can be compared on the same test set. The annotations and weights are released under the CC BY-NC 4.0 license; the code is released under the MIT license.

\end{document}